\documentclass[12pt,reqno]{amsart}

\usepackage{fullpage}
\usepackage{bm}
\usepackage{amsmath,amssymb,amsthm}
\usepackage{graphicx}
\usepackage{subfigure}

\begin{document}

\newtheorem{thm}{Theorem}
\newtheorem{cor}{Corollary}
\newtheorem{lem}{Lemma}
\newtheorem{prop}{Proposition}
\def\Ref#1{Ref.~\cite{#1}}

\def\const{\text{const.}}
\def\Rnum{{\mathbb R}}
\def\sgn{{\rm sgn}}

\def\t{\mathrm{t}}

\def\eff{\text{eff.}}
\def\peak{\text{peak}}
\def\lump{\text{lump}}
\def\solitary{\text{solitary}}

\def\sech{{\rm sech}}
\def\coth{{\rm coth}}
\def\csch{{\rm csch}}
\def\arctanh{{\rm arctanh}}
\def\arccosh{{\rm arccosh}}
\def\tzero{t_0}

\tolerance=80000
\allowdisplaybreaks[4]

\title{Long-time behaviour of sphalerons\\ in $\phi^4$ models with a false vacuum} 

\author{
Stephen C. Anco$^1$ \\
Danial Saadatmand$^2$
\\\\
${}^1$
D\lowercase{\scshape{epartment}} \lowercase{\scshape{of}} M\lowercase{\scshape{athematics and}} S\lowercase{\scshape{tatistics}}\\
B\lowercase{\scshape{rock}} U\lowercase{\scshape{niversity}}\\
S\lowercase{\scshape{t.}} C\lowercase{\scshape{atharines}}, C\lowercase{\scshape{anada}}
\\
${}^2$
S\lowercase{\scshape{chool}} \lowercase{\scshape{of}} M\lowercase{\scshape{athematics and}} S\lowercase{\scshape{tatistics}}\\
C\lowercase{\scshape{arleton}} U\lowercase{\scshape{niversity}}\\
O\lowercase{\scshape{ttawa}}, C\lowercase{\scshape{anada}}
}


\begin{abstract}
Evolution of sphalerons in a class of quartic Klein–Gordon models 
are studied under a growing perturbation. 
Sphalerons are unstable lump-like solutions that arise from a saddle point
between true and false vacua in the energy functional. 
Numerical simulations are presented which show
the sphaleron evolving into an accelerating kink-antikink pair 
whose separation increases in time and asymptotically approaches the speed of light. 
To explain this behaviour analytically,
a nonlinear collective coordinate method is developed 
which has three dynamical parameters and leads to an explicit asymptotic solution
using a power series expansion.
The solution describes the emergence of a spreading tabletop profile
whose height approaches the true vacuum while its flanks steepen and accelerate outward.
In addition, the energy density is shown to concentrate at the flanks,
indicating the onset of a gradient blow-up at large times.
These results provide a detailed description of 
the long-time dynamics of positively perturbed sphalerons, 
and reveal a universal mechanism for the formation of
relativistically expanding structures in nonlinear field theories.
\end{abstract}

\maketitle

\section{Introduction}

In nonlinear field theories, 
a \emph{sphaleron} is a static, finite energy, unstable solution that has a localized lump-like profile \cite{Shn2018,Man2019}. 
In general, such solutions arise whenever there is
a saddle point between true and false vacua in the energy functional of a theory. 

Understanding the nonlinear dynamics of sphalerons is important in many physical models, 
the most prominent being 
the Standard Model of particle physics \cite{Kun.Bri.1989,Yaf.1989,Kli.1990,Bri.Kun}
and the Skyrme model \cite{Isl.LeT.Par.1989,Kru.Sut.2004,Shn.Tch.2011,Shn.Zhi.2013}. 
In the electroweak sector of the Standard Model,
sphalerons are known to mediate phase transitions between topologically distinct vacua,
leading to baryon and lepton number violation,
particularly in high-temperature environments such as the early universe
\cite{Kli.Man.1984,Arn.McLer,Kri.Rin,Mor.Ram}.
One kind of phase transition involves the decay of a sphaleron into oscillon,
which is long-lived oscillatory configuration in the gauge theory.
Formation of oscillons in the early universe has been invoked as 
possible sources of gravity waves \cite{Ami,Zho},
candidates for dark matter \cite{Oll}, 
and seeds of primordial black holes \cite{Aur.Clo.Mui}. 

A different kind of phase transition consists of formation of
a bubble of true vacuum inside a false vacuum.
The profile of the bubble arises as a stationary point of the electroweak action functional
and thus is a sphaleron. 
It describes an expanding spherical wall whose dynamics has 
a fundamental importance for bariogenesis \cite{Kuz.Rub.Sha,Ell.Flo.Rud.Sec}. 
However, due to the complexity of the underlying gauge theory,
the actual evolution of such bubbles is difficult to analyze.
This motivates investigating lower-dimensional scalar field models 
for studying the essential dynamical features of sphaleron evolution. 

A simple relativistic field theory in 1+1 dimensions
exhibiting exact sphaleron solutions is the class of nonlinear Klein-Gordon (KG) models
with a non-symmetric quartic potential \cite{Ave.Baz.Los.Men,Man-review}.
The profile of the sphalerons is symmetric with a single peak,
while asymptotically the sphaleron goes to a background value given by the false vacuum,
with the peak of the profile being the value of the true vacuum.
Instability of a sphaleron can be triggered by either a perturbation
or an interaction with some inhomogeneity. 
There are two different channels for the subsequent evolution of the sphaleron.
One channel consists of the sphaleron collapsing to an oscillon,
which is a long-lived oscillatory solution that has a localized profile
\cite{Bog.Mak,Gle,Gle.Sic}. 
This decay channel occurs when the perturbation is negative \cite{Nav-Obr.Que}. 
The other channel arises from a positive perturbation \cite{Nav-Obr.Que}
and causes the sphaleron to increase in height 
until its peak reaches the value of the true vacuum,
after which its width will increase, 
producing an expanding region of true vacuum
due to conservation of energy. 
Its profile resembles a kink-antikink pair whose flanks steepen as they separate.

The present paper is devoted to exploring this latter long-time behaviour analytically,
which provides a simple model of bubble formation and evolution. 
No previous results in this direction have been obtained in the literature
(see \Ref{Ada.Ciu.Ole.Rom.Wer.2021,Ole.Que.Rom.Wer.2023,Alo-Izq.Nav-Obr.Ole.Que.Rom.Wer.2023} for related work in other directions).

As a main new result, an explicit power series approximate solution
will be derived which describes the evolution of the sphaleron.  
This solution is found by a nonlinear collective coordinate method involving three degrees of freedom.
The asymptotic behaviour of the profile for large time consists of a tabletop
whose height is the value of the true vacuum
and whose width increases
such that the flanks become vertical as their speed approaches light speed. 
The energy of the solution is conserved
while the energy density asymptotically concentrates at the flanks.
This behaviour qualitatively matches what is seen in numerical solutions of the model.

These novel results provide an analytically tractable prototype
for studying the dynamical realization of a bubble transition between vacua,
in which the physical system evolves from
an unstable configuration around a false vacuum 
into an expanding region of true vacuum bounded by topological defects. 
In particular, the energy of the bubble gets redistributed
into propagating localized structures at the wall of the bubble.

In section~\ref{sec:sphaleron}, 
the sphaleron solution in the non-symmetric Klein-Gordon model is reviewed. 
Its linear instability is briefly summarized,
based on recent mathematical analysis in \Ref{Anc2025}. 

In section~\ref{sec:numeric},
the evolution of a perturbed sphaleron is studied numerically. 
At time $t=0$, a positive perturbation is applied to initial data for the sphaleron, 
where the perturbation is chosen to excite the unstable (ground-state) mode.
This is carried out for two distinct types of perturbation.
For one type, 
a perturbation is made to the sphaleron's profile,
with a rate of change that is zero. 
The other type involves a perturbation that produces initial growth
while the sphaleron profile is undisturbed. 
In both scenarios the numerical solution is computed for large $t$, 
and its qualitative features are summarized.

In section~\ref{sec:analytic}, 
the approximate analytical solution describing the evolution of the perturbed sphaleron 
is derived by a nonlinear collective coordinate method
combined with a power series technique.
The method starts with making an ansatz for the solution,
which is given by modulation of the parameters in a general sphaleron solution,
namely the parameters are replaced by unknown functions of $t$. 
This ansatz is substituted into the nonlinear KG action principle,
yielding an effective action principle which produces variational equations for the modulated parameters.
The main step consists of solving these differential equations 
via a series in inverse powers of $t$.
An asymptotic analysis of this series solution is performed,
which shows how an accelerating kink-antikink profile emerges for long times.
This result fully captures the behaviour seen in the numerical solutions. 

In section~\ref{sec:approx},
the series solution is considered for early times and shown to provide
a close approximation for the initial evolution of the perturbed sphaleron
in the scenario when the perturbation is caused by an initial kick. 
Additional features of the resulting solution are discussed.
In particular, the energy density is shown to concentrate at the flanks,
and the flank speed is found to increase asymptotically to light speed. 
The error of the solution is shown to be uniformly small for long times. 

In section~\ref{sec:remarks},
some concluding remarks are given.

Two appendices contain some derivations needed in the collective coordinate method
and in the early time approximation.

\section{Sphaleron in a false vacuum}
\label{sec:sphaleron}

The most general class of quartic KG models of a scalar field $\phi(x,t)$
exhibiting a sphaleron solution in 1+1 dimensions is described by the action principle 
\begin{equation}\label{KG.action}
S[\phi] = \int_{-\infty}^\infty\int_{-\infty}^\infty\big( {-}\tfrac{1}{2}\phi_t^2 + \tfrac{1}{2}\phi_x^2 +V(\phi) \big)dx\,dt . 
\end{equation}
with the non-symmetric potential
\begin{equation}\label{potential}
V(\phi) = 2 \phi^2 (\phi - \tanh(a))(\phi -\coth(a)), 
\quad
a>0 
\end{equation}
in terms of a parameter $a$. 
In particular, every quartic potential with a false vacuum 
belongs to this $1$-parameter family,
up to a shift, scaling, and reflection on $\phi$, and a dilation on $(t,x)$.
(See e.g. \Ref{Ave.Baz.Los.Men,Anc2025}.)

\begin{figure}
\includegraphics[width=0.5\textwidth,trim=2cm 12cm 2cm 1cm, clip]{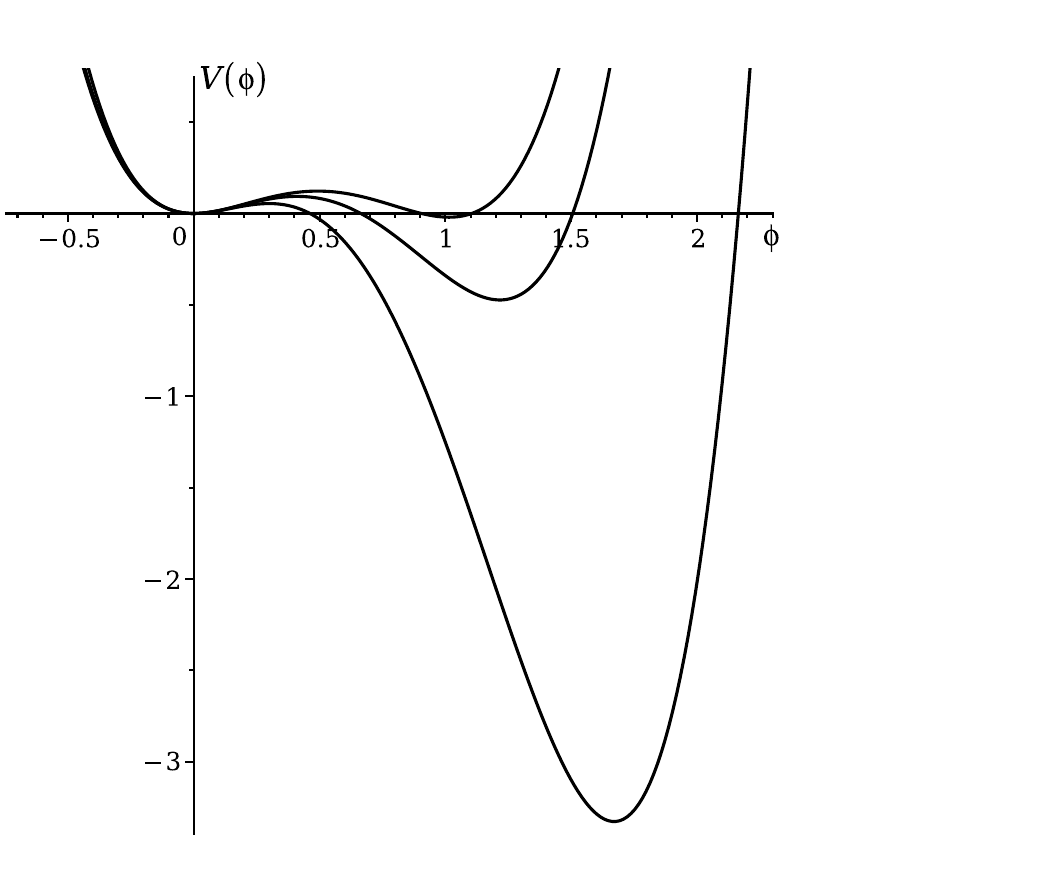}
\caption{Potential with a false vacuum at $\phi=0$: 
$a=$ 0.5, 0.8, 1.5}
\label{fig:potential}
\end{figure}

The false vacuum is $V=0$ at $\phi=0$,
and the true vacuum has a minimum $V_{\min}<0$ at 
\begin{equation}\label{true.vacuum}
\phi = \tfrac{3}{4} \coth(2a) \Big( 1 + \sqrt{1 - \tfrac{8}{9}\tanh(2a)^2} \Big)
\end{equation}
which lies between $\phi=\tanh(a)$ and $\phi=\coth(a)$.
See the plot in Fig.~\ref{fig:potential}.
This family \eqref{potential} is equivalent, up to scaling, to the potential 
\begin{equation}\label{potential1}
V(\phi) = 2\phi^2 (\phi -\tilde a)(\phi-\tilde a/\tilde b)
\end{equation}
with $1>\tilde b>0$, $\tilde a>0$ studied in \Ref{Ave.Baz.Los.Men},
and also to the potential 
\begin{equation}\label{potential2}
V(\phi) = 4 \phi^2 (1 -\phi/\tilde b^2)(1-\phi/\tilde a^2)
\end{equation}
with $\tilde b >\tilde a >0$ listed in \Ref{Man-review}.

The equation of motion in this potential is given by 
\begin{equation}\label{eom.potential}
\phi_{tt} - \phi_{xx} + 4(2\phi^2 -3\coth(2a)\phi + 1)\phi =0 . 
\end{equation}
Here the field and the space-time coordinates are taken to be dimensionless 
(i.e.\ relativistic units are employed). 
There is freedom of adding an arbitrary constant $V_0$ to the interaction potential $V(\phi)$ 
without changing the equation of motion.
The action principle is invariant under time-translations, space-translations, and Lorentz boosts,
which yield respective conservation laws for energy
\begin{equation}\label{ener}
E[\phi] = \int_{-\infty}^{\infty} \Big( \tfrac{1}{2} \phi_t^2 + \tfrac{1}{2} \phi_x^2 + V(\phi) \Big)\, dx , 
\end{equation}
linear momentum 
\begin{equation}\label{mom}
P[\phi] = \int_{-\infty}^{\infty}\Big( {-}\phi_t \phi_x\Big) \,dx ,
\end{equation}
and boost momentum
\begin{equation}\label{boostmom}
J[\phi] = \int_{-\infty}^{\infty} \Big( t \phi_t \phi_x + x\big(\tfrac{1}{2} \phi_t^2 + \tfrac{1}{2} \phi_x^2 + V(\phi) \big) \Big)\,dx . 
\end{equation}
These three integrals are conserved for all solutions $\phi(x,t)$
with sufficient asymptotic decay as $x\to\pm\infty$.

Note that the boost momentum can be expressed as
\begin{equation}
J[\phi] = \chi[\phi;t] E[\phi] -t P[\phi]
\end{equation}
where
\begin{equation}
\chi[\phi;t] = \frac{1}{E[\phi]} \int_{-\infty}^{\infty} x\Big(\tfrac{1}{2} \phi_t^2 + \tfrac{1}{2} \phi_x^2 + V(\phi) \Big)\,dx
\end{equation}
defines the center of energy. 
Conservation of $E$ and $P$ thereby implies $\frac{d}{dt}\chi = P/E$,
showing that the center of energy moves at constant speed.
Through the relativistic particle relation $E=m$, this can be viewed equivalent to
center of mass of the field configuration $\phi(x,t)$. 
Under Lorentz boosts,
the conserved integrals $(E[\phi],P[\phi])$ transform like a 2-vector.

\subsection{Sphalerons}

A ground state of the KG equation \eqref{eom.potential}
is a static solution $\phi=\phi(x)$ 
satisfying the ODE 
\begin{equation}\label{groundstate.eqn}
\phi_{xx} = V'(\phi) . 
\end{equation}
This is an oscillator equation with effective potential $V_\eff(\phi)= -V(\phi)$. 
Integration gives 
\begin{equation}\label{groundstate.energy.eqn}
\tfrac{1}{2}\phi_x^2  - V(\phi) =E_0
\end{equation}
where $E_0=\const$ is the oscillator energy. 

The sphaleron solution corresponds to $E_0=0$. 
Its profile as obtained by integration of the oscillator equation \eqref{groundstate.energy.eqn}
has the shape of a lump 
\begin{equation}\label{lump}
\phi(x) = \sinh(2a)/\big( \cosh(2a) + \cosh(2(x-x_0)) \big) . 
\end{equation}
A very useful alternative form of the sphaleron is given by
\begin{equation}\label{kinkantikink}
\phi(x) = \tfrac{1}{2} \big(\tanh(x-x_0+a) - \tanh(x-x_0 -a)\big) 
\end{equation}
which shows that it describes the superposition of a kink and an antikink.
It has a peak height
\begin{equation}\label{height}
\phi(x_0)= \tanh(a)
\end{equation}
and its full-width is 
\begin{equation}\label{width}
\Delta x = \arccosh\big(2\cosh(2a) + \sqrt{3\cosh(2a)^2 + 6}\big)
\end{equation} 
as defined by where the convexity of $\phi(x)$ is maximum.

The energy of a ground state is given by the integral 
\begin{equation}
E[\phi] = \int_{-\infty}^{\infty} \Big( \tfrac{1}{2} \phi_x^2 + V(\phi) \Big)\, dx 
= 2 \int_{-\infty}^{\infty} V(\phi) \, dx . 
\end{equation}
Since the ground-state lump \eqref{lump} has a single peak and decays to zero,
its energy can be expressed as a Bogomolny integral 
\begin{equation}
E[\phi] = 2\int_{0}^{\phi_{\max}} \sqrt{2V(\phi)} \, d\phi
\end{equation}
where $\phi_{\max} = \tanh(a)$.
This yields 
\begin{equation}\label{lump.E}
E_\lump = \tfrac{2}{3} + 2\big(1 - 2a\coth(2a)\big)/\sinh(2a)^2
\end{equation}
for the lump energy. 
Since it is stationary, the lump has zero momentum, $P_\lump=0$, 
and zero boost-momentum, $J_\lump=0$. 
Thus the center of mass is located at $x=x_0$.

\subsection{Linear Instability}
\label{sec:stability}

\begin{figure}
\includegraphics[width=0.75\textwidth,trim=2cm 12cm 2cm 1cm, clip]{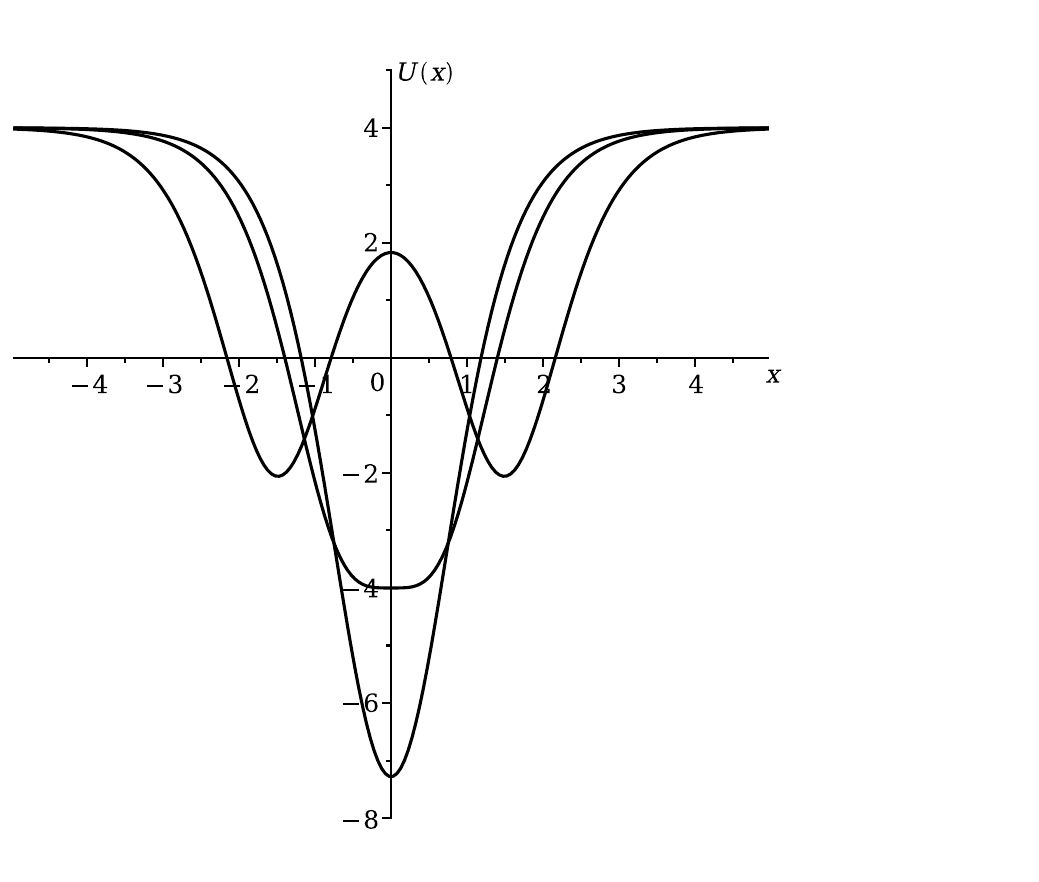}
\caption{Perturbation potential: 
$a=$ 0.25, $\arctanh\tfrac{1}{\sqrt{3}}$ (=0.658), 1.5}
\label{fig:U.well}
\end{figure}

Sphalerons are well known to exhibit a linear instability.
Specifically,
consider a small perturbation with frequency $\omega$, 
\begin{equation}\label{perturbed}
\varphi(x,t)=\phi(x)+ \epsilon \eta(x)\exp(i\omega t)
\end{equation}
where $|\epsilon|\ll 1$. 
Linearization of the equation of motion \eqref{eom.potential} 
for $\varphi(x,t)$ around $\epsilon=0$ yields
a linear Schr\"{o}dinger equation for $\eta(x)$, 
\begin{equation}\label{perturbation.eqn}
-\eta_{xx} +U(x)\eta =\omega^2\eta
\end{equation}
with the potential
\begin{equation}\label{perturbation.potential}
U(x)= V''(\phi(x)) = 4 - 24\frac{1 +\cosh(2x)\cosh(2a)}{(\cosh(2x)+\cosh(2a))^2} 
\end{equation}
This is an eigenfunction problem for the linear operator $-\partial_x^2 + U(x)$, 
with eigenvalue $\lambda = \omega^2$.
Note that the potential is symmetric in $x$,
such that $x=0$ is the minimum when $\coth(a) \geq \sqrt{3}$ 
or the local maximum when $\coth(a) < \sqrt{3}$,
which gives a double-well in the latter case and a single-well in the former case. 
See Fig.~\ref{fig:U.well}. 

The derivative of the sphaleron
\begin{equation}\label{zeromode} 
\eta(x)= \phi'(x) = -2\sinh(2a) \sinh(2x)/(\cosh(2x) + \cosh(2a))^2,
\quad
\lambda_0 =0
\end{equation}
satisfies the linearized field equation
and thus constitutes an eigenfunction with zero eigenvalue, 
called a zero mode. 
This eigenfunction has a single node, which is located at the peak of the sphaleron,  
$\eta=0$ at $x=0$.
From a general result in eigenfunction theory
(Sturm oscillation theorem, see e.g. \Ref{Gey.Pel-book}),
however, the ground-state eigenfunction is always nodeless.
Therefore, the ground state must have a negative eigenvalue.

This implies that the sphaleron is linearly unstable when perturbed by the ground-state mode. 
A detailed analysis of the eigenfunction equation \eqref{perturbation.eqn}
shows \cite{Anc2025} that it can be solved explicitly in terms of special functions
known as local Heun functions.

The ground state eigenfunction is explicitly given by 
\begin{equation}\label{-1.eigenfunct.x}
\eta_{-1}(x) = 
\sech(x)^{\sqrt{4-\lambda_{-1}}} (\sinh(a)^2 \sech^2(x) + 1)^3 
H\ell(1-p,\alpha\beta - q;\alpha,\beta,\delta,\gamma;\tanh^2(x))
\end{equation}
in terms of the parameters 
\begin{equation}\label{Heun.ode.params}
\alpha = \tfrac{1}{2}\sqrt{4-\lambda_{-1}} + 3,
\quad
\beta = \tfrac{1}{2}(\sqrt{4-\lambda_{-1}} + 7),
\quad
\gamma = \sqrt{4-\lambda_{-1}} + 1,
\quad
\delta =\tfrac{1}{2},
\quad
\epsilon = 6
\end{equation}
and
\begin{equation}\label{Heun.ode.p.q}
p = -\frac{1}{\sinh^2(a)},
\quad
q = \frac{6\cosh(2a)(\sqrt{4-\lambda_{-1}} +3) -(\sqrt{4-\lambda_{-1}} +6)(\sqrt{4-\lambda_{-1}} +1)}{4\sinh^2(a)} . 
\end{equation}
Here $H\ell$ is a local Heun function,
which satisfies the Heun equation 
\begin{equation}\label{heun.ode}
H''(z) 
+  \big( \gamma/z + \delta/(z-1) +\epsilon/(z-p) \big)H'(z) 
+\big( (\alpha\beta z -q)/(z(z-1)(z-p) \big) H(z)
=0
\end{equation}
with $\epsilon = \alpha + \beta -\gamma -\delta + 1$.
This is a second order linear differential equation 
which has regular singular points $z=0,1,p,\infty$, 
where $\alpha$, $\beta$, $\gamma$, $\delta$, and $q$ are constant parameters.
Local Heun functions are given by an analytic Frobenius series around any one of the points $z=0,1,p$. 
The main properties of local Heun functions are summarized in \Ref{Ron-book}.

\begin{figure}
\includegraphics[width=0.6\textwidth,trim=2cm 12cm 2cm 1cm, clip]{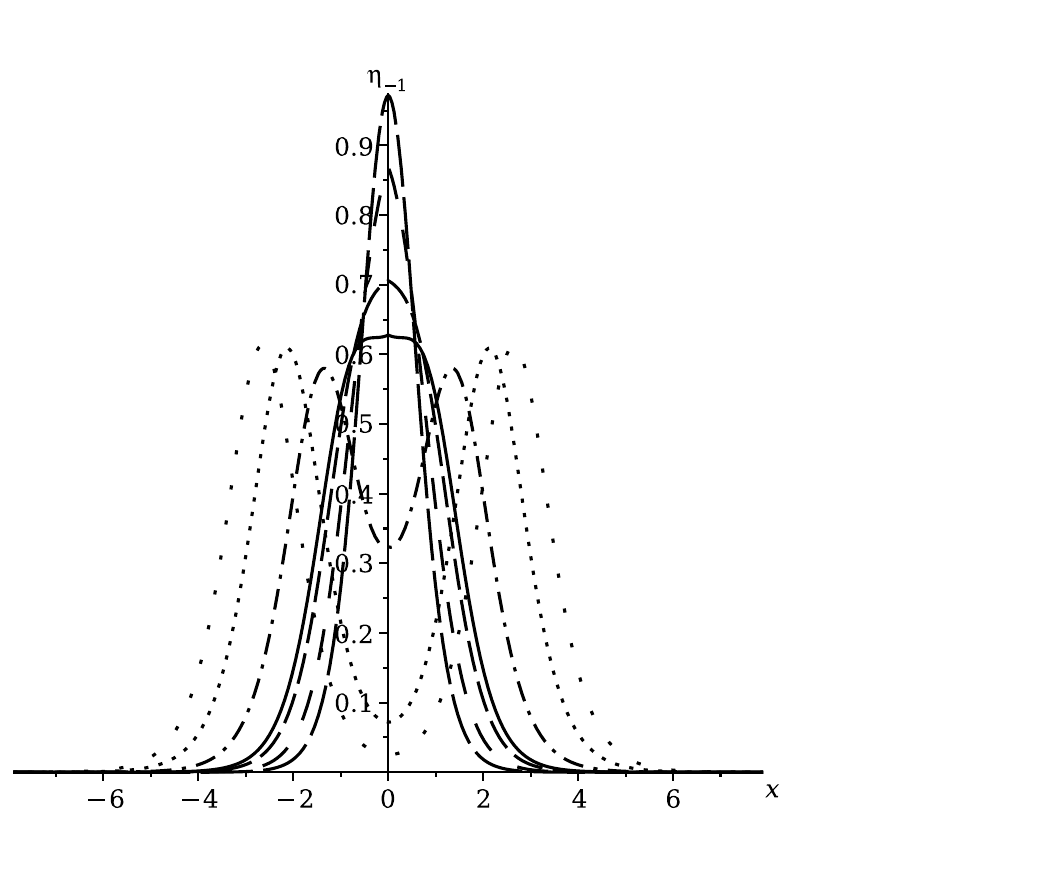}
\caption{Ground-state eigenfunction, normalized in $L^2$, 
for $a=$ 2.65, 2.14, 1.44, 1.00, 0.881, 0.647, 0.100}
\label{fig:neg.eigenfunction.eta.x}
\end{figure}

A plot of the eigenfunction $\eta_{-1}(x)$ is shown in Fig.~\ref{fig:neg.eigenfunction.eta.x}.
Its profile has a single peak at $x=0$ for $a\lesssim 1.00$ and a double peak for $a\gtrsim 1.00$.
The height of the single peak is simply 
\begin{equation}\label{singlepeak}
\eta_{-1}|_\peak =1 .
\end{equation}
In comparison, the double peak can be shown to have the approximate height
\begin{equation}\label{doublepeak}
\eta_{-1}|_\peak \simeq 16(1+2(2e^{-2a})^{1/3}),
\end{equation}
which is significantly higher than the height at $x=0$. 

\begin{figure}
\includegraphics[width=0.6\textwidth,trim=2cm 12cm 2cm 1cm, clip]{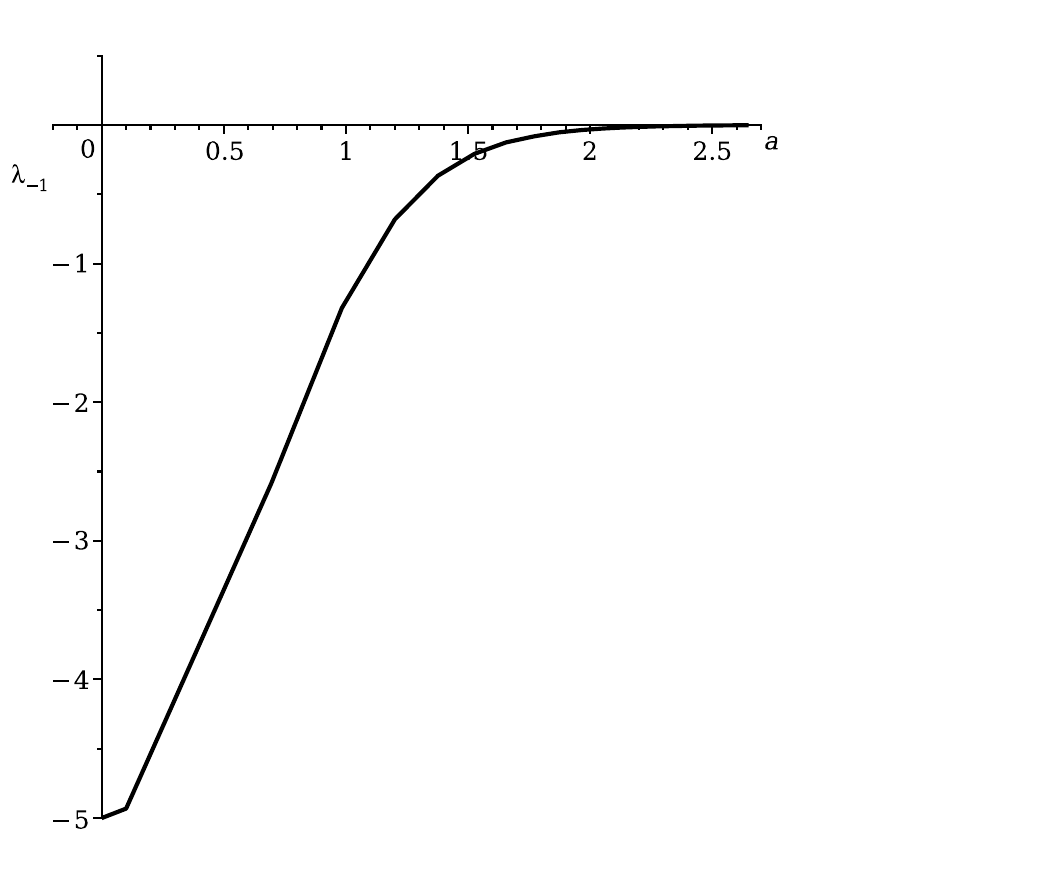}
\caption{Ground-state eigenvalue: $\lambda_{-1}$}
\label{fig:neg.eigenvalue}
\end{figure}

The corresponding eigenvalue $\lambda_{-1}$ is determined by the equation 
\begin{equation}\label{eigenvalue.deteqn}
H\ell(p,q;\alpha,\beta,\gamma,\delta;1) H\ell(1-p,\alpha\beta - q;\alpha,\beta,\delta,\gamma;1) 
-1 =0
\end{equation}
The numerical solution is plotted as a function of the parameter $a$
in Fig.~\ref{fig:neg.eigenvalue}.
Note that $\lambda_{-1}$ increases monotonically with $a$
from $-5$ at $a=0$ to $0$ as $a\to \infty$.
For $a \simeq 1.00$, when the eigenfunction has no convexity at $x=0$,
$\lambda_{-1}\simeq -1.30$. 

\Ref{Anc2025} also develops approximate expressions
for both the eigenvalue and eigenfunction.
Here we will adapt them to get simpler, albeit rougher, approximations:
\begin{equation}\label{groundstate.approx}
\eta_{-1}(x) \approx 
\begin{cases}
\sech(x)^3\big( 1 + \tfrac{2}{7} a^2 \big(7\tanh(x)^2 - 4\ln(\sech(x))\big) \big),
& a\lesssim 1.00
\\
\big(2-\sech(x)^2\big)^4/\big(2 e^{-2a}\sinh(x)^2 + 1\big),
& a\gtrsim 1.00
\end{cases} 
\end{equation}
The time scale for the growth of this unstable mode is given by
\begin{equation}\label{timescale.approx}
\tau = 1/\sqrt{|\lambda_{-1}|}
\approx 
\begin{cases}
\tfrac{1}{\sqrt{5}}(1 +\tfrac{24}{35} a^2) , 
& a\lesssim 1.00
\\
\tfrac{1}{4\sqrt{6}} e^{2a} , 
& a\gtrsim 1.00
\end{cases} 
\end{equation}  

In \Ref{Anc2025} the complete spectrum of the problem has been found to consist of
two positive eigenvalues, $0<\lambda_1<3$ and $3<\lambda_2<4$,
with their respective eigenfunctions $\eta_{1}(x)$ and $\eta_{2}(x)$ 
having two and three nodes.
These describe two internal vibrational modes. 
The continuous spectrum comprises the interval $[4,\infty)$
(by Weyl's theorem, see e.g. \Ref{Gey.Pel-book}).

\section{Numerical solution}
\label{sec:numeric}

The sphaleron in the quartic potential \eqref{potential} with a false vacuum
is given by the lump solution \eqref{lump}. 
We will now numerically examine the long-time behaviour 
when this solution is perturbed via exciting its unstable mode \eqref{-1.eigenfunct.x}
in the positive channel. 
As remarked in \Ref{Man-review},
qualitatively the lump is expected to increase in width and produce an expanding region of true vacuum. 

Consider initial data given by the lump \eqref{lump} with $x_0=0$ 
plus a small positive perturbation which is given by the unstable mode \eqref{groundstate.approx}
scaled by the lump's height \eqref{height}.
This ensures that the instability is triggered at $t=0$. 

Thus, we take 
\begin{equation}\label{perturb.lump.initialdata}
\phi(x,0) =
\sinh(2a)/\big( \cosh(2a) + \cosh(2x) \big)
+ \epsilon\, \tanh(a)\, \eta_{-1}(x), 
\quad
\phi_t(x,0) = 0 
\end{equation}
as initial data,
where $\eta_{-1}(x)$ is the approximate expression \eqref{groundstate.approx}. 
The equation of motion \eqref{eom.potential} in the potential \eqref{potential}
is solved numerically to obtain $\phi(x,t)$ for $t>0$. 

We discretize the partial differential equation \eqref{eom.potential} as follows:
\begin{equation}\label{discrete.eqn}
\begin{aligned}
& \frac{d^2 \phi_n}{dt^2}
- \frac{1}{2h} \left( \phi_{n-1} - 2\phi_n + \phi_{n+1} \right)
+ \frac{1}{12h^2} \left( \phi_{n-2} - 4\phi_{n-1} + 6\phi_n - 4\phi_{n+1} + \phi_{n+2} \right)
\\&\qquad
+ 4(2\phi^2_n -3\coth(2a)\phi_n + 1)\phi_n =0
\end{aligned}
\end{equation}
where $h$ is the lattice spacing,
$\phi_n(t) = \phi(nh, t)$ is the discretized solution,
with $n \in \mathbb Q$. 
Here a fourth-order accurate finite-difference scheme has been used to approximate $\phi_{xx}$,
resulting in an overall spatial accuracy of $\mathcal{O}(h^4)$,
which helps to keep spatial discretization effects small. 

We use a spatial grid with 6000 points,
corresponding to the following spatial domain: 
$[-300, 300]$ for $h = 0.1$,
and $[-150, 150]$ for $h = 0.05$.
In addition,
absorbing boundary conditions are implemented at the edges of the spatial domain, 
which helps ensure there is no reflection of outgoing radiation. 
The time integration is carried out by 
an explicit St\"ormer method with a time step $\Delta t = 0.005$
and temporal accuracy of $\mathcal{O}(\Delta t^4)$.

We monitor energy conservation in the simulation and find that it holds
for parameters $a\gtrsim 1.15$.
For smaller values of $a$,
deviation from energy conservation becomes noticeable,
which is primarily due to the increasingly steep gradients that develop as $a$ decreases,
leading to stronger localization of energy at the flanks. 
Such features require higher spatial resolution to be accurately resolved,
and thus place greater demands on the numerical scheme.
To ensure that the observed long-time behaviour is not a numerical artifact,
we have performed additional checks by varying the spatial and temporal resolution. 
In particular, decreasing both the lattice spacing and time step
improves energy conservation without altering the qualitative evolution of
the numerical solution. 
This indicates that the features of the numerical solution are physically robust
and not sensitive to small violations of energy conservation.

\begin{figure}[h]
\centering
\includegraphics[width=0.9\textwidth]{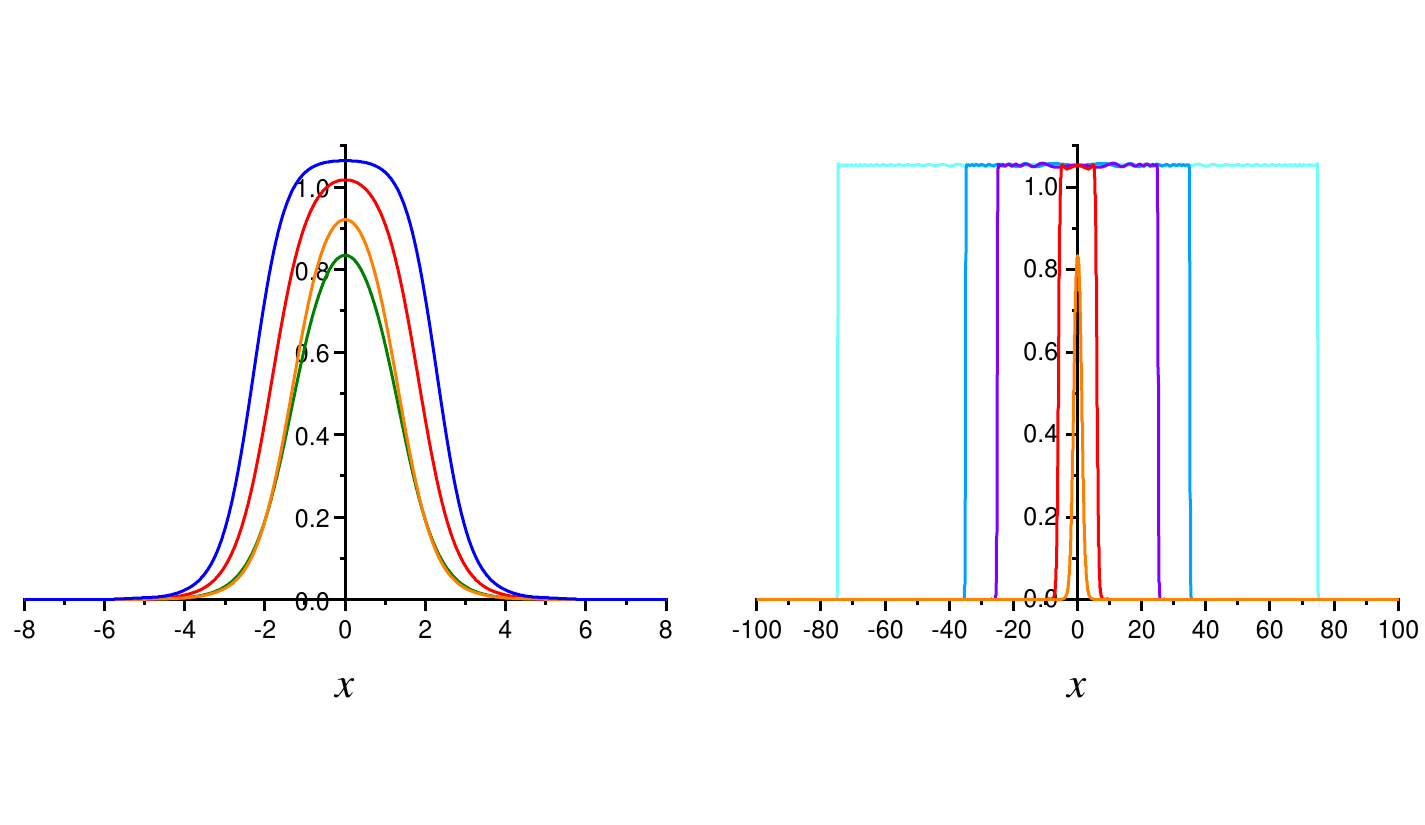} 
\caption{
Numerical solution for $a=1.174$. \\
(Left)\ Short times $t=0,2,4,5$;
\quad
(Right)\ Long times $t=10,30,40,80$. }
\label{fig:Numerics_a_1.174}
\end{figure}

\begin{figure}[h]
\centering
\includegraphics[width=0.9\textwidth]{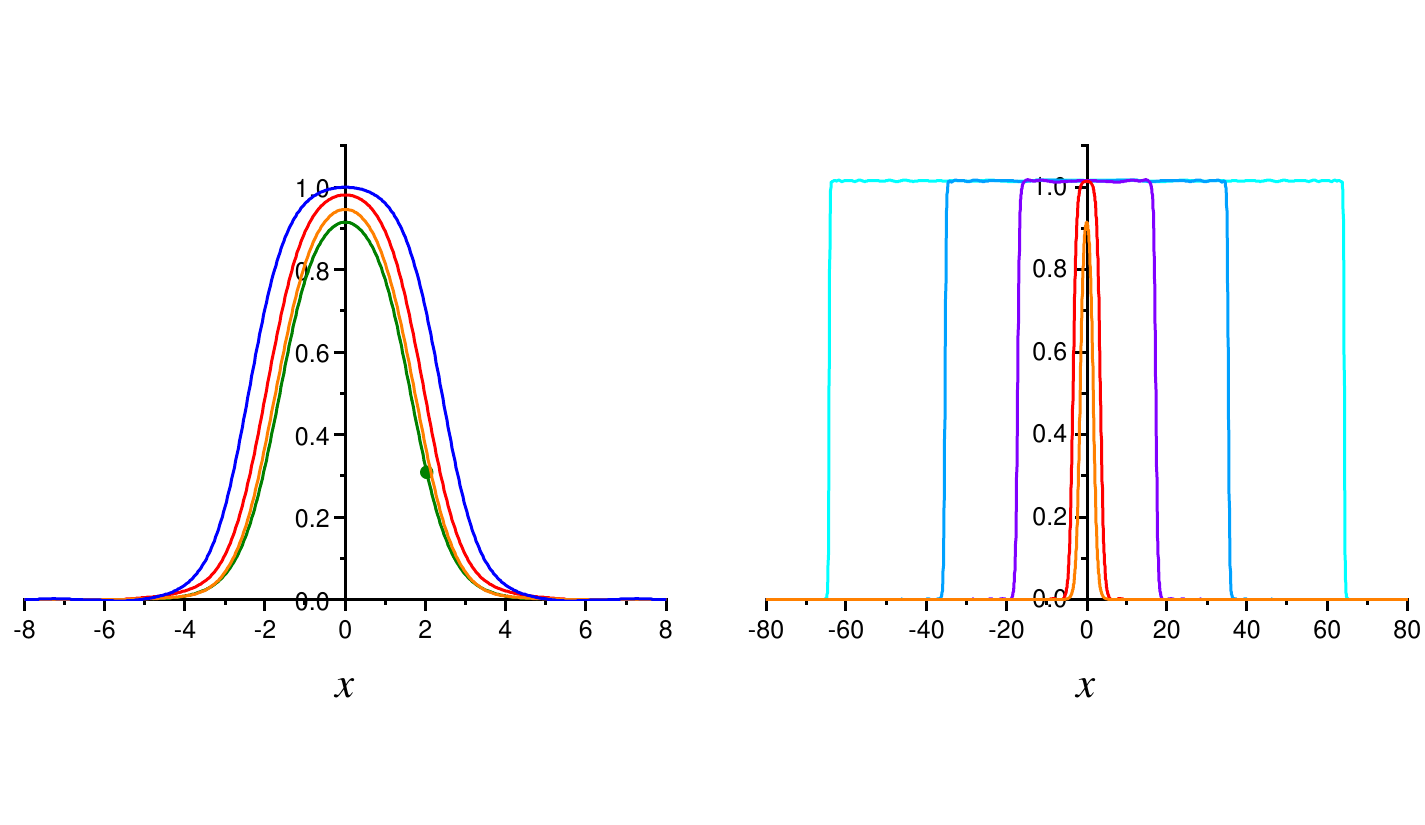} 
\caption{
Numerical solution for $a=1.5$. \\
(Left)\ Short times $t=0,3,5,7$;
\quad
(Right)\ Long times $t=10,30,50,80$. }
\label{fig:Numerics_a_1.5}
\end{figure}

\begin{figure}[h]
\centering
\includegraphics[width=0.9\textwidth]{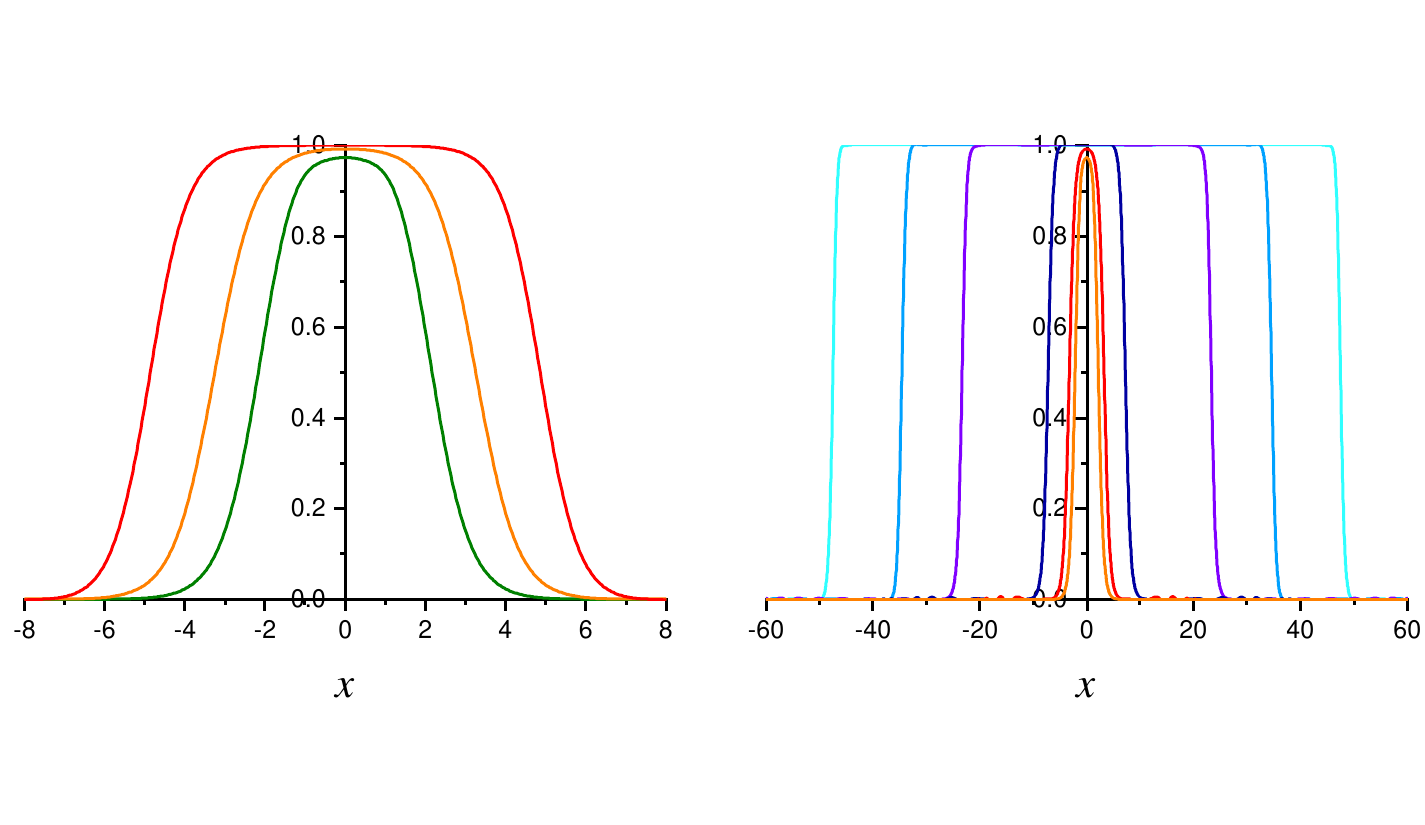} 
\caption{
Numerical solution for $a=2.0$. \\
(Left)\ Short times $t=0,20,30$;
\quad
(Right)\ Long times $t=20,40,80,100,120$. }
\label{fig:Numerics_a_2.0}
\end{figure}

The numerical results are shown in Figs.~\ref{fig:Numerics_a_1.174} to~\ref{fig:Numerics_a_2.0} respectively for parameters $a=$ 1.174, 1.5, 2.0,
using perturbation parameter $\epsilon = 0.01$. 
These solutions exhibit the following behaviour:
\begin{itemize}
\item
the height first rises rapidly to the value given by the true vacuum \eqref{true.vacuum};
\item
next the flanks start to steepen and move outward; 
\item
the profile becomes a spreading tabletop shape,
which describes a kink-antikink pair; 
\item
asymptotically, the flanks become vertical and accelerate to light speed.
\end{itemize}

\begin{figure}[h]
\centering
\includegraphics[width=0.9\textwidth]{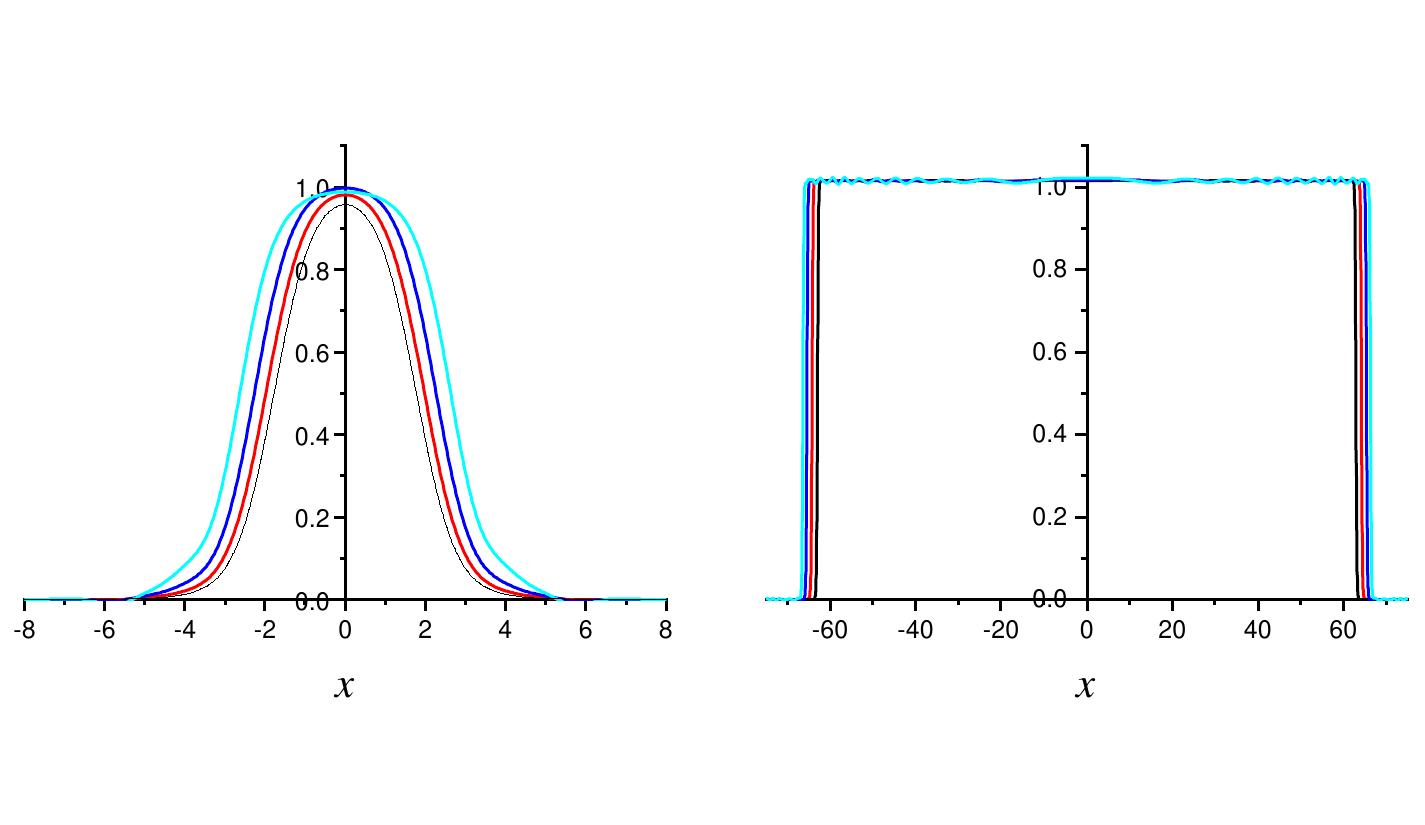} 
\caption{
Numerical solution for $a=1.5$ using $\epsilon=$ 0.01, 0.03, 0.05, and 0.10. \\
(Left)\ $t=5$
\quad
(Right)\ $t=80$
}
\label{fig:Numerics_a_1.5-epsilon}
\end{figure}

An important remark is that the details of the evolution of the instability
depend sensitively on the size of the perturbation parameter $\epsilon$
as illustrated in Figs.~\ref{fig:Numerics_a_1.5-epsilon}
for $a=1.5$ and $\epsilon=$ 0.01, 0.03, 0.05, and 0.10.
In general we see that the tabletop forms more quickly
as the perturbation parameter increases in size.

\subsection{Numerical solutions with initial growth}

An alternative way to excite the unstable growing mode is by taking initial data at $t=0$
that has non-zero growth of $\phi$ 
while matching the lump profile \eqref{kinkantikink} for $\phi$ exactly. 
Specifically, we take
\begin{subequations}\label{perturb.lump.initialdata.kick}
\begin{align}
\phi(x,0) & = \sinh(2a)/\big( \cosh(2a) + \cosh(2x) \big),
\label{lump.initialdata}
\\
\phi_t(x,0) & = \dfrac{\epsilon}{\tau} \tanh(a)\, \eta_{-1}(x)
\label{kick.initialdata}
\end{align}
\end{subequations}
where $\tau$ is the approximate time scale \eqref{timescale.approx} of the growing mode
and, again, $\eta_{-1}(x)$ is the approximate expression \eqref{groundstate.approx}.
Here $\epsilon=0.01$ is chosen as the perturbation parameter. 

\begin{figure}[h]
\centering
\includegraphics[width=0.9\textwidth]{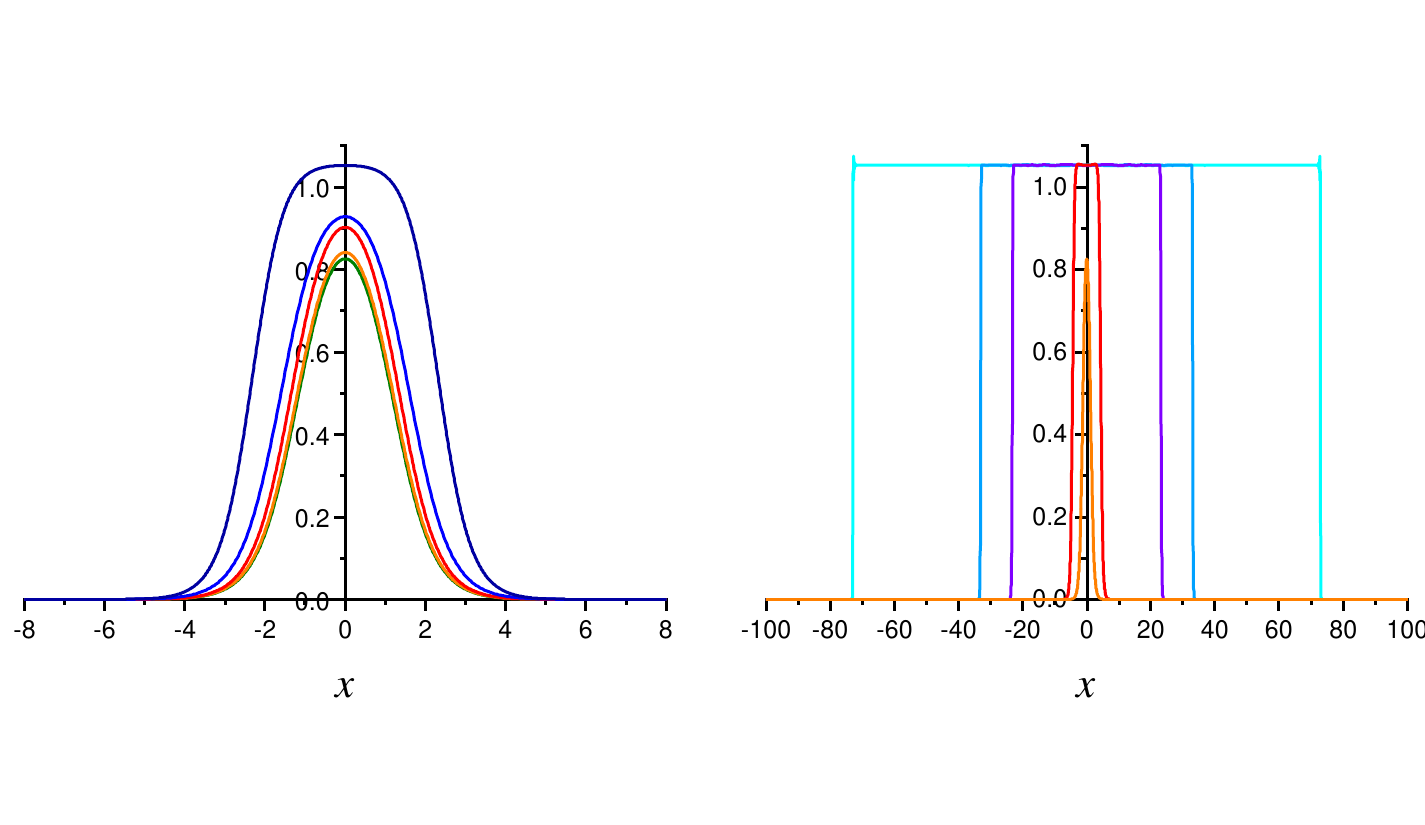} 
\caption{
Numerical solution for $a=1.174$ with initial data \eqref{perturb.lump.initialdata.kick}. \\
(Left)\ Short times $t=0,2,4,5,7$;
\quad
(Right)\ Long times $t=0,10,30,40,80$.
}
\label{fig:Numerics_a_1.174_kick}
\end{figure}

\begin{figure}[h]
\centering
\includegraphics[width=0.9\textwidth]{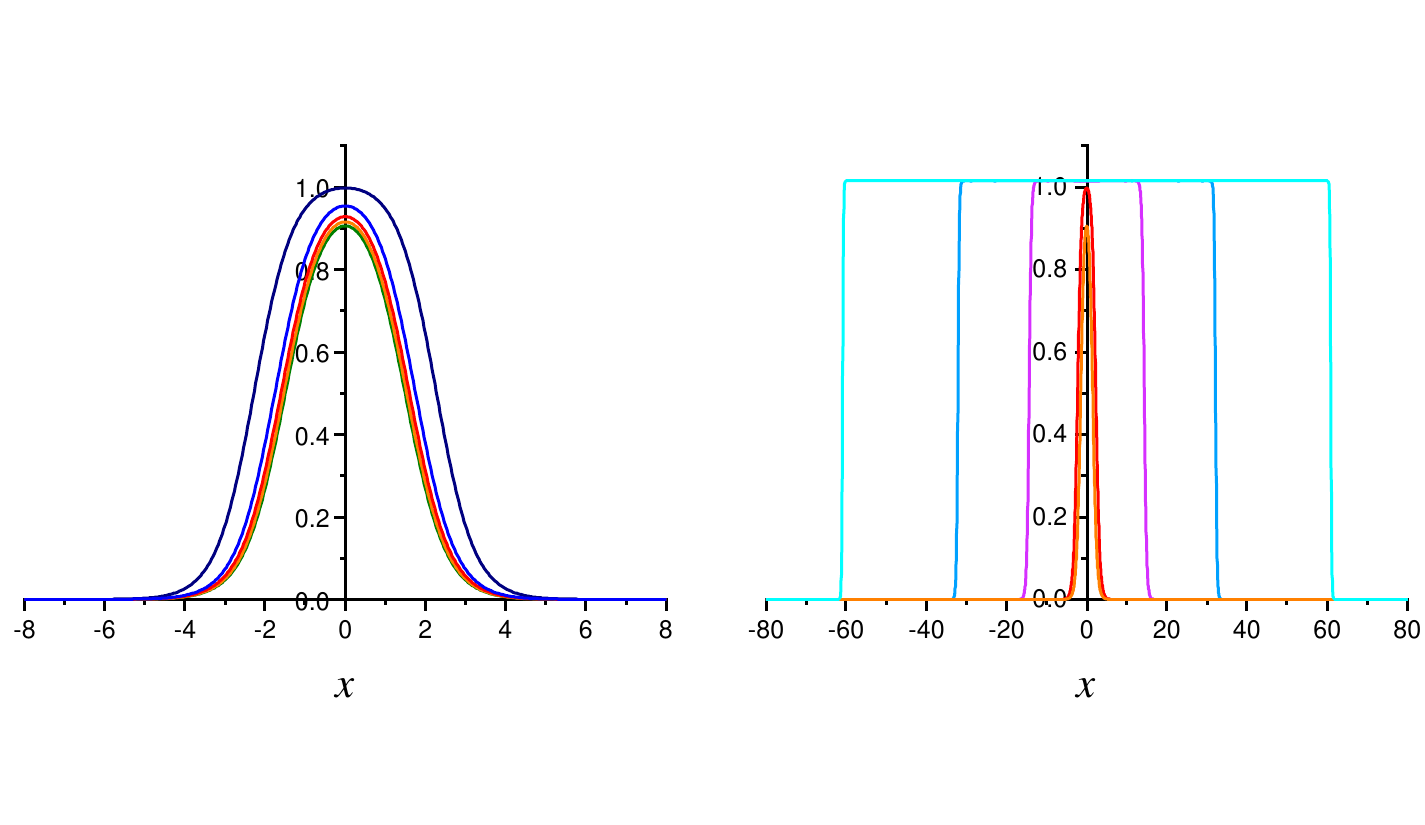} 
\caption{
Numerical solution for $a=1.5$ with initial data \eqref{perturb.lump.initialdata.kick}. \\
(Left)\ Short times $t=0,3,5,7,10$;
\quad
(Right)\ Long times $t=0,10,30,50,80$.
}
\label{fig:Numerics_a_1.5_kick}
\end{figure}

Solving the equation of motion \eqref{eom.potential} 
numerically gives the results shown in Figs.~\ref{fig:Numerics_a_1.174_kick} and~\ref{fig:Numerics_a_1.5_kick}.
These solutions have the same qualitative behaviour as seen in the previous solutions
(cf Figs.~\ref{fig:Numerics_a_1.174} and~\ref{fig:Numerics_a_1.5}).
An evident difference is that the profile reaches its maximum height at an earlier time,
as would be expected due to the initial kick at $t=0$.

\section{Collective coordinates for sphalerons}
\label{sec:analytic}

The goal now will be to obtain an approximate analytical expression for the behaviour
seen in the numerical solutions at large times. 

We will start from the kink-antikink form \eqref{kinkantikink} for the lump solution with $x_0=0$ 
and consider a collective-coordinate modulation for its perturbed evolution
\begin{equation}\label{ansatz}
\phi(x,t) = \tfrac{1}{2} A(t)\big( \tanh(B(t)x +C(t)) - \tanh(B(t)x-C(t)) \big)
\end{equation}
with unknown functions $A(t)>0$, $B(t)>0$, $C(t)>0$. 
This ansatz exactly matches the lump profile \eqref{kinkantikink}, where $x_0=0$, 
when 
\begin{equation}\label{ansatz.lump}
A = 1,
\quad
B=1,
\quad
C=a . 
\end{equation}
Note that the time derivative of expression \eqref{ansatz} is given by
\begin{equation}\label{time.derivative.ansatz}
 \begin{aligned}
\phi_t(x,t) = &
\tfrac{1}{2} A'(t)\big( \tanh(B(t)x +C(t)) - \tanh(B(t)x-C(t)) \big)
\\&\quad 
+ \tfrac{1}{2} A(t)\big( (B'(t)x+C'(t)) \sech^2(B(t)x +C(t))
\\&\qquad 
- (B'(t)x-C'(t)) \sech^2(B(t)x -C(t)) \big)
\end{aligned}
\end{equation}
which vanishes when the condition \eqref{ansatz.lump}
for matching the lump profile holds.

At any fixed $t$,
the modulation profile \eqref{ansatz} has a tabletop shape which is symmetric in $x$,
with a height equal to $A$,
and with right and left flanks located at $x=\pm C/B$.
The steepness of the flanks is proportional to $B$,
and the flanks spread outward with speed $\nu = \dfrac{d (C/B)}{dt}$.
This qualitatively captures the behaviour seen in the numerical solutions
described in section~\ref{sec:numeric}. 

From the perspective of collective coordinates, 
the profile \eqref{ansatz} contains excitations of 
the translation mode, governed by $C/B$,
as well as two internal modes, governed by $A$ and $B$.
This parameterization is a dynamical modulation of the exact sphaleron solution,
which itself possesses a kink–antikink representation \eqref{kinkantikink}. 
Higher-order deformation modes,
such as radiation tails or asymmetric distortions,
are not included explicitly in the ansatz.
This is justified by the observation that the dominant dynamics at long times
is governed by the large-scale motion of the kink–antikink pair,
while radiation carries just a small fraction of the total energy and disperses over time.
Moreover, the numerical solutions remain approximately symmetric in $x$,
indicating that asymmetric modes are not significantly excited
in the evolution considered here. 
Therefore,
the leading-order behaviour can be accurately captured by the symmetric, three-parameter modulation \eqref{ansatz}. 

We now proceed to determine the three unknown functions $A(t)$, $B(t)$, $C(t)$ 
by the standard variational procedure using the action principle \eqref{KG.action} 
for the potential \eqref{potential}:
\begin{equation}\label{KG.action.potential}
S[\phi] =
\int_{-\infty}^\infty\int_{-\infty}^\infty\big(
{-}\tfrac{1}{2}\phi_t^2 +\tfrac{1}{2} \phi_x^2 +2 \phi^2 (\phi - \tanh(a))(\phi -\coth(a)) \big) dx\,dt . 
\end{equation}
Substitution of the collective-coordinate ansatz \eqref{ansatz} 
followed by integration over $x$
yields an effective action principle
\begin{equation}\label{ansatz.action}
S[A,B,C] = \int_{-\infty}^\infty \Big( \coth(2C)^3 I_3 +\coth(2C)^2 I_2 + \coth(2C) I_1 + I_0 \Big)\,dt
\end{equation}
which governs the collective coordinate dynamics of $(A,B,C)$, 
where (as shown in Appendix~\ref{appendix:action})
\begin{subequations}
\begin{align}
& I_3 = 
(10 A^2 - 2 B^2 - 2 C'{}^2)A^2 C/B
  + \tfrac{1}{6} (\pi^2 + 4 C^2) A^2 C B'{}^2/B^3
  ,
\\
& I_2 = 
2 AC A'C'/B
- (12 \coth(2a) A C+ 5 A^2 - B^2  - C'{}^2)A^2/B
- \tfrac{1}{12} (\pi^2 + 12 C^2) A^2 B'^2/ B^3
,
\\
& \begin{aligned}
I_1 = & 
A C A' B'/B^2
- (C A'^2 + A A'C' - 2 A^2 C C'^2)/B
-\tfrac{1}{6} A^2 C (\pi^2 + 4 C^2) B'^2/B^3
\\&
+ (6 \coth(2a) A  -(6 A^2 - 2 B^2 - 4) C) A^2/B
,
\end{aligned}
\\
& \begin{aligned}
I_0 = &
(\tfrac{1}{2} A'^2 - 2 A C A' C' - \tfrac{4}{3} A^2 C'{}^2)/B    
-(\tfrac{1}{2} A A' B' - \tfrac{2}{3} A^2 C C')/ B^2
\\&
+ \tfrac{1}{18} (\pi^2 + 3 +12 C^2) A^2 B'^2/ B^3 
+ \tfrac{2}{3} (6 \coth(2a) A C + 2 A^2 - B^2 - 3)A^2/ B
.
\end{aligned}
\end{align}
\end{subequations}
The Euler-Lagrange equations of this action principle give a coupled system of three second-order ODEs: 
\begin{equation}\label{ansatz.eom}
\frac{\delta S[A,B,C]}{\delta A} =0,
\quad
\frac{\delta S[A,B,C]}{\delta B} =0,
\quad
\frac{\delta S[A,B,C]}{\delta C} =0 . 
\end{equation}
Explicit expressions are shown in Appendix~\ref{appendix:eom}.
Solutions $(A(t),B(t),C(t))$ are referred to as the moduli space of the collective coordinate ansatz. 

Since the action principle is clearly invariant under time-translations,
Noether's theorem can be applied to obtain a conserved integral.
This simply yields the nonlinear KG energy \eqref{ener}
evaluated for the modulation ansatz \eqref{ansatz}:
\begin{equation}\label{ener.lump}
E =  \coth(2C)^3 H_3 +\coth(2C)^2 H_2 + \coth(2C) H_1 + H_0 =\const
\end{equation}
where 
\begin{subequations}
\begin{align}
& H_3 = 
(10 A^2 - 2 B^2 + 2 C'{}^2)A^2 C/B
  - \tfrac{1}{6} (\pi^2 + 4 C^2) A^2 C B'{}^2/B^3
  ,
\\
& H_2 = 
-2 AC A'C'/B
- (12 \coth(2a) A C+ 5 A^2 - B^2  +C'{}^2)A^2/B
+ \tfrac{1}{12} (\pi^2 + 12 C^2) A^2 B'^2/ B^3
,
\\
& \begin{aligned}
H_1 = & 
-A C A' B'/B^2
+ (C A'^2 + A A'C' - 2 A^2 C C'^2)/B
+\tfrac{1}{6} A^2 C (\pi^2 + 4 C^2) B'^2/B^3
\\&
+ (6 \coth(2a) A  -(6 A^2 - 2 B^2 - 4) C) A^2/B
,
\end{aligned}
\\
& \begin{aligned}
H_0 = & 
(-\tfrac{1}{2} A'^2 + 2 A C A' C' + \tfrac{4}{3} A^2 C'{}^2)/B
+(\tfrac{1}{2} A A' B' - \tfrac{2}{3} A^2 C C')/ B^2
\\&
- \tfrac{1}{18} (\pi^2 + 3 +12 C^2) A^2 B'^2/ B^3 
+ \tfrac{2}{3} (6 \coth(2a) A C + 2 A^2 - B^2 - 3)A^2/ B
.
\end{aligned}
\end{align}
\end{subequations}
Therefore, energy conservation holds automatically. 
Note that, as the profile \eqref{ansatz} is reflection symmetric in $x$, 
both the linear and boost momenta vanish, $P=J=0$.

The lump profile \eqref{ansatz}--\eqref{ansatz.lump}
is readily checked to be a solution of the variational equations \eqref{ansatz.eom}, 
while its energy \eqref{lump.E} can be verified to equal to the energy constant 
$E|_{A=1,B=1,C=a}$ evaluated for this solution.

\subsection{Series solution} 

To understand the behaviour of the solution of 
the collective coordinate dynamics \eqref{ansatz.eom} for $(A(t),B(t),C(t))$, 
we make two changes of variables, which will simplify the subsequent analysis.
The first is that we write
\begin{equation}\label{C.BX.rel}
C = B X
\end{equation}
so that the location of the flanks is at $x = \pm X$
and their outward speed is given by $\nu= \dfrac{dX}{dt}$. 
The second is that we put
\begin{equation}\label{a.b.rel}
  \tanh(2a) = \frac{3b}{2+ b^2}  , 
\end{equation}
whereby the false vacuum is located at $\phi = 1/b$, while the true vacuum remains at $\phi=0$,
and the parameter range $0<a<\infty$ becomes finite, namely
\begin{equation}\label{b.range}
0< b < 1 .
\end{equation}
Note that in the parameterization \eqref{a.b.rel} the KG equation \eqref{eom.potential}
has the form 
\begin{equation}\label{b.KG}
\phi_{tt} - \phi_{xx} + 8\phi (\phi - 1/b)(\phi - b/2) =0 . 
\end{equation}
in terms of parameter $b$. 
The inverse of this parameterization is given by
\begin{equation}\label{b.a.rel}
b = \big( \tfrac{3}{2}\cosh(2a) -\tfrac{1}{2}\sqrt{\cosh(2a)^2 + 8} \big)/\sinh(2a) ,
\quad
0 < a < \infty. 
\end{equation}

We will now derive the long-time behaviour for $(A(t),B(t),X(t))$
by seeking a power series solution in terms of $1/t$.

Since the numerical solution has the flanks separating as $t$ increases, we assume that
\begin{equation}\label{X.series}
X = c t + X_0 + X_1/t + X_2/t^2 + X_3/t^3 + \cdots 
\end{equation}
with $c>0$. 
Likewise, because the steepness of the flanks increases with $t$, we assume that
\begin{equation}\label{B.series}
B = k t + B_0 + B_1/t + B_2/t^2 + B_3/t^3 +\cdots 
\end{equation}
with $k>0$.
Finally, the height in the numerical solution goes to a constant, and so we also assume that
\begin{equation}\label{A.series}
A = h +A_1/t + A_2/t^2 + A_3/t^3 +\cdots
\end{equation}
with $h>0$.

The form of this series can be slightly simplified by taking advantage of
the invariance of the variational equations \eqref{ansatz.eom} under time translation.
If we replace $t\to t+T$, with $T$ being an arbitrary constant,
and expand $t+T = t(1 + T/t)$, 
then $X_0\to X_0+ T$ and $B_0\to B_0+ T$,
and so we can put $X_0=0$ or $B_0=0$.
We will take
\begin{equation}\label{X0.gauge}
  X_0=0
\end{equation}

To determine the unknown coefficients in the series, 
we first substitute expressions \eqref{X.series}--\eqref{X0.gauge}
into the variational equations \eqref{ansatz.eom}. 
We expand each term in powers of $1/t$ and observe that
$\coth(2BX) = 1 + O(e^{-4kc/t^2})$ as $t\to\infty$.
Ignoring exponentially small terms,
we find that the three equations \eqref{ansatz.eom} respectively have degrees
$-1,0,0$ in $1/t$. 

At lowest degree, we have:
\begin{gather}
\tfrac{2}{3} bk (c^2 -1) + 8 c (bh - 1) (b - 2 h)   =0 ,
\\
c^2 - 1  =0 , 
\\
12 b (h^2+1) -8 (b^2+ 2)  h   + 2b k c  =0 . 
\end{gather}  
From the second equation, we obtain
\begin{equation}\label{c}
c=1 . 
\end{equation}
Then the first equation gives $h=1/b$ or $h = b/2$,
which is the asymptotic height of the solution. 
We expect this height to equal the value $\phi = 1/b$ of the false vacuum. 
Hence we will take
\begin{equation}\label{h}
h = 1/b . 
\end{equation}
The third equation now yields
\begin{equation}\label{k}
k = 2 (1-b^2)/b^2 . 
\end{equation}

We proceed to the next lowest order. 
Substituting the preceding expressions into the variational equations,
we find that only the first equation is non-trivial: 
\begin{equation}
(b^2 - 2)  A_1= 0 . 
\end{equation}
Hence, since $b$ is desired to be arbitrary, we obtain
\begin{equation}\label{A1}
A_1= 0 . 
\end{equation}

Continuing to the next order,
we obtain the following three equations:
\begin{align}
& 8 (b^2 - 1) \big( b(b^2-2)A_2 + 8 (b^2 -1) X_1 \big)  + b^2(3 b^2 - 4) =0 ,
\\
& 8 (b - 1)^2 (b + 1)^2 X_1 -b^2 =0 , 
\\
& 8 b(b^2 -1) A_2 -(b^2 -1)X_1 - 2b^2 B_1 =0 .
\end{align}
They directly yield
\begin{equation}\label{X1.B1.A2}
X_1 = \frac{b^2}{8 (b^2 - 1)^2} , 
\quad
B_1 = - \frac{b^2(2 b^2 - 3)}{4 (b^2 -1)(b^2 -2)} , 
\quad
A_2 = -\frac{b}{8 (b^2 - 2)} .
\end{equation}

This process can be repeated to higher orders.
At the next two orders,
we get
\begin{align}\label{X2.B2.A3}
X_2 &= \frac{b^4 B_0}{16 (b^2 - 1)^3} 
,
\\
B_2 & = -\frac{b^2 B_0}{8 (b^2 - 1)^2} 
,
\\
A_3 & = -\frac{b^3 B_0}{16 (b^2 - 1)(b^2 - 2)} 
,
\end{align}
and
\begin{align}
&\begin{aligned}
X_3 = &
\frac{b^6 B_0^2}{32 (b^2 - 1)^4}
+ \frac{\pi^2 b^4 }{96 (b^2 - 1)^2}
- \frac{b^4 (12 b^6 - 44 b^4 + 53 b^2 - 22)}{128 (b^2 - 2) (b^2 - 1)^4} , 
\end{aligned}
\label{X3}
\\
&\begin{aligned}
B_3 = &
-\frac{b^4 B_0^2}{16 (b^2 - 1)^3}
- \frac{b^2 (13 b^2 - 22) \pi^2}{144  (b^2 - 2)}
\\&\qquad
+ \frac{b^2( 80 b^{10} - 686 b^8 + 2213 b^6 - 3368 b^4 + 2450 b^2 - 692 )}{192 (b^2 - 1)^3 (b^2 - 2)^3} , 
\end{aligned}
\label{B3}
\\
&\begin{aligned}
A_4 = &
- \frac{b^3 B_0^2}{32 (b^2 - 2)(b^2 - 1)^2}
-\frac{\pi^2 b^3}{144 (b^2 - 1)(b^2 - 2)}
+ \frac{b^3( 17 b^6 - 91 b^4 + 148 b^2 - 62 ) }{192 (b^2 - 1)^2 (b^2 - 2)^3} . 
\end{aligned}
\label{A4}
\end{align}

Clearly, the series solution can be continued to any desired order. 

Next, 
we turn to the energy constant \eqref{ener.lump}. 
After substitution of the series \eqref{X.series}--\eqref{A.series},
along with the expressions \eqref{X0.gauge}, \eqref{c}--\eqref{k} and \eqref{A1},
we consider the lowest order term.
This yields the relation 
$E = \tfrac{2}{3} B_0/b^2$.  
In addition, we find that the higher order terms vanish
upon substitution of expressions \eqref{X1.B1.A2}--\eqref{A4},
which follows from the fact that the energy constant is conserved for all solutions.

Finally, we can put $t\to t+T$ in the series solution due to time-translation invariance. 
This gives the following result.

\begin{thm}
The variational equations \eqref{ansatz.eom} have a lump solution 
\begin{equation}\label{series.lump}
\phi(x,t) = \tfrac{1}{2} A(t)\big( \tanh\big(B(t)(x +X(t))\big) - \tanh\big(B(t)(x-X(t))\big) \big)
\end{equation}
given by an explicit power series 
\begin{align}
&\begin{aligned}
X= &
t + T 
+ \frac{b^2}{8 (1 -b^2)^2} \Big(
(t+T)^{-1}
- \Big(
\frac{b^2 K_0}{2 (1-b^2)} \Big) (t+T)^{-2}
\\&\qquad\quad
+ \Big(
\frac{b^4 K_0^2}{4 (1 -b^2)^2}
+ \frac{\pi^2 b^2}{12 (1 -b^2)^2}
+ \frac{b^2( 12 b^6 - 44 b^4 + 53 b^2 - 22 )}{16 (2 -b^2) (1 -b^2)^2} \Big) (t+T)^{-3}
+\cdots \Big) , 
\end{aligned}
\label{series.X}\\
&\begin{aligned}
B = &
\frac{2 (1 -b^2)}{b^2}  (t +T)
+ B_0
+ \frac{b^2}{4(1 -b^2)} \Big(
\Big(\frac{3 -2 b^2}{2 -b^2}\Big) (t+T)^{-1}
\\&\qquad\quad
-\Big( 
\frac{B_0}{2 (1 -b^2)} \Big) (t+T)^{-2}
+ \Big( 
\frac{b^2 B_0^2}{4 (1 -b^2)^2}
+ \frac{\pi^2(22-13 b^2)}{36 (2 -b^2)}
\\&\qquad\qquad\qquad
+ \frac{80 b^{10} - 686 b^8 + 2213 b^6 - 3368 b^4 + 2450 b^2 - 692}{48 (1 -b^2)^2 (2 -b^2)^3} \Big) (t+T)^{-3}
+\cdots \Big) , 
\end{aligned}
\label{series.B}\\
&\begin{aligned}
A = &
\frac{1}{b} +\frac{b}{8 (2 -b^2)} \Big(
(t+T)^{-2} - \Big( 
\frac{b^2 B_0}{2 (1 -b^2)} \Big) (t+T)^{-3}
\\&\qquad\quad
+\Big( 
\frac{b^4 B_0^2}{4 (1 -b^2)^2} 
-\frac{\pi^2 b^2}{18 (1 -b^2)}
- \frac{b^2( 17 b^6 - 91 b^4 + 148 b^2 - 62 )}{24 (1-b^2)^2 (2-b^2)^2} \Big) (t+T)^{-4}
+\cdots \Big) , 
\end{aligned}
\label{series.A}
\end{align}
involving two free parameters $T$ and $B_0$. 
The conserved energy of the solution is given by
\begin{equation}\label{series.E}
E = \tfrac{2}{3} B_0/b^2 
.
\end{equation}
\end{thm}

This solution will be treated as an asymptotic series,
leaving aside the difficult question of its convergence as a power series. 

As a main result,
the series solution will now be shown to have the same qualitative behaviour
as the numerical solutions for large $t$.

\subsection{Asymptotic behaviour}
\label{sec:asymptotics}

The asymptotic behaviour of $\phi(x,t)$ as $t\to\infty$
can be obtained from expanding expressions \eqref{series.X}--\eqref{series.E}
in powers of $1/t$. 
To order $1/t$, we have 
\begin{align}
X= & 
t +T +\frac{b^2}{8 (1-b^2)^2} t^{-1}
+ O( t^{-2} ),
\\
B = & 
\frac{2 (1 -b^2)}{b^2}  (t + T) + \frac{3b^2 E}{2}
+ \frac{4b^2(3 -2 b^2)}{(1 -b^2)(2 -b^2)} t^{-1} 
+ O( t^{-2} ),
\\
A = & 
\frac{1}{b} +O( t^{-2} ) , 
\end{align} 
which yields
\begin{equation}\label{asympt}
\begin{aligned}
\phi(x,t) \sim & 
\frac{1}{2b} \big( \tanh(Mx+N) - \tanh(Mx-N) \big)
\\& \qquad
+ \Big(
\frac{b (3-2 b^2)}{8(2-b^2) (1-b^2)} x (\sech(Mx+N)^2 - \sech(Mx-N)^2) 
\\& \qquad\qquad
-\frac{3b E}{2(1 - b^2)^2} (\sech(Mx-N)^2 + \sech(Mx+N)^2) 
\Big) t^{-1}
+O( t^{-2} )
\end{aligned}
\end{equation}
where
\begin{equation}
\begin{aligned}  
M = & \Big( \frac{2 (1 - b^2)}{b^2} (t +T) + \frac{3E}{2} \Big) , 
\\
N = & \Big( \frac{2 (1 - b^2)}{b^2}  (t + T) + \frac{3E}{2}  \Big) (t + T)
+ \frac{1 - b^2(1+ b^2)}{2(1 -b^2)(2-b^2)} . 
\end{aligned}
\end{equation}
Note that
$\sech(Mx-N)^2 -\sech(Mx+N)^2 = 4 \sinh(2 Mx) \sinh(2 N)/(\cosh(2 Mx) + \cosh(2 N))^2$
and
$(\sech(Mx-N)^2 + \sech(Mx+N)^2) = 4 (\cosh(2 Mx) \cosh(2 N) + 1)/(\cosh(2 Mx) + \cosh(2 N))^2$.

Therefore, this asymptotic profile \eqref{asympt} approaches a spreading tabletop shape
with height
\begin{equation}
\phi(0,t) \sim  \frac{1}{b} +O( t^{-2} )
\end{equation}
and width
\begin{equation}
\Delta x \sim 2(t + T)   +\frac{b^2}{4 (1-b^2)^2} t^{-1} + O( t^{-2} ) . 
\end{equation}
given by the positions of its flanks, 
\begin{equation}
x=\pm X= \pm \Big( t +T +\frac{b^2}{8 (1-b^2)^2} t^{-1} \Big) + O( t^{-2} ) . 
\end{equation}
The flanks steepen to become vertical and accelerate outward to light-cone speed, 
\begin{equation}
\frac{dX}{dt} = \nu \sim 1 -\frac{b^2}{8 (1-b^2)^2} t^{-2} + O( t^{-3} )
<1 . 
\end{equation}
In particular, the energy density, $\mathcal{E}$, becomes concentrated at the flanks:
\begin{equation}
\begin{aligned}
\mathcal{E} = & 
\frac{(1-b^2)^2}{b^6}\Big( 
\big(
\tfrac{5}{2} \sech(Mx-N)^4 + 3 \sech(Mx-N)^2 \sech(Mx+N)^2 + \tfrac{5}{2} \sech(Mx+N)^4
\big) (t+T)^2
\\& 
+\Big( 2 x\big(\sech(Mx+N)^4 -\sech(Mx-N)^4\big) 
+\Big( 6 T + \frac{3b^2}{2 (1-b^2)}E \Big) \sech(Mx+N)^2 \sech(Mx-N)^2
\\& 
+ \Big( 5 T+ \frac{9b^2}{4 (1-b^2)}E \Big) (\sech(Mx+N)^4 + \sech(Mx-N)^4)
\Big) (t+T)
\Big)
+O(1) 
\end{aligned}
\end{equation}
as $\sech(M\pm N)^2$ is very sharply peaked around $x=\pm X$,
with the full width being approximately $b^2/((1 - b^2) t)$, as $t\to\infty$.

\subsection{Dependence on the parameter in the Klein-Gordon model}

The dynamics of the perturbed sphaleron depends essentially on the parameter $b$
in the nonlinear KG equations \eqref{b.KG}. 
In particular, $b$ controls the depth of the false vacuum $\phi = 1/b$,
which determines the strength of the instability
as a function of $b$ in the range \eqref{b.range}.

As $b\to 1$, 
the false vacuum becomes shallow,
and the instability develops slowly. 
This can be seen from the time scale \eqref{timescale.approx}
which becomes large 
\begin{equation}
\tau \approx 1/(4\sqrt{1-b}) , 
\end{equation}
implying that the sphaleron remains close to its initial profile for a relatively long time
before expanding. 
In contrast, as $b\to 0$,
the false vacuum is deep, 
and the instability develops on a short time scale 
\begin{equation}
\tau \approx (27\sqrt{5}/350) b^2
\end{equation}
This leads to a rapid onset of expansion of the sphaleron.
Similar behaviour is seen in the numerical solutions,
where the formation of the tabletop profile occurs
at early times for values of $b$ near $0$
and at late times for values of $b$ close to $1$. 

The steepness and acceleration of the flanks are given by the respective expressions
\begin{equation}\label{steep}
  B(t) = 2(1/b^2 -1 ) t + O(1)
\end{equation}
and
\begin{equation}\label{accel}
\frac{d^2X}{dt^2} = (b/2)^2(1-b^2)^{-2} t^{-3} + O( t^{-4} )
\end{equation}
as $t\to\infty$. 
Notice that the steepness \eqref{steep} is greater for smaller values of $b$, 
indicating that the transition region between vacua becomes sharper
when the false vacuum is deeper.
Acceleration \eqref{accel} is higher for values of $b$ near $1$,
when the false vacuum is shallower,
which implies that the inertial of the flanks is greater when they are steeper.

\section{Approximate behaviour of the perturbed sphaleron}
\label{sec:approx}

The series solution \eqref{series.lump}--\eqref{series.E} for the collective coordinate dynamics
will now be shown to approximate 
the evolution of the sphaleron \eqref{kinkantikink} with $x_0=0$
after the unstable mode is excited. 
We do not expect the solution to match the perturbation at $t=0$,
which would require matching the initial data
\eqref{perturb.lump.initialdata} or \eqref{perturb.lump.initialdata.kick} 
in one of the two scenarios considered for the perturbation. 
However, we can aim to find a time $t=\tzero>0$ as small as possible
for which $\phi(x,\tzero)$ and $\phi_t(x,\tzero)$ given by the series solution 
approximately match the perturbed sphaleron.

Note that the time derivative of expressions \eqref{series.X}--\eqref{series.A}
is non-vanishing and consequently the series solution has $\phi_t \neq 0$ 
as seen from expression \eqref{time.derivative.ansatz}.
Hence we cannot match the initial data \eqref{perturb.lump.initialdata}
having $\phi_t=0$.
So we consider instead the other scenario in which the initial data \eqref{perturb.lump.initialdata.kick}
has $\phi_t$ proportional to the unstable mode.

\subsection{Perturbation of the sphaleron at early times}

For short times after $t=0$,
the perturbed sphaleron can be expected to have
its profile being close to $\phi_\lump$
and its time derivative being close to the initial kick \eqref{kick.initialdata}. 
We will seek $\tzero$ as small as possible so that in the series solution 
$\phi|_{\tzero}$ and $\phi_t|_{\tzero}$ are close to this initial data. 

Specifically, we aim to match the height of $\phi$ at $x=0$ (cf \eqref{height})
\begin{equation}\label{height.cond}
\phi(0,\tzero) = A(\tzero)\tanh\big(B(\tzero)X(\tzero)\big)
\approx
\phi_\lump(0)
= \frac{1}{3b}\big( 2 + b^2 - \sqrt{(1-b^2)(2-b^2)} \big) , 
\end{equation}
and also  the full-width of $\phi$ (cf \eqref{width})
\begin{equation}\label{width.cond}
\begin{aligned}
\Delta x\big|_{\tzero} &
= \arccosh\Big( 2\cosh(2B(\tzero)X(\tzero)) + \sqrt{3\cosh(2B(\tzero)X(\tzero))^2 +6} \Big)/B(\tzero)
\\& 
\approx
\Delta x_\lump
= \arccosh\Big( \big(4 +2b^2 + 3\sqrt{(1-b^2)^2 +3}\big)/\sqrt{(1-b^2)(2-b^2)} \Big) 
\end{aligned}
\end{equation}
defined by where $\phi(0,\tzero)$ has maximum convexity.
In addition, we want to match the amplitude of $\phi_t$. 
From the approximation \eqref{groundstate.approx},
the peak amplitude occurs at $x_\peak=0$ for $b\gtrsim 0.9$ ($a \gtrsim 1)$,
and at the double peak, $x_\peak\neq0$, for $b\lesssim 0.9$ ($a \lesssim 1)$. 
Thus, we require 
\begin{equation}\label{kick.cond}
\begin{aligned}
\phi_t(x_\peak,\tzero) & =
\tfrac{1}{2} A'(\tzero)\big( \tanh(B(\tzero)x_\peak +C(\tzero)) - \tanh(B(\tzero)x_\peak -C(\tzero)) \big)
\\&\quad 
+ \tfrac{1}{2} A(\tzero)\Big( (B'(\tzero)x_\peak +C'(\tzero)) \sech^2(B(\tzero)x_\peak +C(\tzero))
\\&\qquad\qquad\quad
- (B'(\tzero)x_\peak -C'(\tzero)) \sech^2(B(\tzero)x_\peak -C(\tzero)) \Big)
\\& \approx 
\dfrac{\epsilon}{\tau} \phi_\lump(0) \eta_{-1}|\peak
\end{aligned}
\end{equation}
where $\eta_{-1}|_\peak$ is the peak amplitude \eqref{singlepeak}--\eqref{doublepeak}, 
and $\tau$ is the approximate time scale \eqref{timescale.approx} of the mode $\eta_{-1}$.
(Recall that this mode was normalized to satisfy $\eta_{-1}(0)=1$
and that it is scaled by the height of $\phi_\lump$
in the perturbation
\eqref{kick.initialdata}.)
We will also require that the outward speed of the flanks
is initially positive and increasing, 
\begin{equation}\label{lightcone.cond}
\nu(\tzero) = X'(\tzero) >0,
\quad
\nu'(\tzero) = X''(\tzero) >0
\end{equation}

We now put $E=E_\lump + E_\text{perturb}$ in the series solution \eqref{series.lump}--\eqref{series.A}, 
where $E_\lump$ is the sphaleron energy \eqref{lump.E} 
and $E_\text{perturb}$ is the correction to this energy
due to the initial kick \eqref{kick.initialdata}.
(See Appendix~\ref{appendix:ener.correction}.)
This determines
\begin{equation}\label{B0}
B_0 = \tfrac{3}{2} b^2 (E_\lump + E_\text{perturb})
\end{equation}
so the resulting series has just one free parameter, namely $T$.

To impose all of the preceding conditions \eqref{height.cond}--\eqref{lightcone.cond} in practice, 
we first expand the series \eqref{series.X}--\eqref{series.A} in powers of $1/t$,
and truncate the expansion at 5 or 6 terms,
and then we search for values of $\tzero$ and $T$ so that each of the quantities
\begin{equation}
|1- \phi(0,\tzero)/\phi_\lump(0)| , 
\quad
|1- \Delta x|_{\tzero}/\Delta x_\lump| , 
\quad
|1- \phi_t(x_\peak,\tzero) \tau/(\epsilon\phi_\lump(0)\eta_{-1}|_\peak)|
\end{equation}
is smaller than some small bound $\varepsilon$,
namely $\varepsilon \simeq$ 0.05 to 0.15. 

The behaviour seen in all of the series solutions, presented next, 
qualitatively agrees with the behaviour in the numerical solutions.
Accuracy of the series solutions is verified by plotting their error
as defined by evaluation of the left-hand side of
the actual equation of motion \eqref{eom.potential} in the potential \eqref{potential}
(which would be $0$ for an exact solution).
Hereafter, $t'=t-\tzero$ denotes the time elapsed from $t=\tzero$.

\subsubsection{Example $b=0.998$ ($a=2.0$)}
\begin{figure}
\includegraphics[width=0.48\textwidth,trim=2cm 12cm 2cm 1cm,clip]{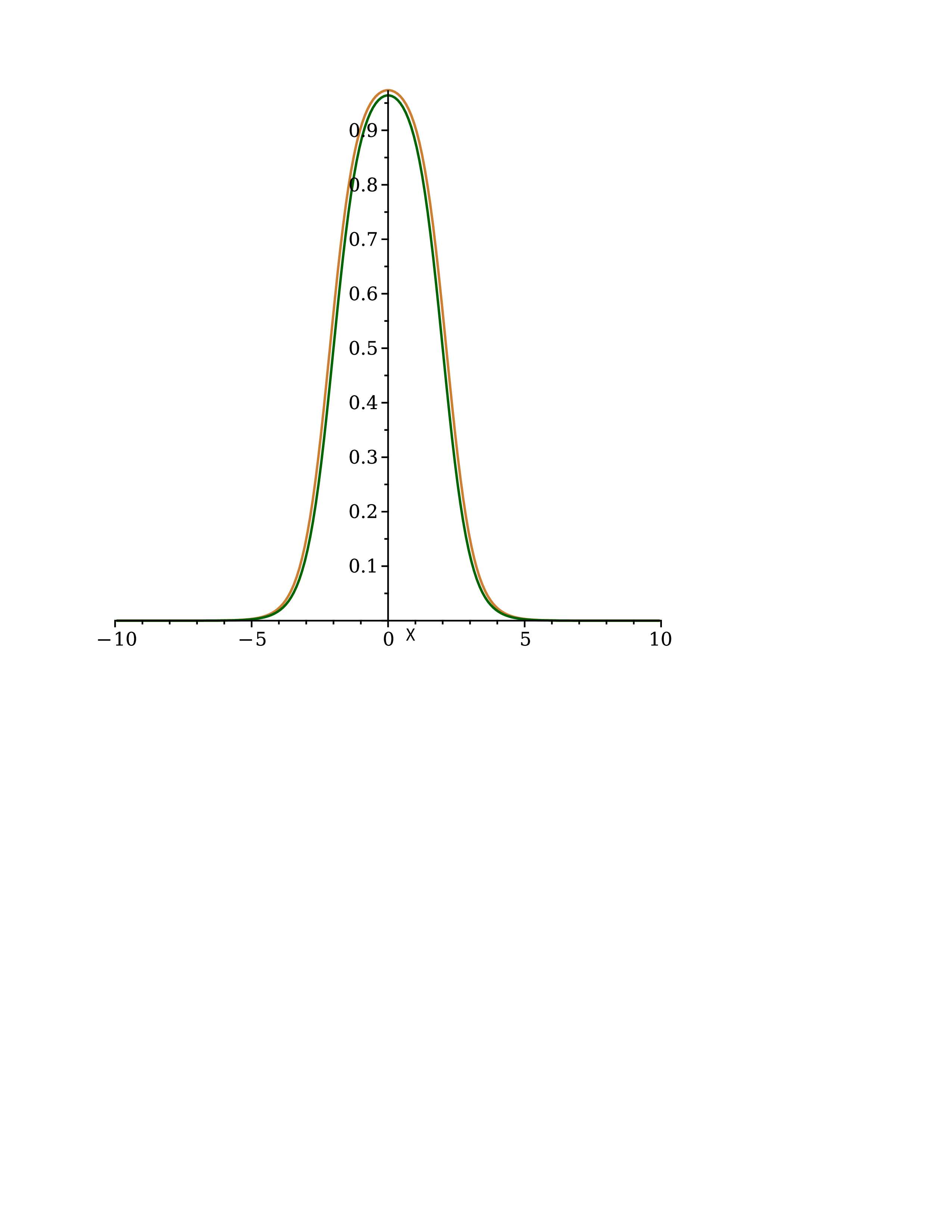}
\includegraphics[width=0.48\textwidth,trim=2cm 12cm 2cm 1cm,clip]{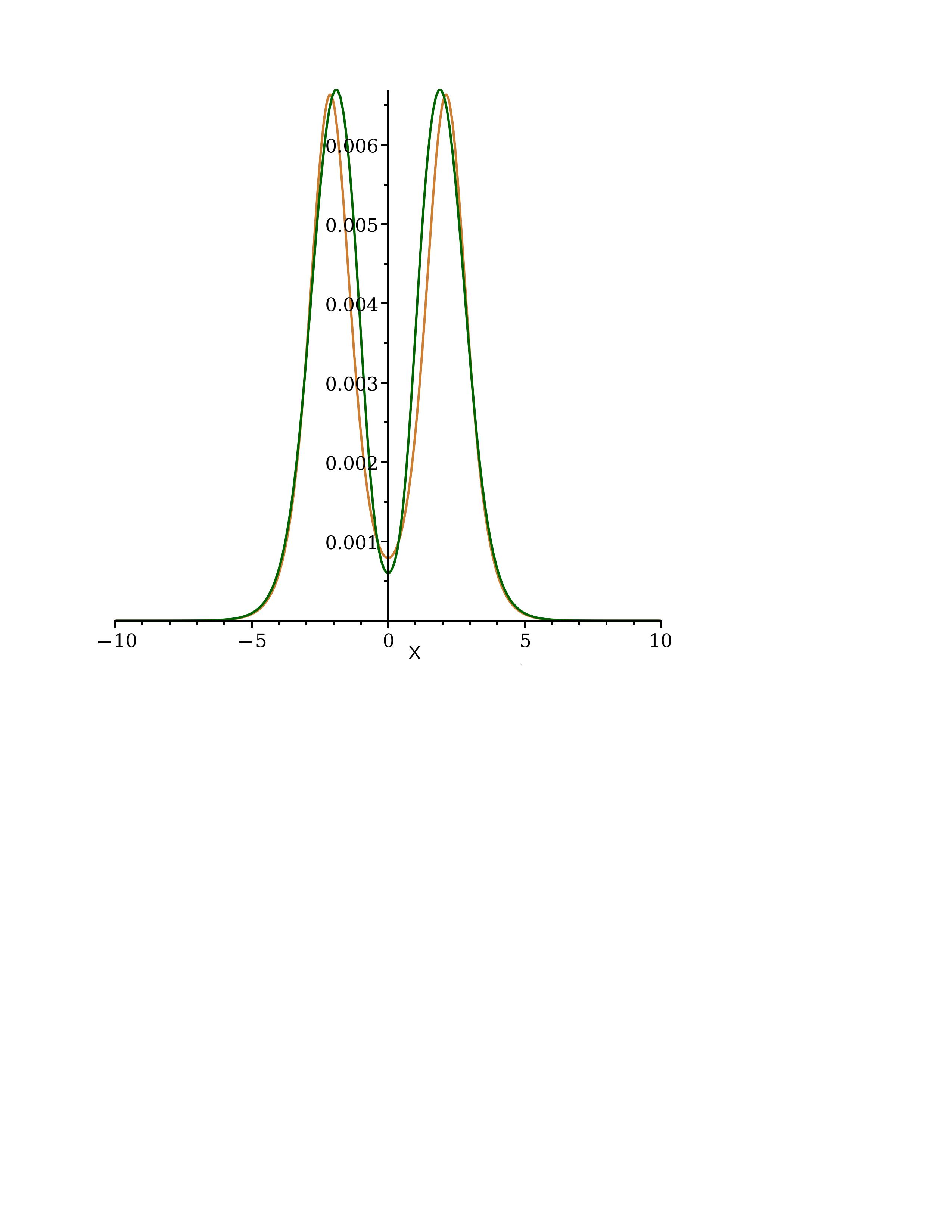}
\caption{
$b=0.998$ ($a=2.0$):
\quad  
(Left)\ $\phi$; 
\quad
(Right)\ $\phi_t$.
\quad
Dark green represents the initial data of the sphaleron; 
Brown represents the series solution.
}
\label{fig:match.b=0.998}
\end{figure}

We find $T=-110.12$ and $\tzero=53.88$.
Fig.~\ref{fig:match.b=0.998} shows the matching to initial data \eqref{perturb.lump.initialdata.kick} with $\epsilon =0.005$ (and $\varepsilon = 0.01$). 

\begin{figure}
\includegraphics[width=0.50\textwidth,trim=2cm 9cm 2cm 4cm, clip]{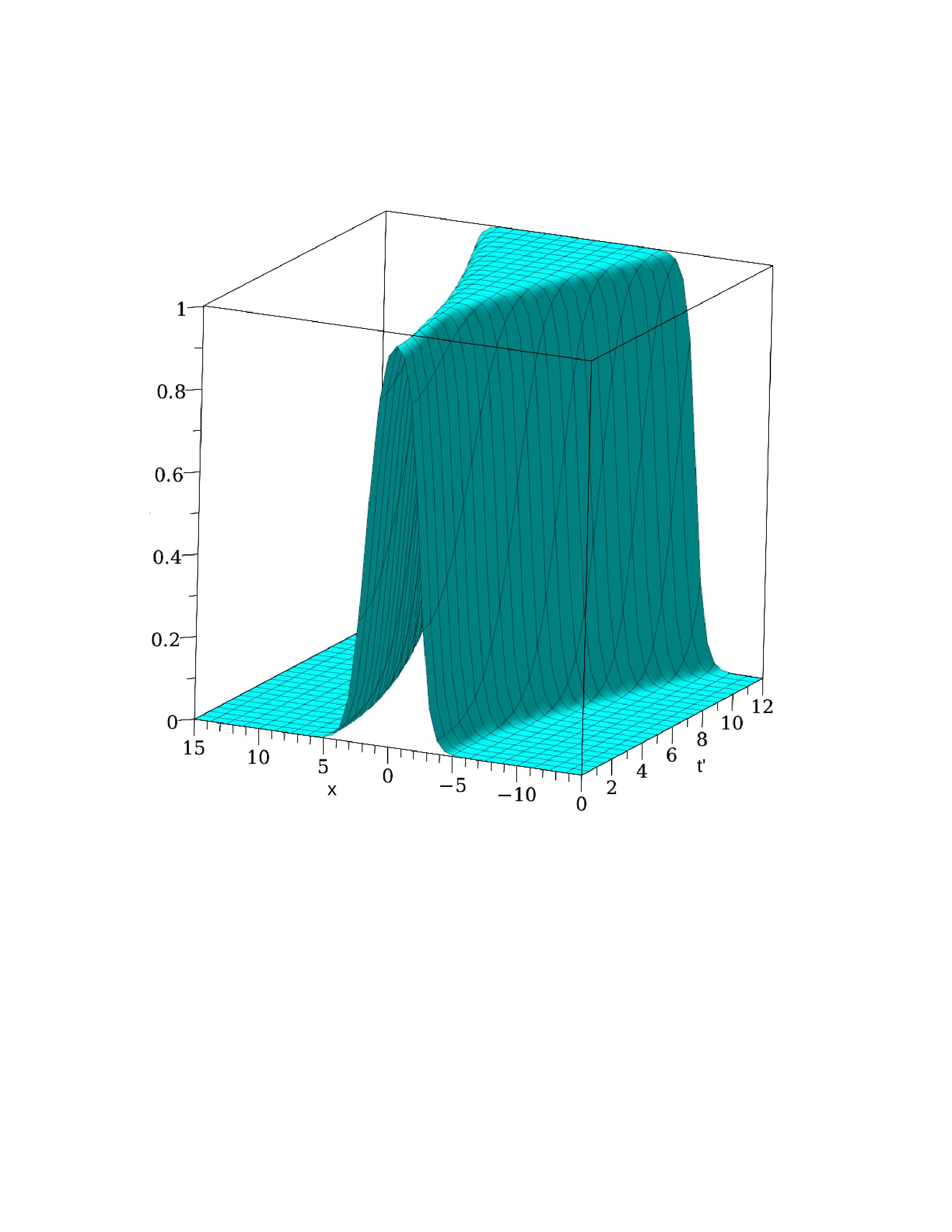}
\includegraphics[width=0.45\textwidth,trim=2cm 12cm 2cm 1cm, clip]{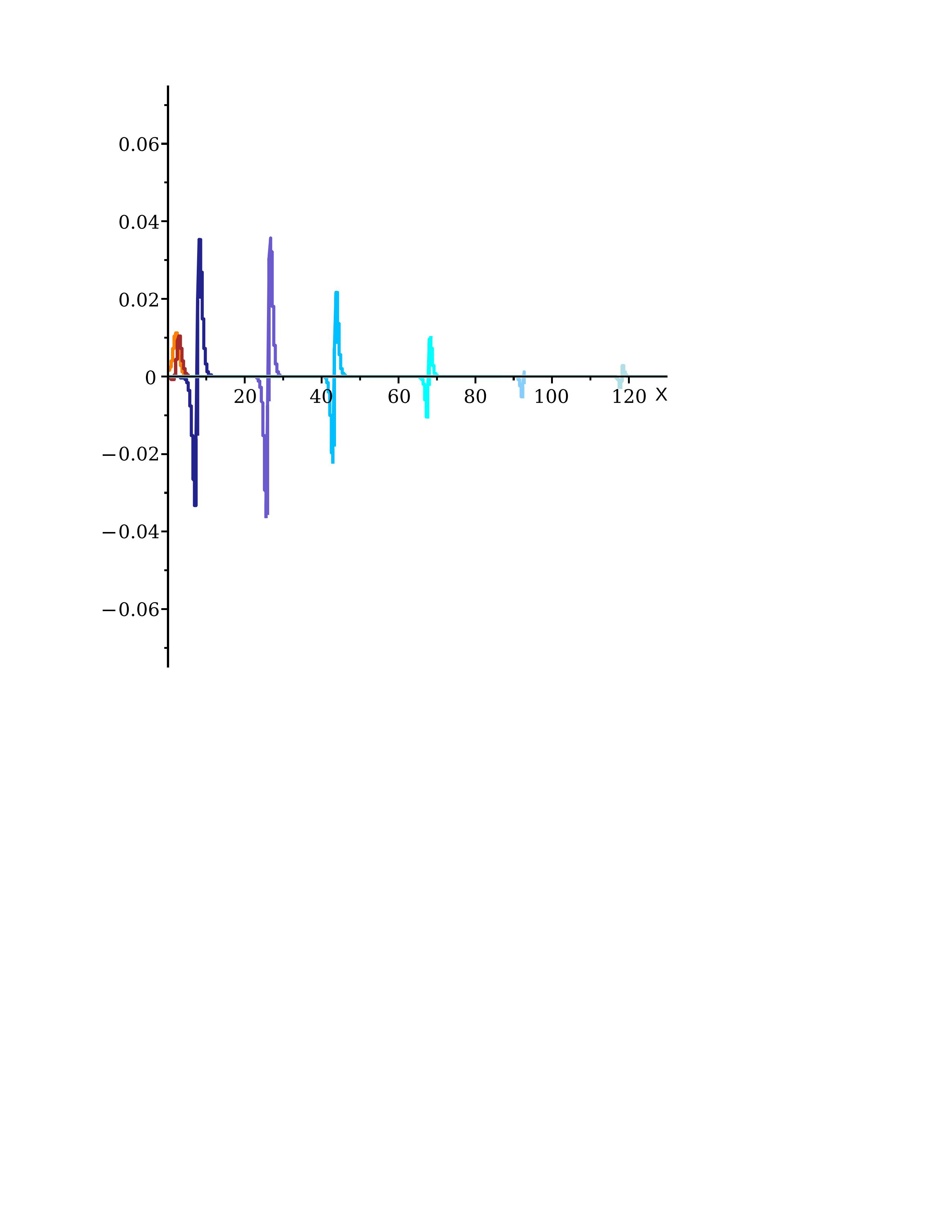}
\caption{
(Left)\ 
Series solution for $b=0.998$ ($a=2.0$). 
\quad
(Right)\
Error plot at times $t'=0,2,10,30,50,80,110$.
}  
\label{fig:soln.b=0.998}
\end{figure}

\begin{figure}
\includegraphics[width=0.48\textwidth,trim=2cm 12cm 2cm 1cm,clip]{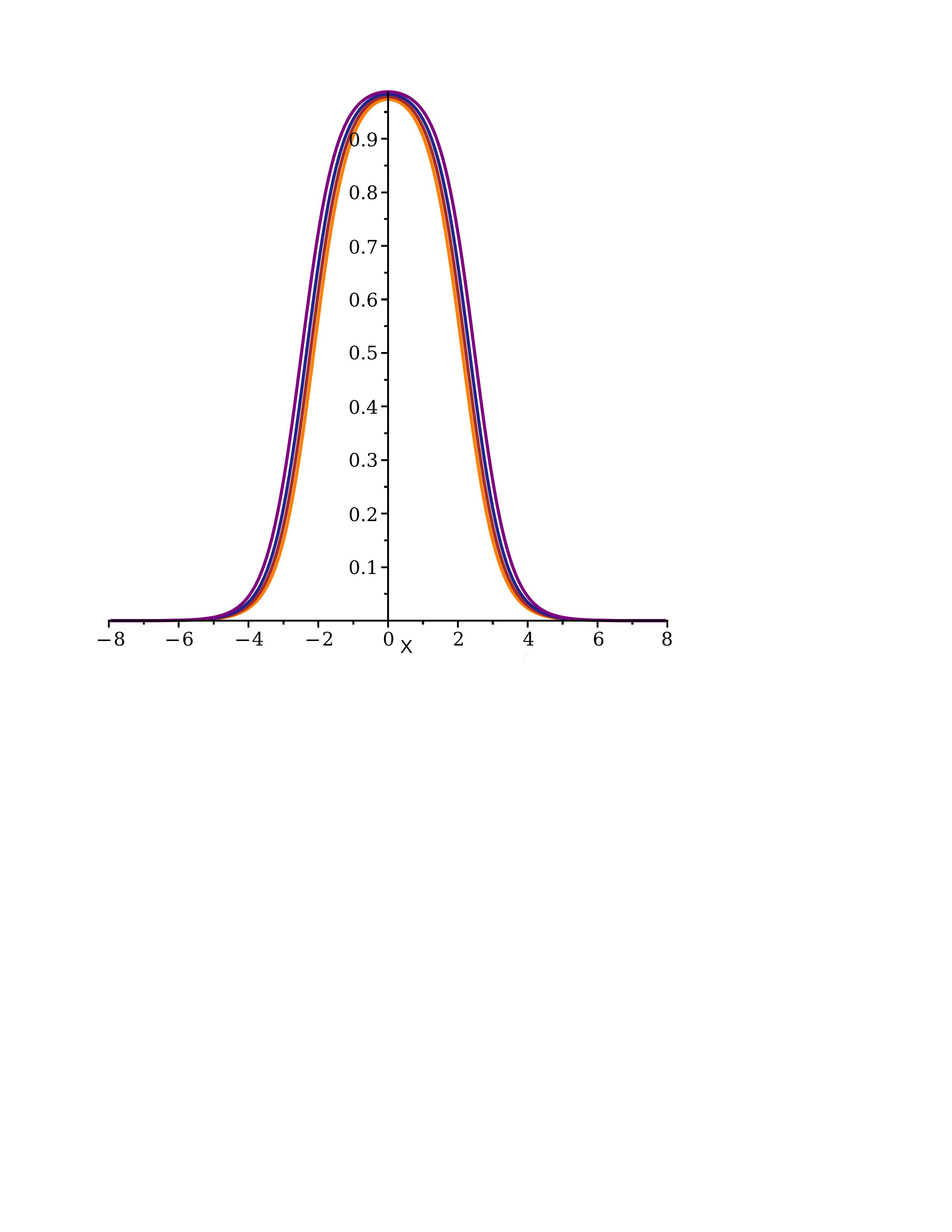}
\includegraphics[width=0.48\textwidth,trim=2cm 12cm 2cm 1cm,clip]{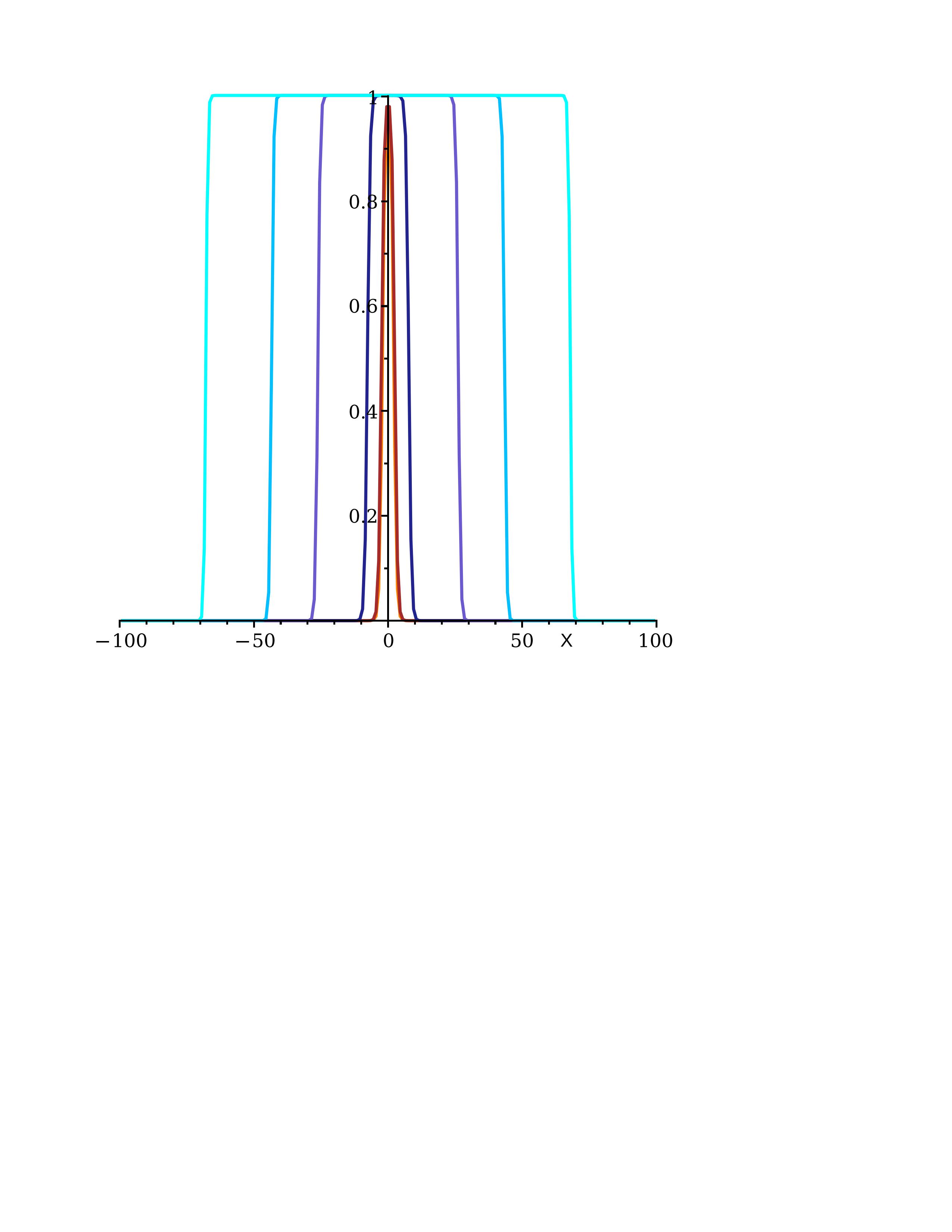}
\caption{
$b=0.998$ ($a=2.0$):
\quad
(Left)\ Short times $t'=0,0.2,0.5,0.75$
\quad
(Right)\ Long times $t'=0,2,10,30,50,80$ 
  }
\label{fig:shortlong.time.b=0.998}
\end{figure}

The perturbed sphaleron solution and its error are is shown in Fig.~\ref{fig:soln.b=0.998}.
Notice that the error decreases steadily to less than $1\%$.
Profiles of this solution at short times $t'=0,1.0,1.5,2.0$ (up to peak height)
and long times $t'=2,10,30,50,80$ 
are shown in Fig.~\ref{fig:shortlong.time.b=0.998}.

\subsubsection{Example $b=0.95$ ($a=1.174$)}
\begin{figure}
\includegraphics[width=0.48\textwidth,trim=2cm 12cm 2cm 1cm,clip]{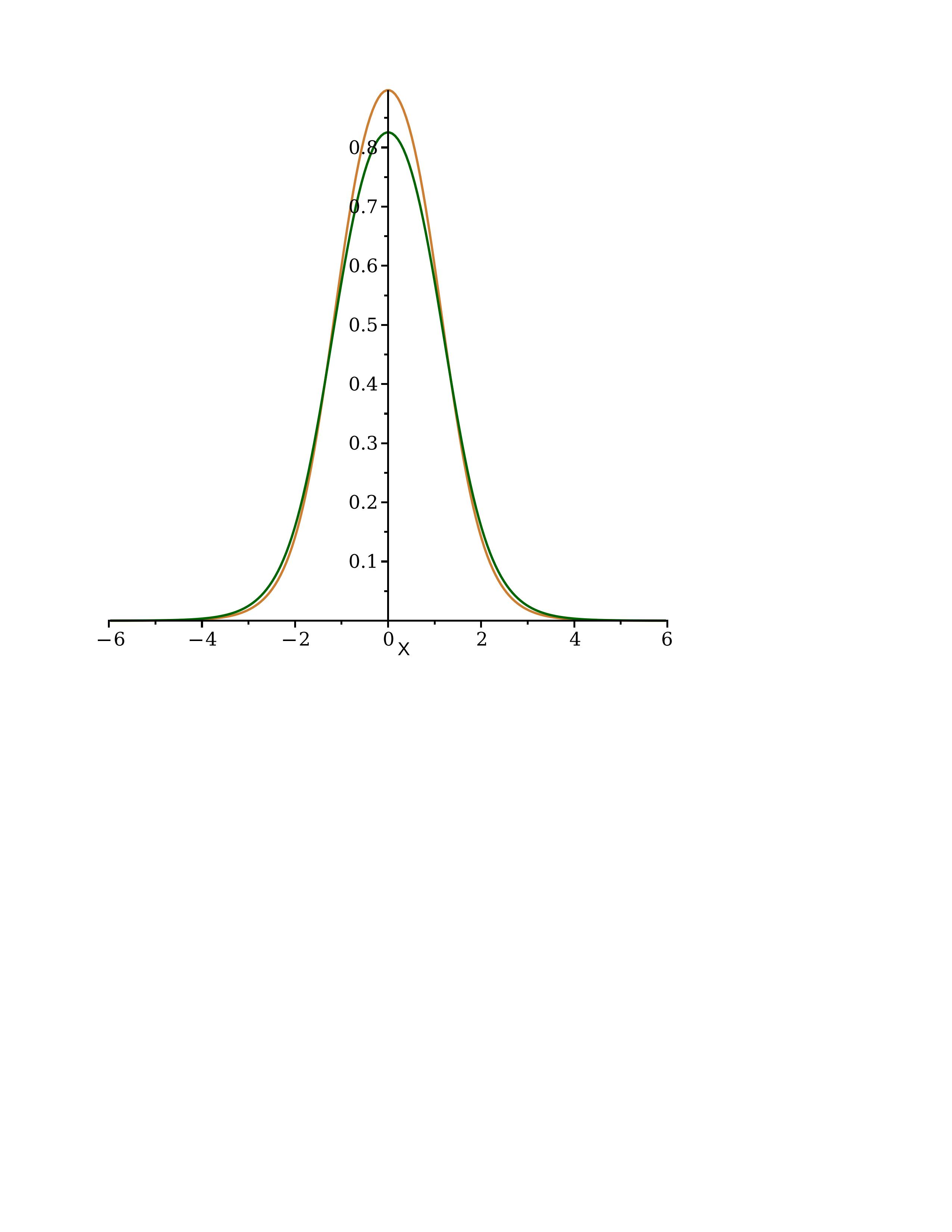}
\includegraphics[width=0.48\textwidth,trim=2cm 12cm 2cm 1cm,clip]{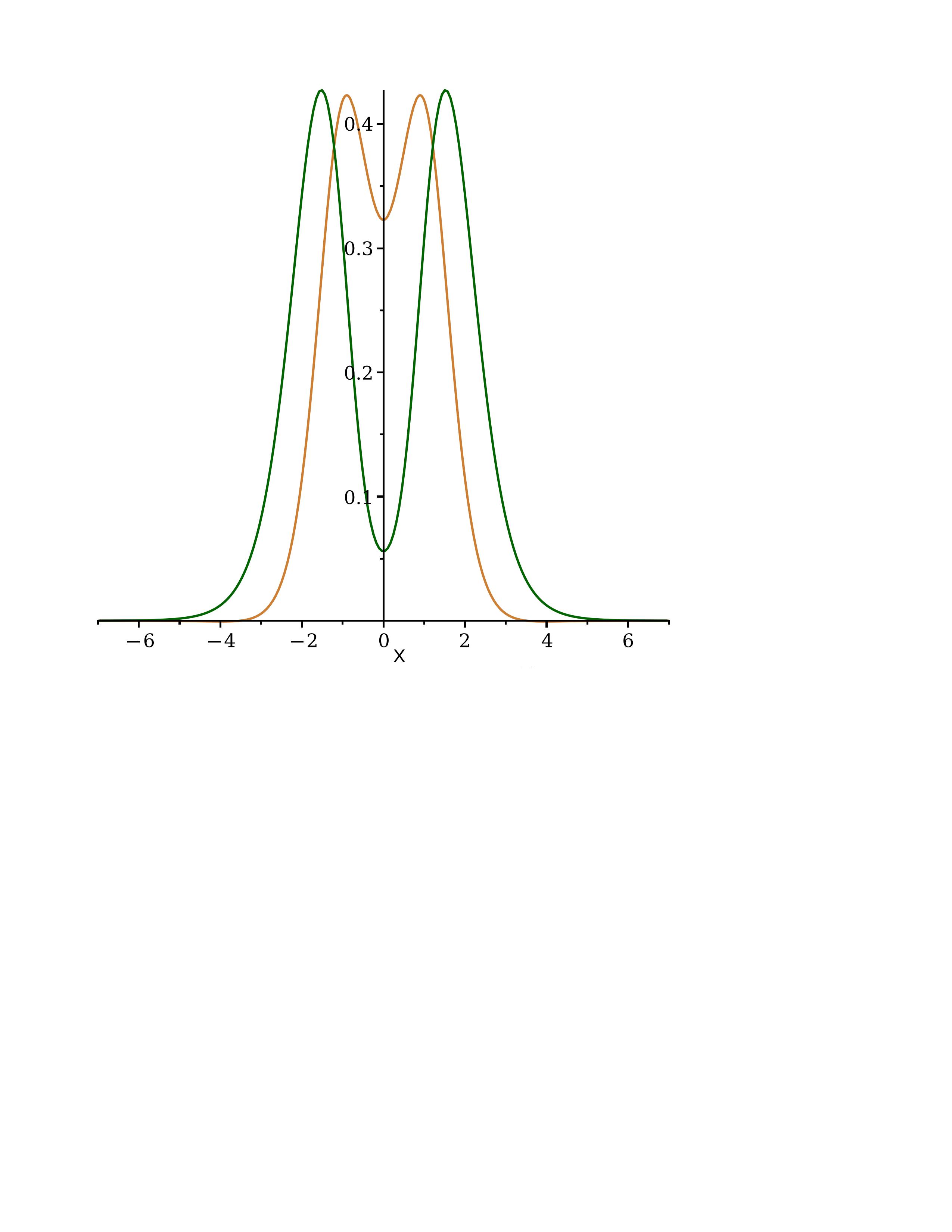}
\caption{
$b=0.95$ ($a=1.174$):
\quad
(Left)\ $\phi$; 
\quad
(Right)\ $\phi_t$.
\quad
Dark green represents the initial data of the sphaleron; 
Brown represents the series solution.
}
\label{fig:match.b=0.95}
\end{figure}

We find $T=-6.09
$ and $\tzero=4.67$.
Fig.~\ref{fig:match.b=0.95} shows the matching to initial data \eqref{perturb.lump.initialdata.kick}, with $\epsilon =0.07$ and ($\varepsilon=0.07$).

\begin{figure}
\includegraphics[width=0.50\textwidth,trim=2cm 9cm 2cm 4cm, clip]{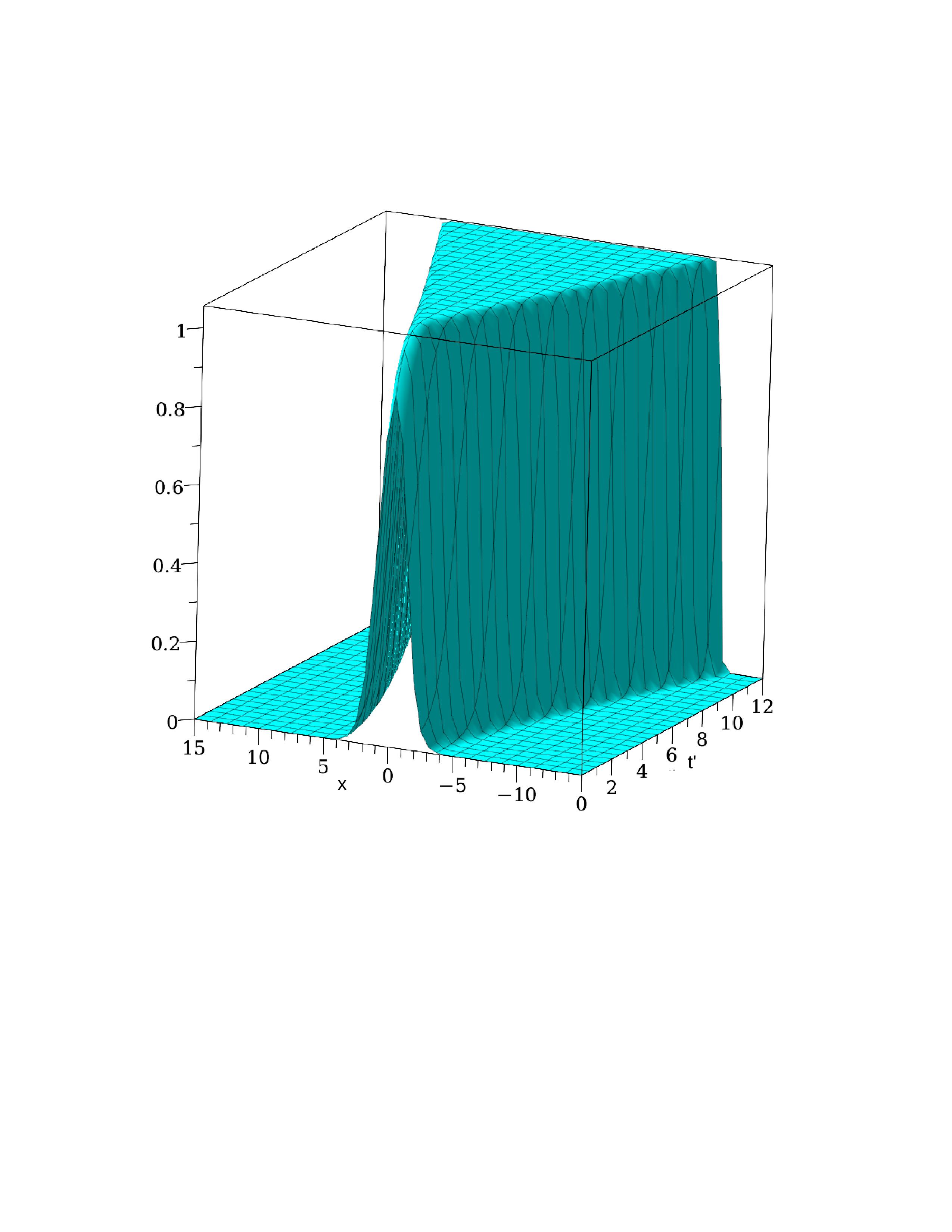}
\includegraphics[width=0.45\textwidth,trim=2cm 12cm 2cm 1cm, clip]{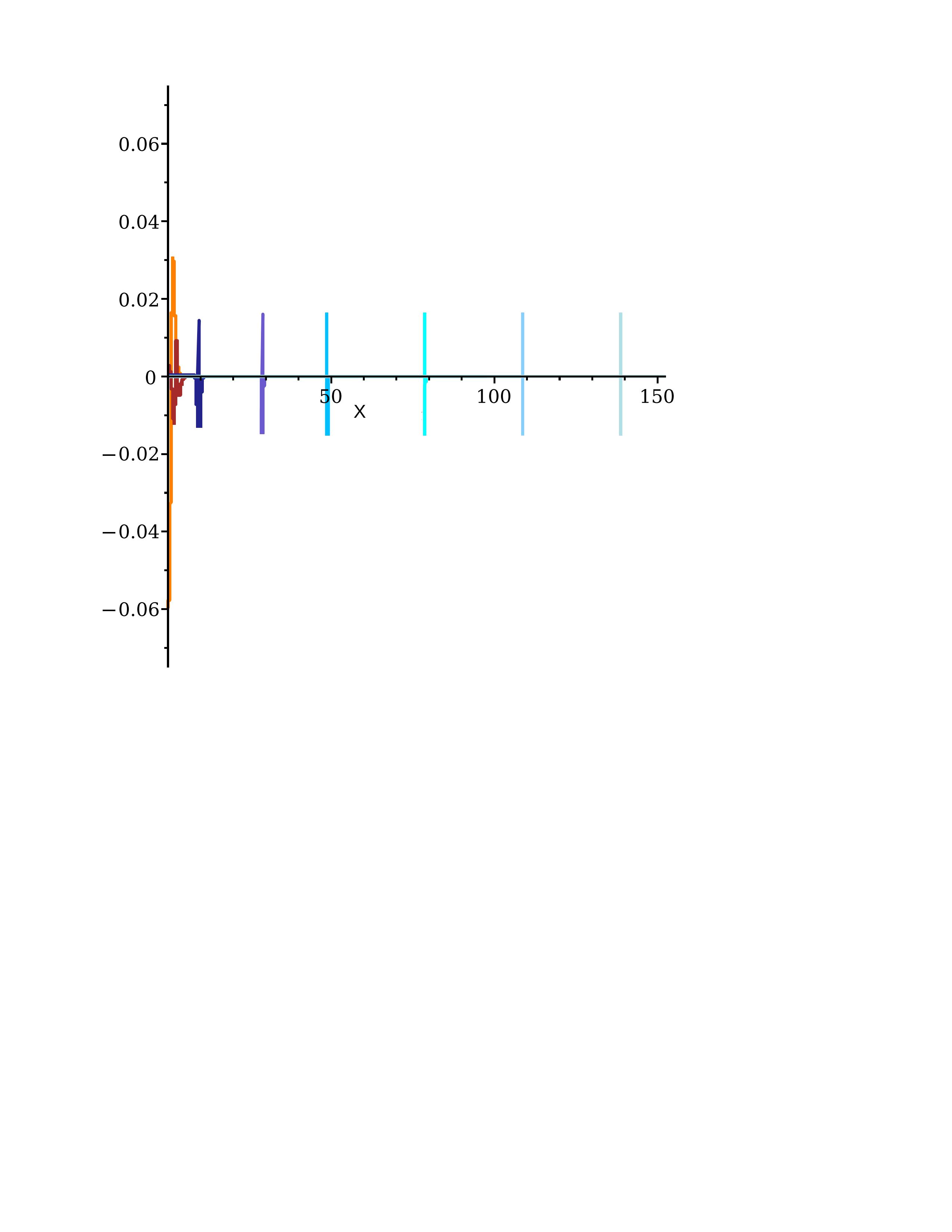}
\caption{
(Left)\ 
Series solution for $b=0.95$ ($a=1.174$). 
\quad
(Right)\
Error plot at times $t'=0,2,10,30,50,80,110$.
}  
\label{fig:soln.b=0.95}
\end{figure}

\begin{figure}
\includegraphics[width=0.48\textwidth,trim=2cm 12cm 2cm 1cm,clip]{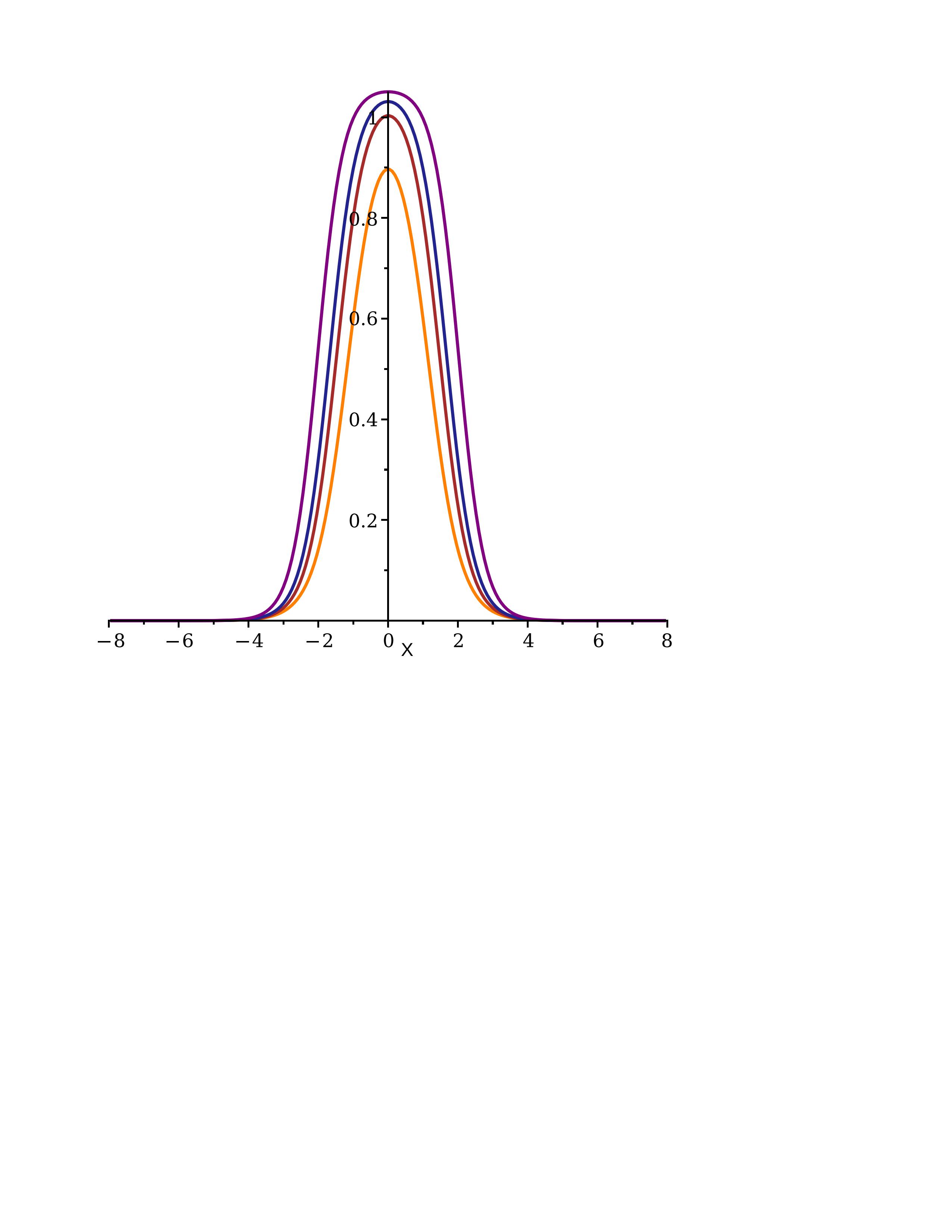}
\includegraphics[width=0.48\textwidth,trim=2cm 12cm 2cm 1cm,clip]{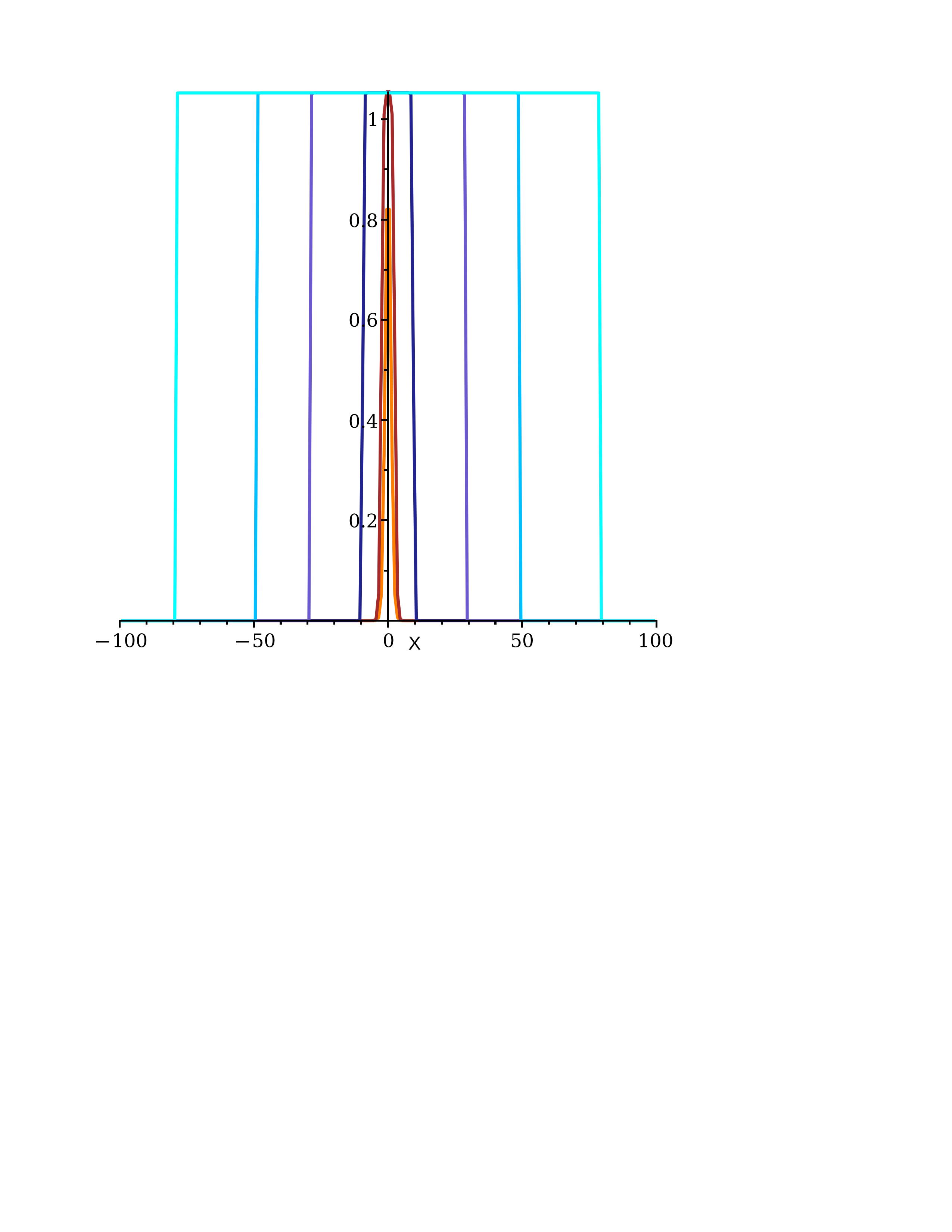}
\caption{
$b=0.95$ ($a=1.174$):
\quad
(Left)\ Short times $t'=0,0.5,0.8,1.3$; 
\quad
(Right)\ Long times $t'=0,2,10,30,50,80$
}
\label{fig:shortlong.time.b=0.95}
\end{figure}

The perturbed sphaleron solution and its error are shown in Fig.~\ref{fig:soln.b=0.95}.
Notice that the error is roughly constant $2\%$.
Profiles of this solution at short times $t'=0,0.5,0.8,1.3$ (up to peak height)
and long times $t'=0,2,10,30,50,80$ 
are shown in Fig.~\ref{fig:shortlong.time.b=0.95}.

\subsubsection{Example $b=0.7$ ($a=0.616$)}
\begin{figure}
\includegraphics[width=0.48\textwidth,trim=2cm 12cm 2cm 1cm,clip]{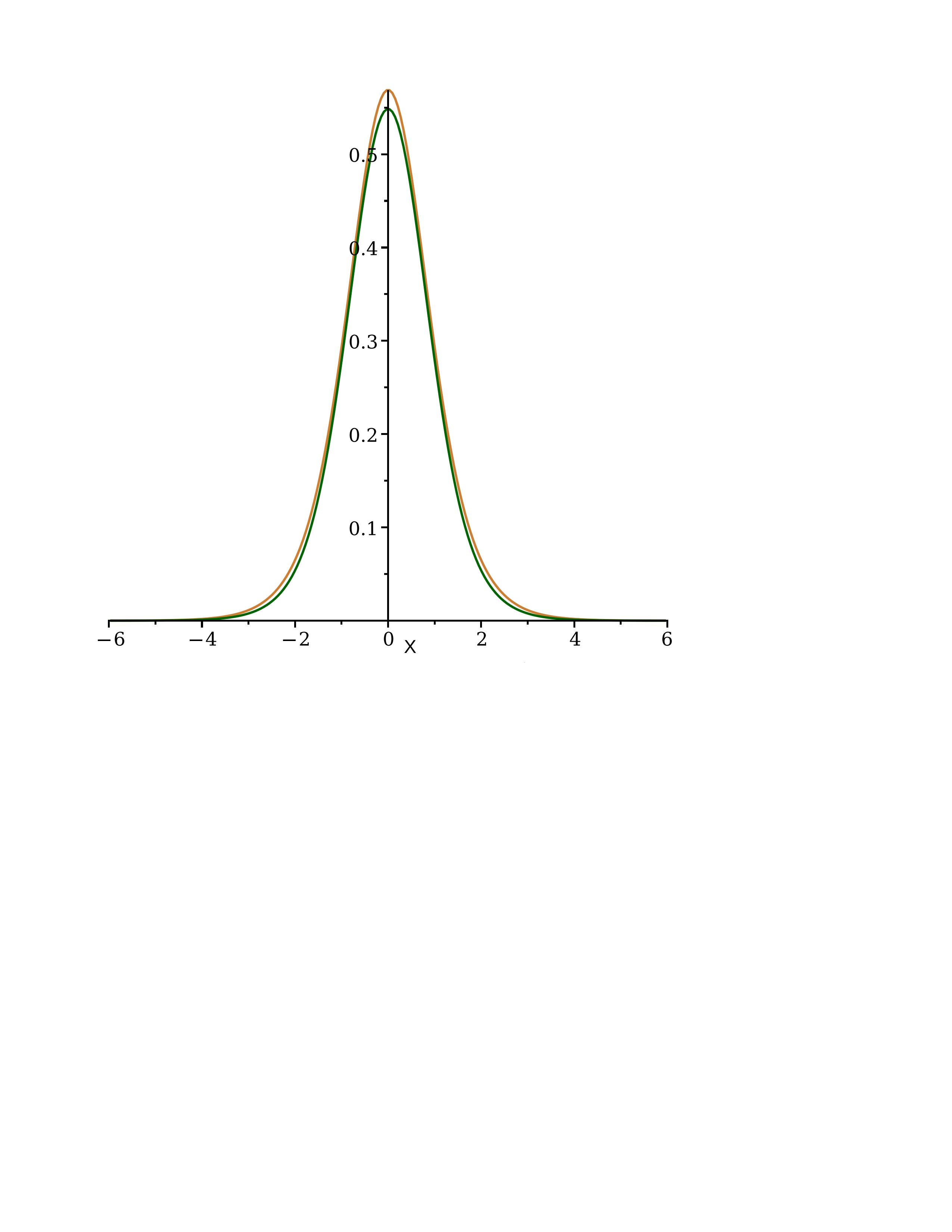}
\includegraphics[width=0.48\textwidth,trim=2cm 12cm 2cm 1cm,clip]{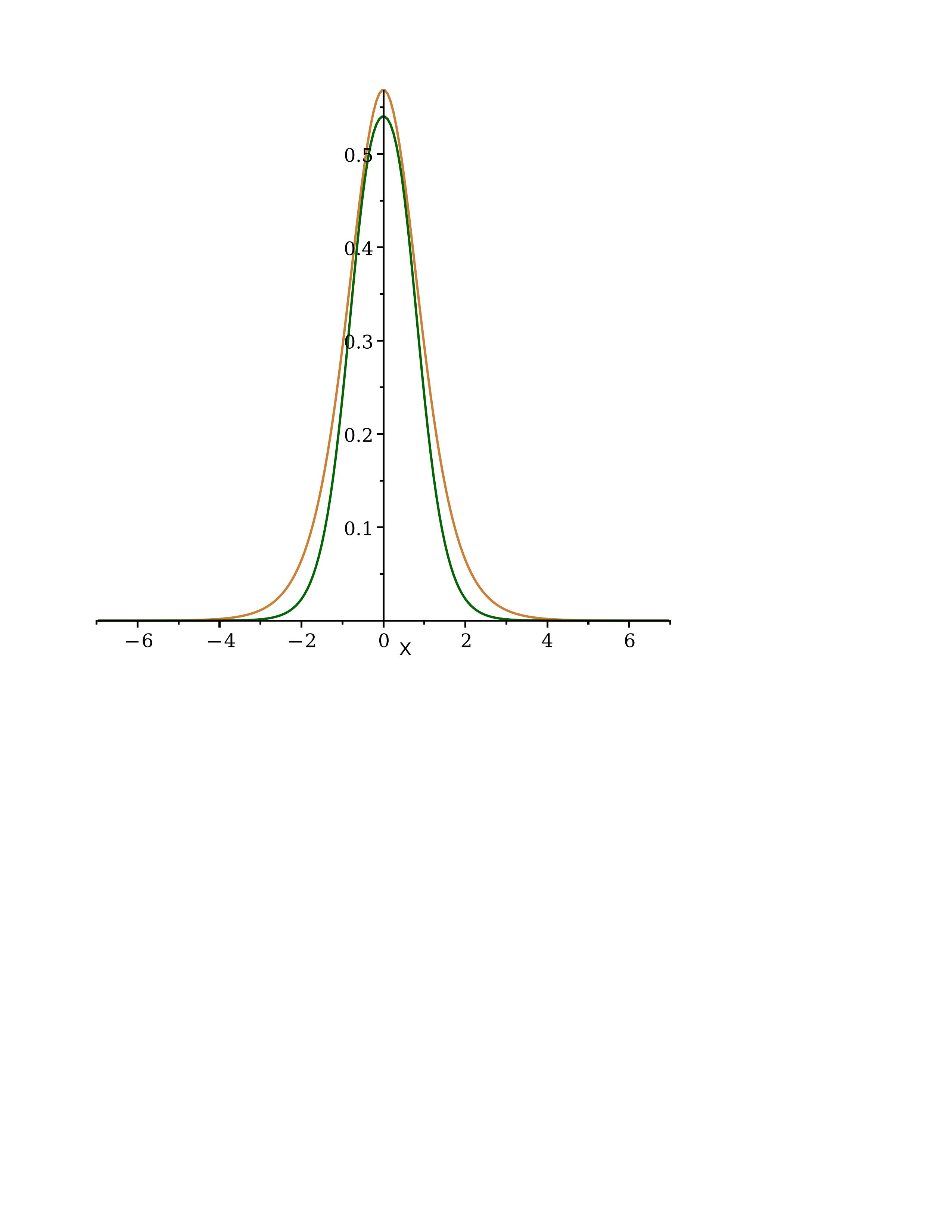}
\caption{
$b=0.7$ ($a=0.616$):
\quad
(Left)\ $\phi$; 
\quad
(Right)\ $\phi_t$.
\quad
Dark green represents the initial data of the sphaleron; 
Brown represents the series solution.
}
\label{fig:match.b=0.7}
\end{figure}

We find $T=-1.835$ and $\tzero=1.617$.
Fig.~\ref{fig:match.b=0.7} shows the matching to initial data \eqref{perturb.lump.initialdata.kick}, with $\epsilon =0.55$ and ($\varepsilon=0.55$).

\begin{figure}
\includegraphics[width=0.50\textwidth,trim=1cm 8cm 1cm 1cm, clip]{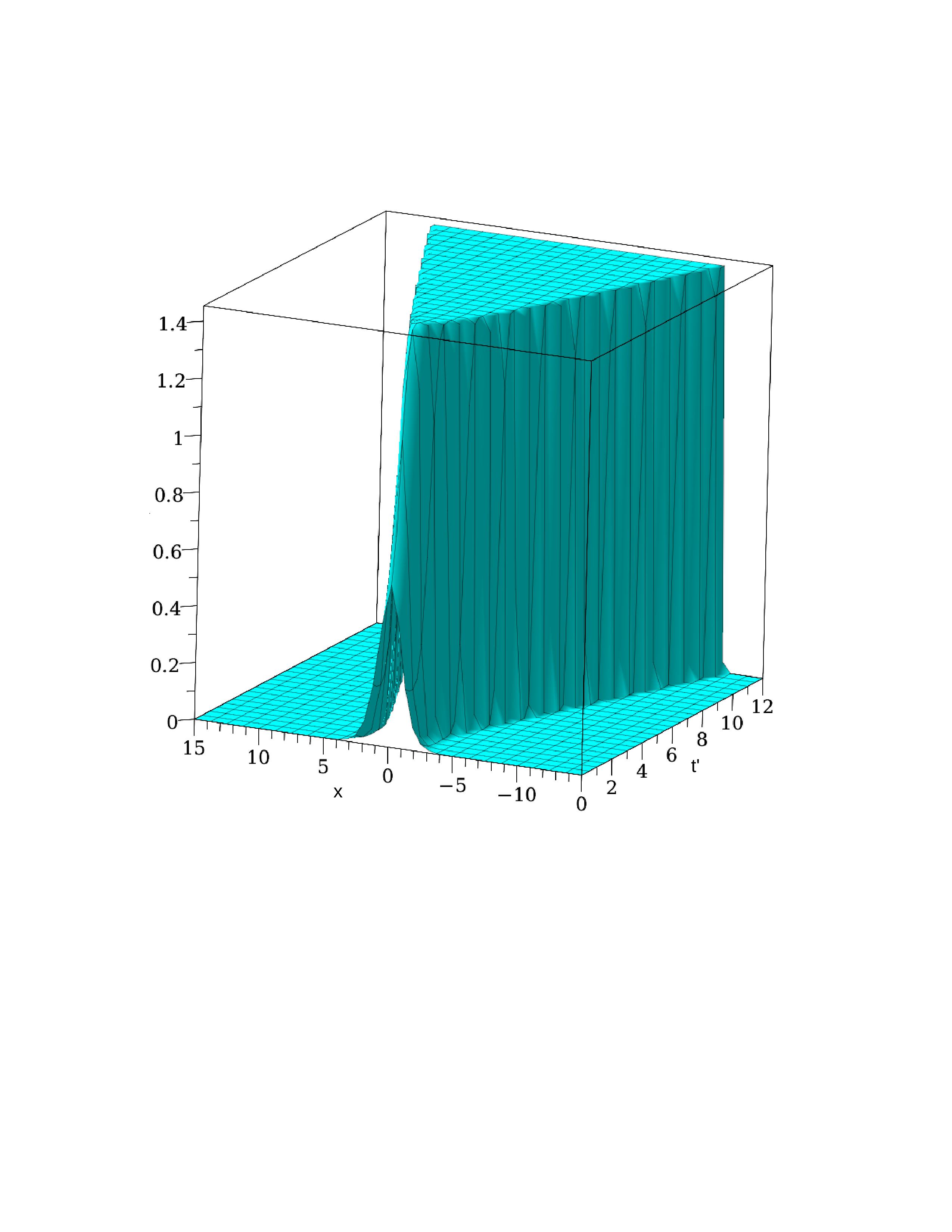}
\includegraphics[width=0.45\textwidth,trim=2cm 10cm 2cm 3cm, clip]{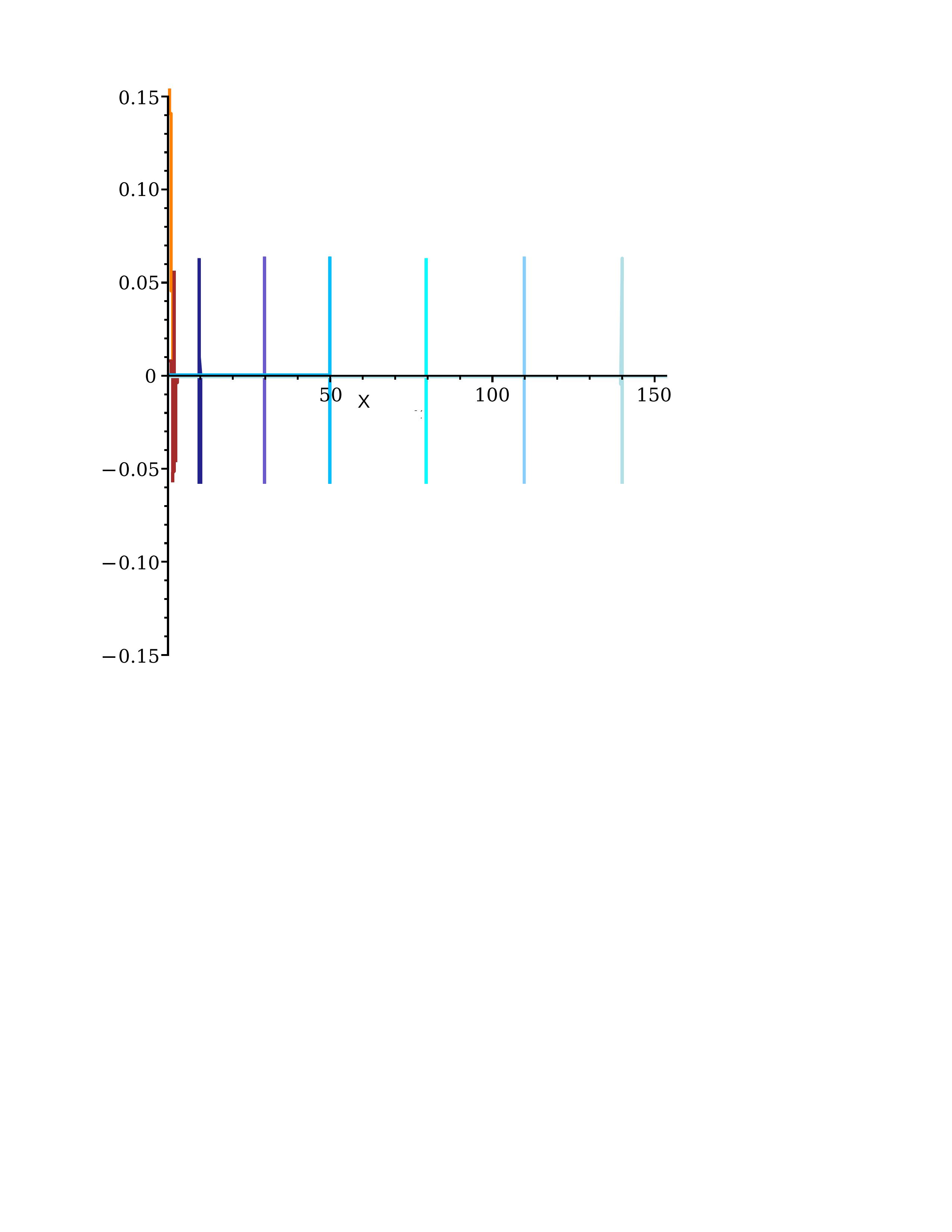}
\caption{$b=0.7$ ($a=0.616$)}
\label{fig:soln.b=0.7}
\end{figure}

\begin{figure}
\includegraphics[width=0.48\textwidth,trim=2cm 12cm 2cm 1cm,clip]{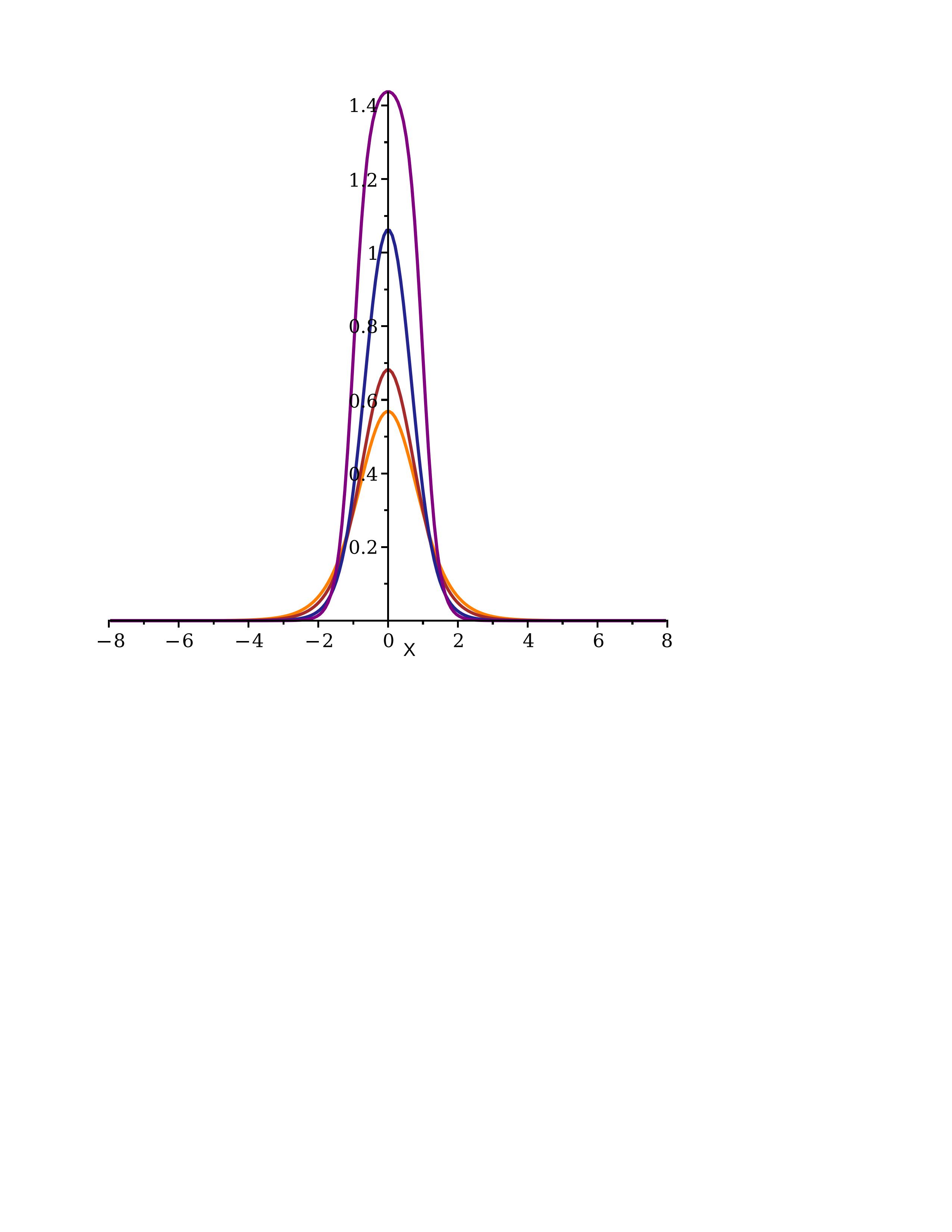}
\includegraphics[width=0.48\textwidth,trim=2cm 12cm 2cm 1cm,clip]{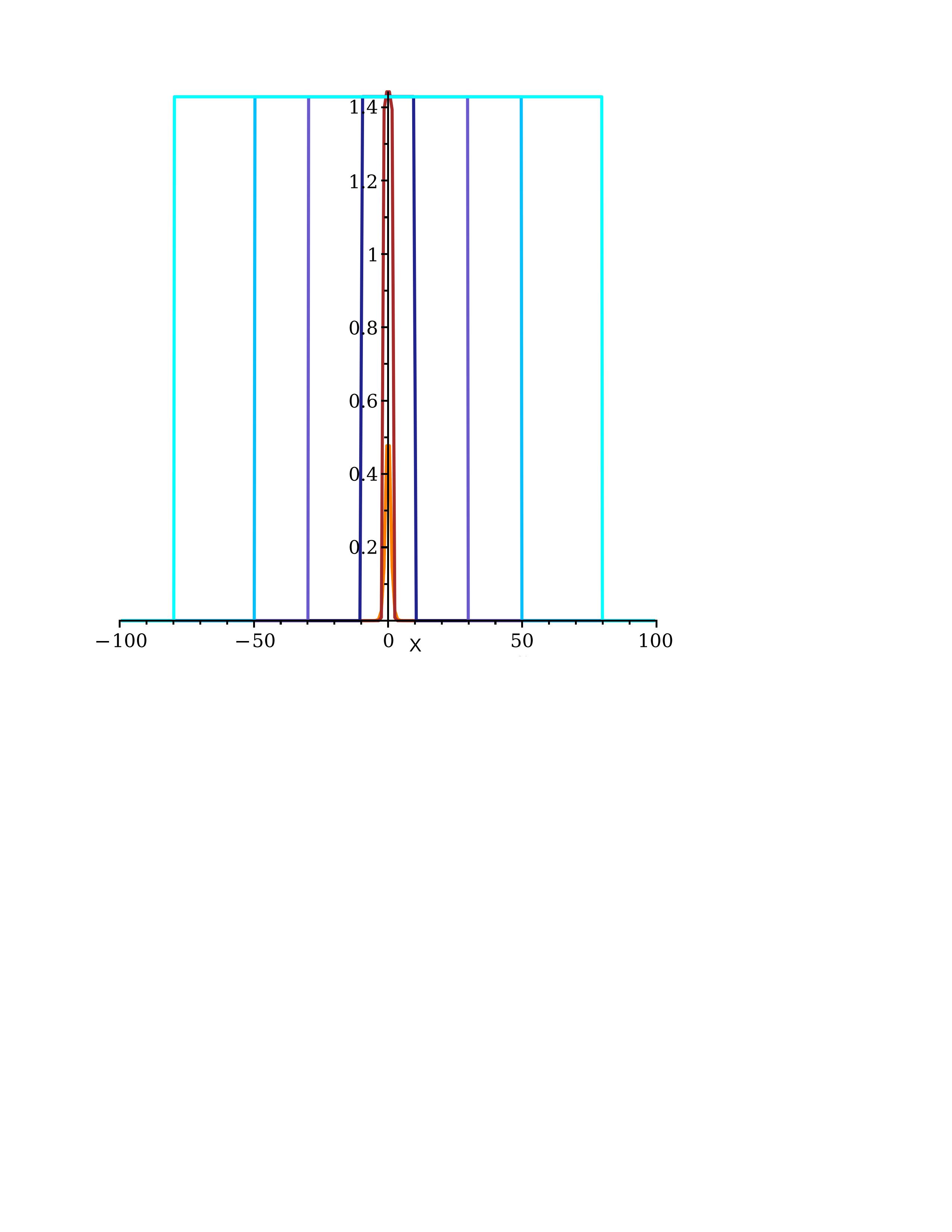}
\caption{(Left)\quad Short times $t'=0,0.2,0.5,1.0$; \quad
(Right) Long times $t'=0,2,10,30,50,80$}
\label{fig:shortlong.time.b=0.7}
\end{figure}

The perturbed lump solution and its error are plotted in Fig.~\ref{fig:soln.b=0.7}.
Notice that the error is roughly constant $7\%$.
Profiles of this solution at short times $t'=0,0.2,0.5,1.0$ (up to peak height)
and long times $t'=0,2,10,30,50,80$ 
are shown in Fig.~\ref{fig:shortlong.time.b=0.7}.

\subsubsection{Example $b=0.5$ ($a=0.402$)}
\begin{figure}
\includegraphics[width=0.48\textwidth,trim=2cm 12cm 2cm 1cm,clip]{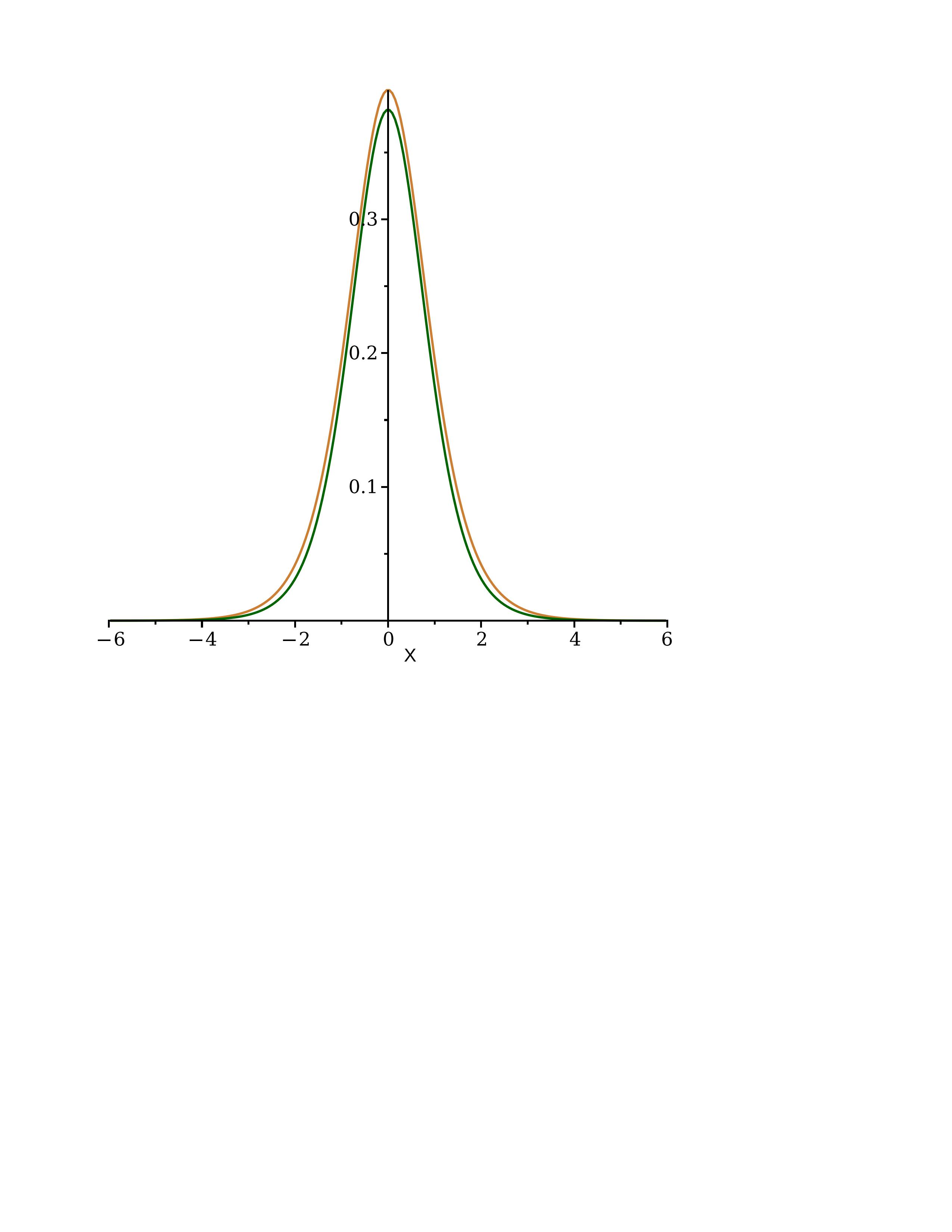}
\includegraphics[width=0.48\textwidth,trim=2cm 12cm 2cm 1cm,clip]{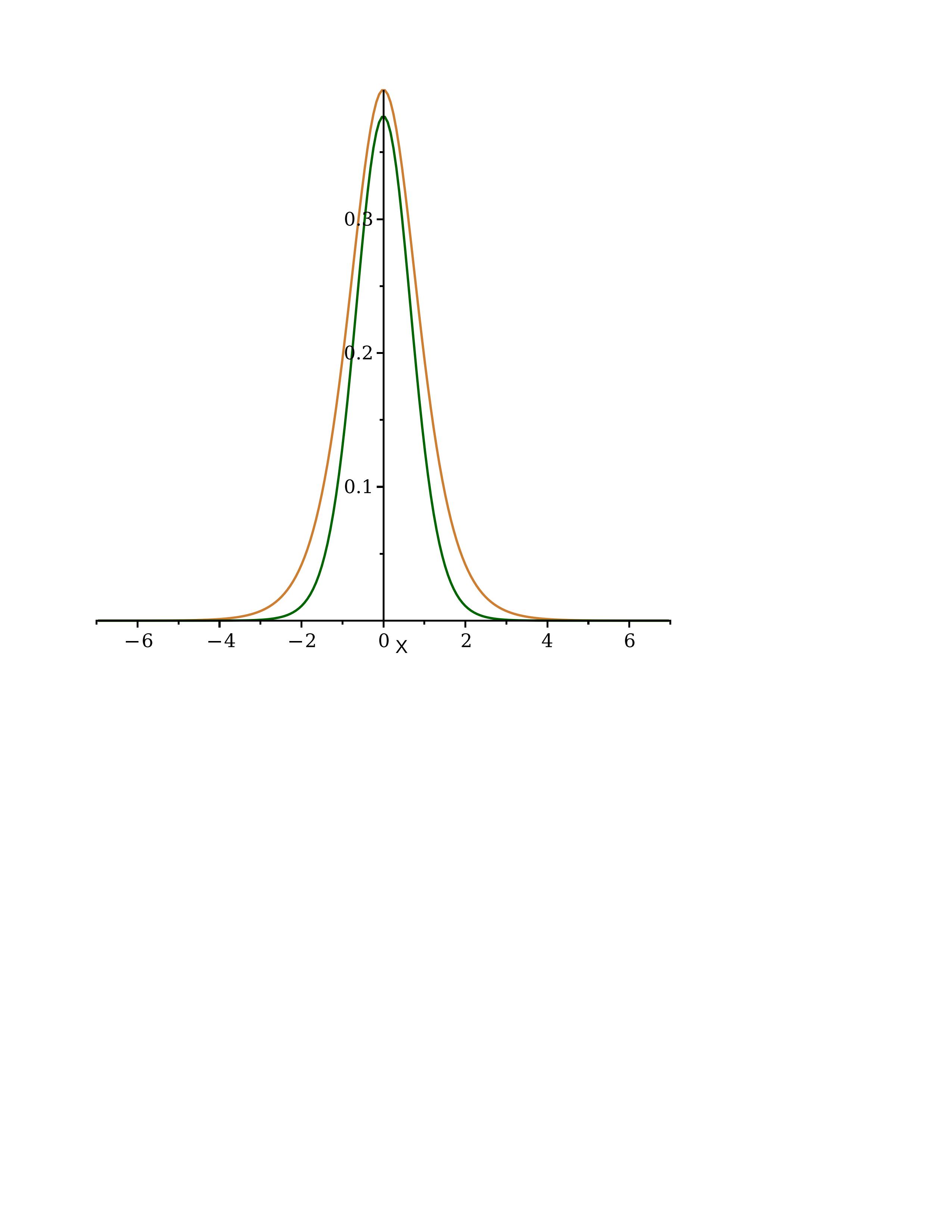}
\caption{
$b=0.5$ ($a=0.402$):
\quad
(Left)\ $\phi$; 
\quad
(Right)\ $\phi_t$.
\quad
Dark green represents the initial data of the sphaleron; 
Brown represents the series solution.
}
\label{fig:match.b=0.5}
\end{figure}

We find $T=-1.711$ and $\tzero=1.8$.
Fig.~\ref{fig:match.b=0.5} shows the matching to initial data \eqref{perturb.lump.initialdata.kick}, with $\epsilon =0.45$ and ($\varepsilon=0.1$).

\begin{figure}
\includegraphics[width=0.50\textwidth,trim=1cm 8cm 1cm 1cm, clip]{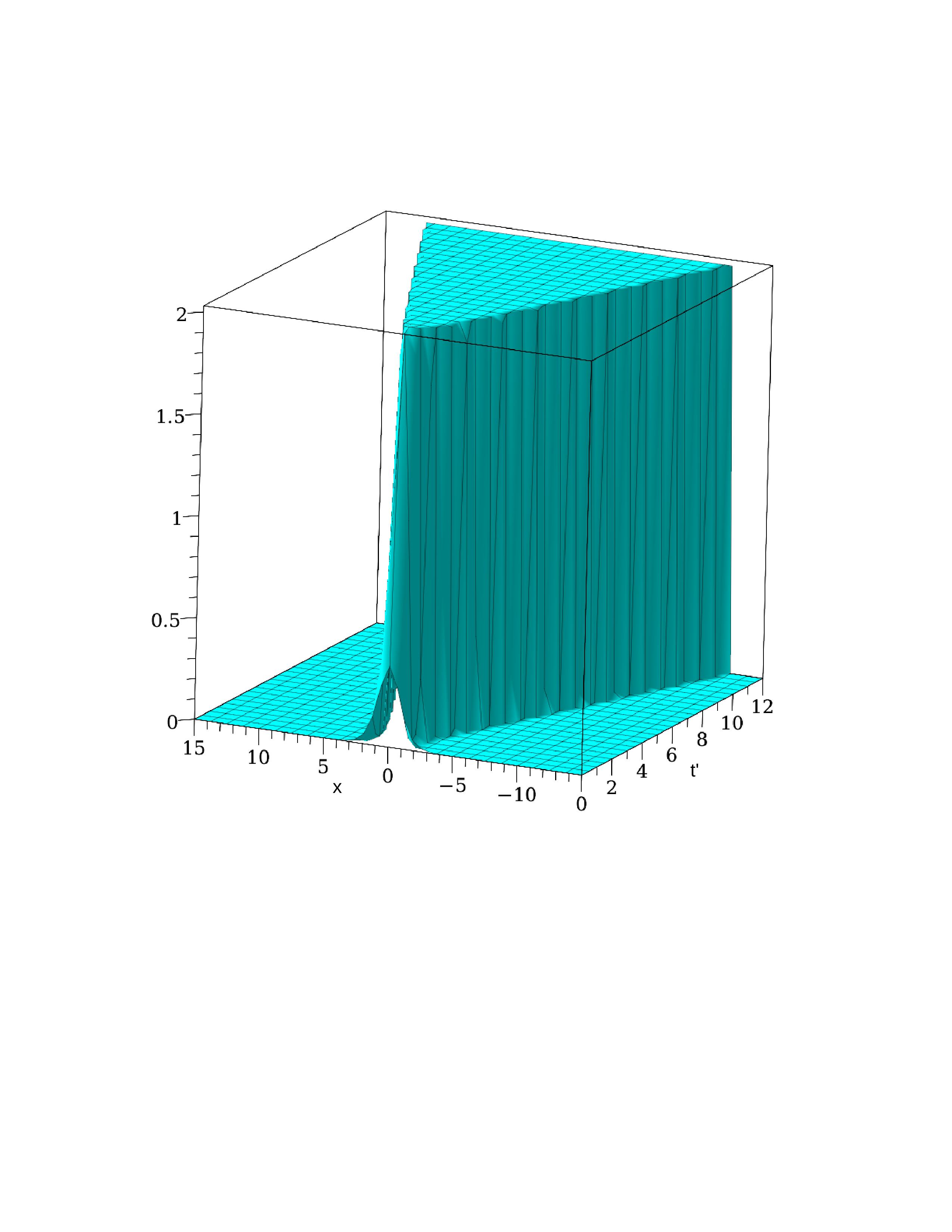}
\includegraphics[width=0.45\textwidth,trim=2cm 10cm 2cm 2cm, clip]{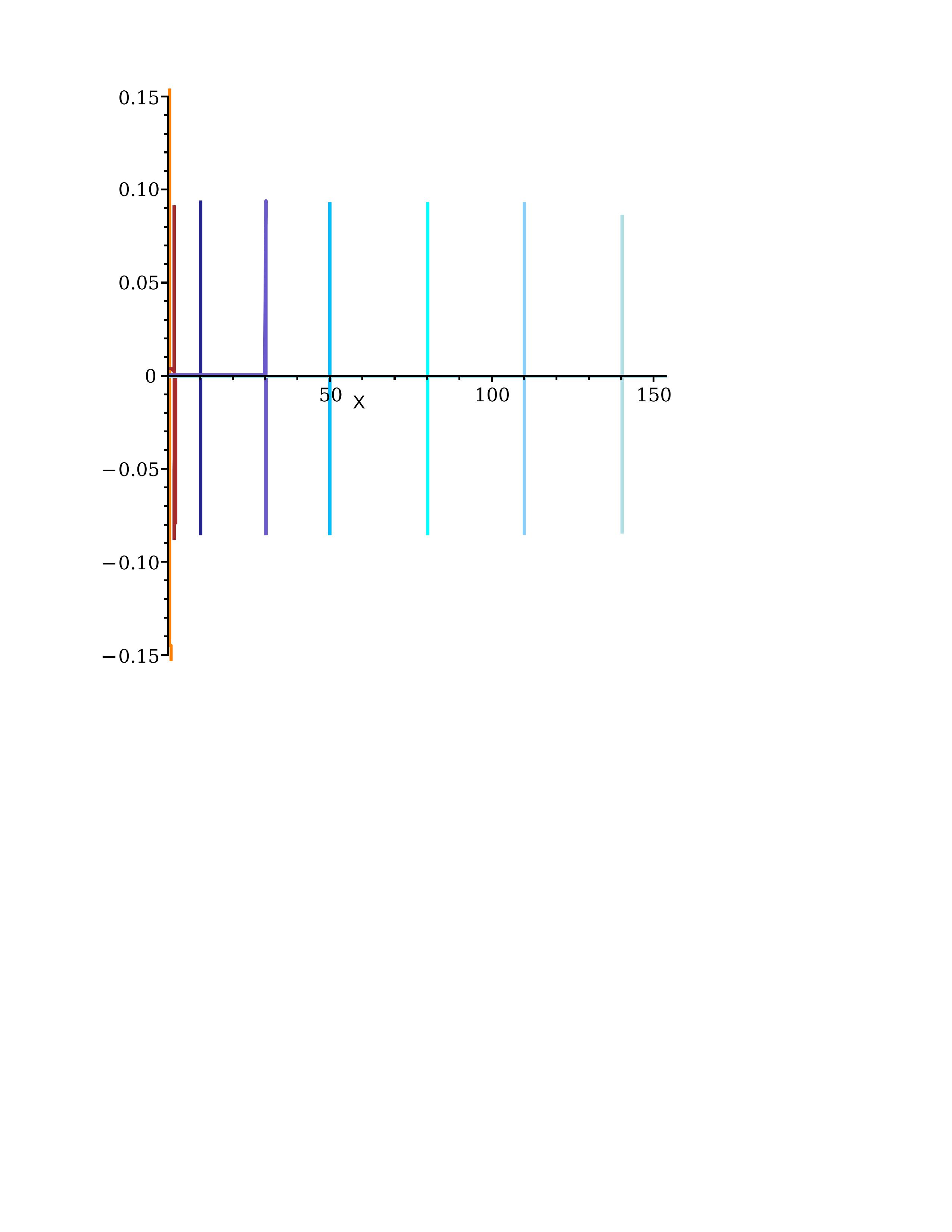}
\caption{$b=0.5$ ($a=0.402$)}
\label{fig:soln.b=0.5}
\end{figure}

\begin{figure}
\includegraphics[width=0.48\textwidth,trim=2cm 12cm 2cm 2cm,clip]{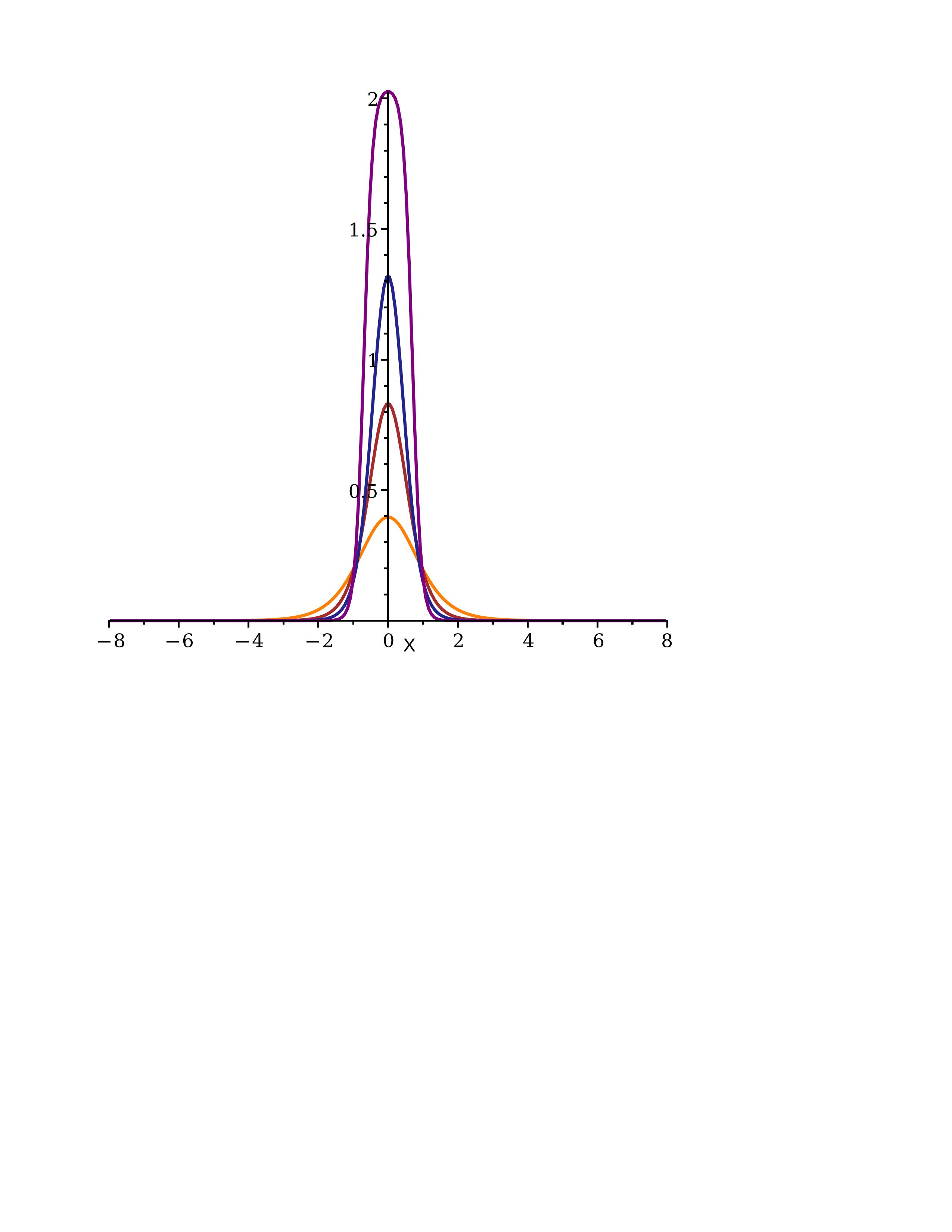}
\includegraphics[width=0.48\textwidth,trim=2cm 12cm 2cm 1cm,clip]{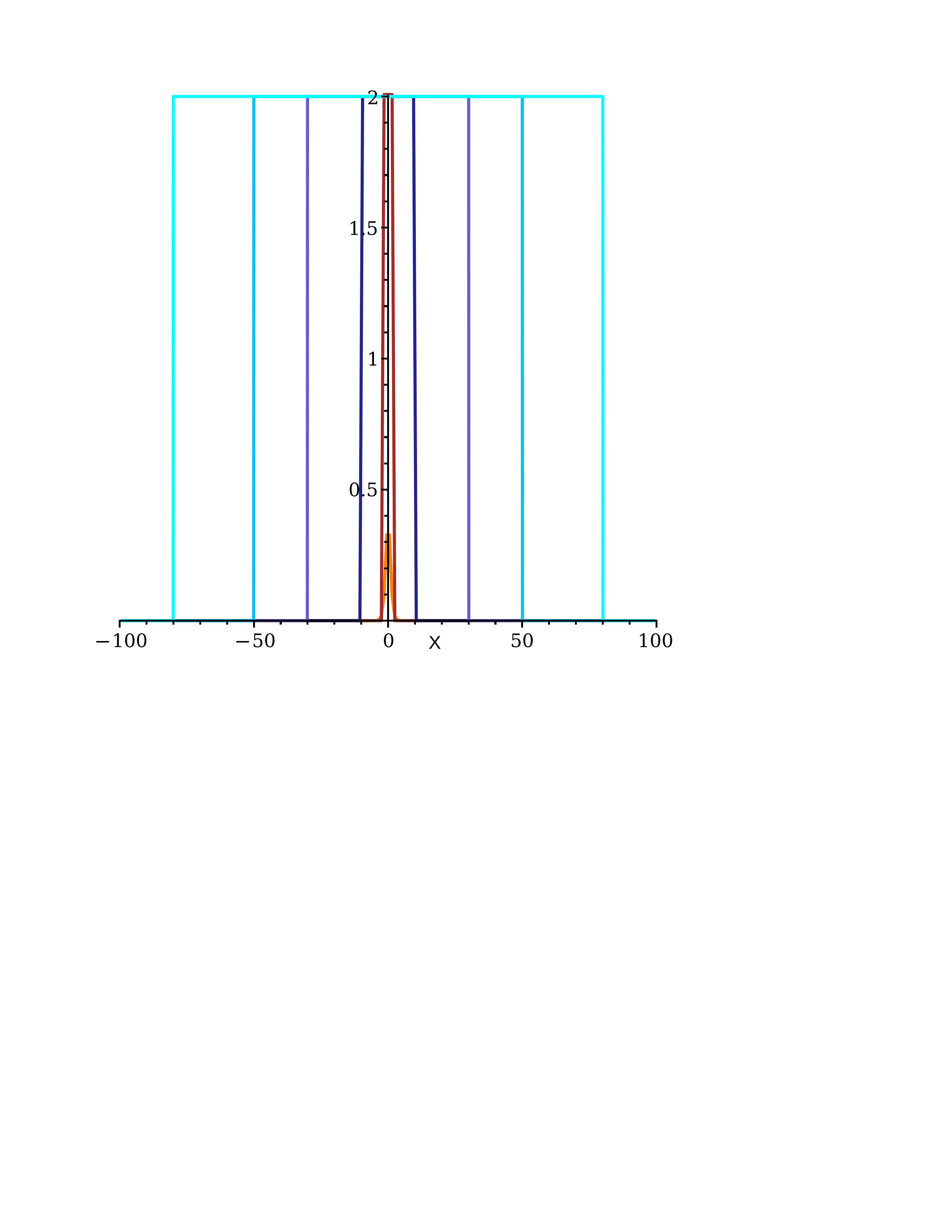}
\caption{(Left)\quad Short times $t'=0,0.1,0.2,0.5$; \quad
(Right) Long times $t'=0,2,10,30,50,80$}
\label{fig:shortlong.time.b=0.5}
\end{figure}

The perturbed lump solution and its error are plotted in Fig.~\ref{fig:soln.b=0.5}.
Notice that the error is roughly constant $9\%$.
Profiles of this solution at short times $t'=0,0.1,0.2,0.5$ (up to peak height)
and long times $t'=0,2,10,30,50,80$ 
are shown in Fig.~\ref{fig:shortlong.time.b=0.5}.

\subsection{Concentration of energy}

\begin{figure}
\includegraphics[width=0.48\textwidth,trim=2cm 12cm 2cm 1cm,clip]{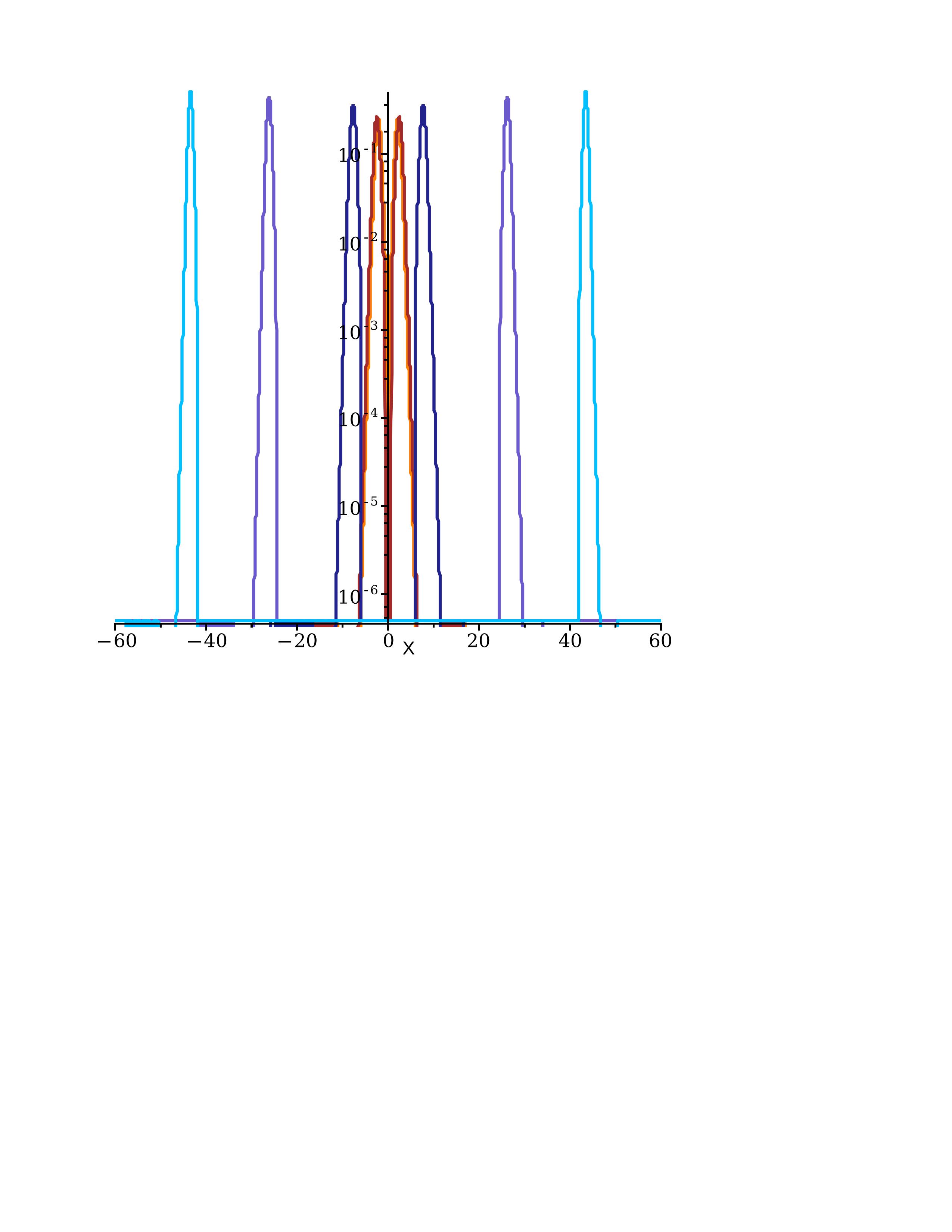}
\includegraphics[width=0.48\textwidth,trim=2cm 12cm 2cm 1cm,clip]{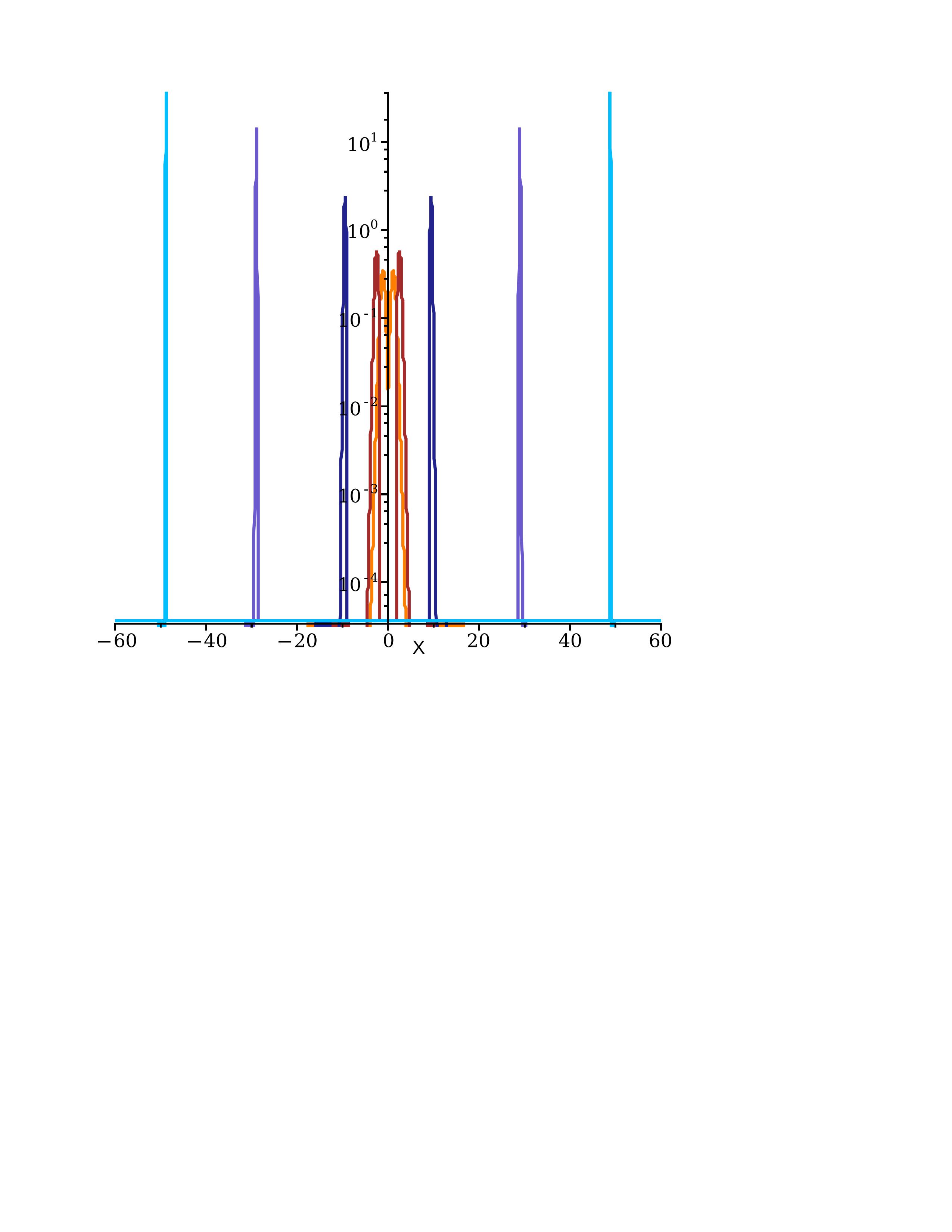}
\caption{Energy density $\ln\mathcal{E}$ for $b=0.998$ (Left) and $b=0.95$ (Right)}
\label{fig:ener.b=0.998.b=0.95}
\end{figure}

\begin{figure}
  \includegraphics[width=0.48\textwidth,trim=2cm 12cm 2cm 1cm,clip]{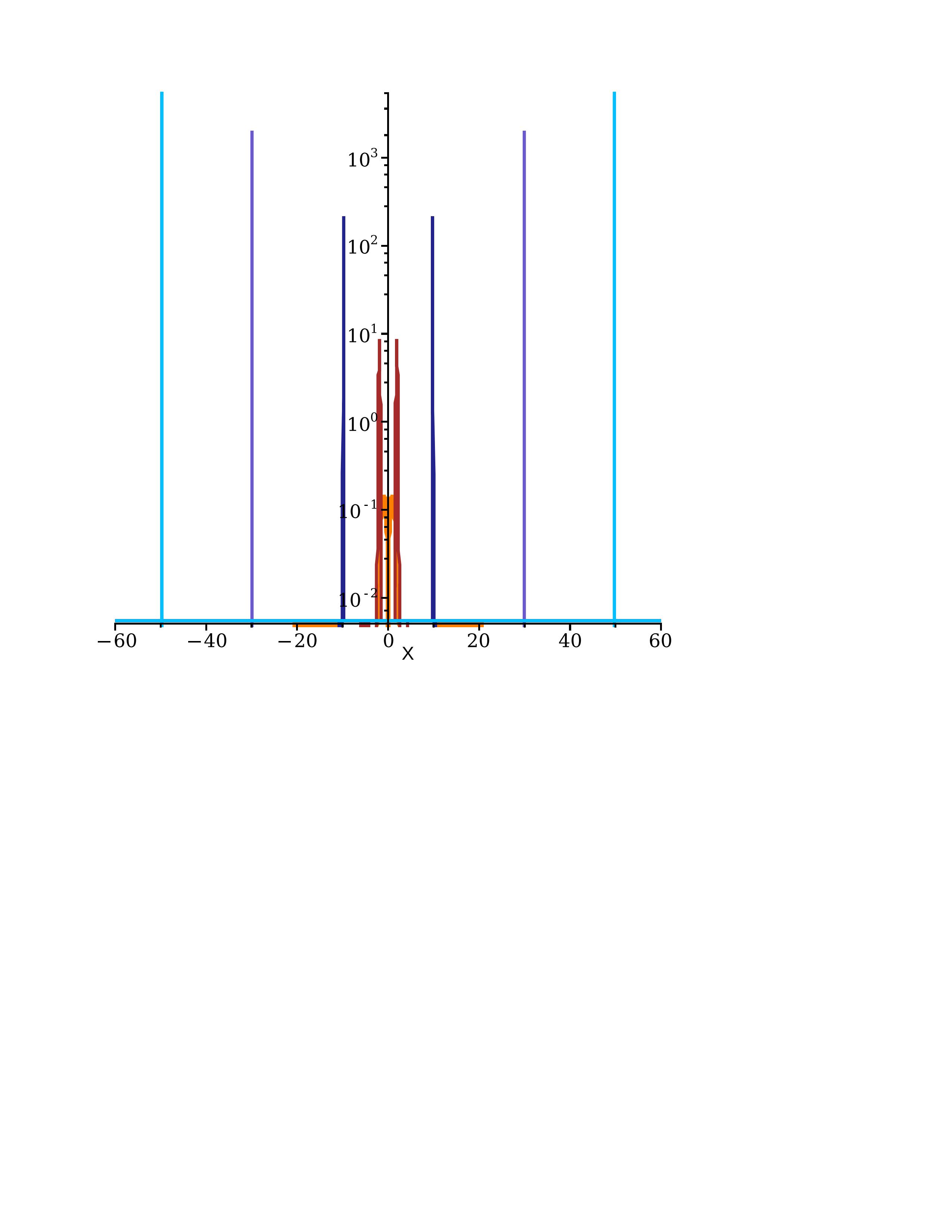}
  \includegraphics[width=0.48\textwidth,trim=2cm 12cm 2cm 1cm,clip]{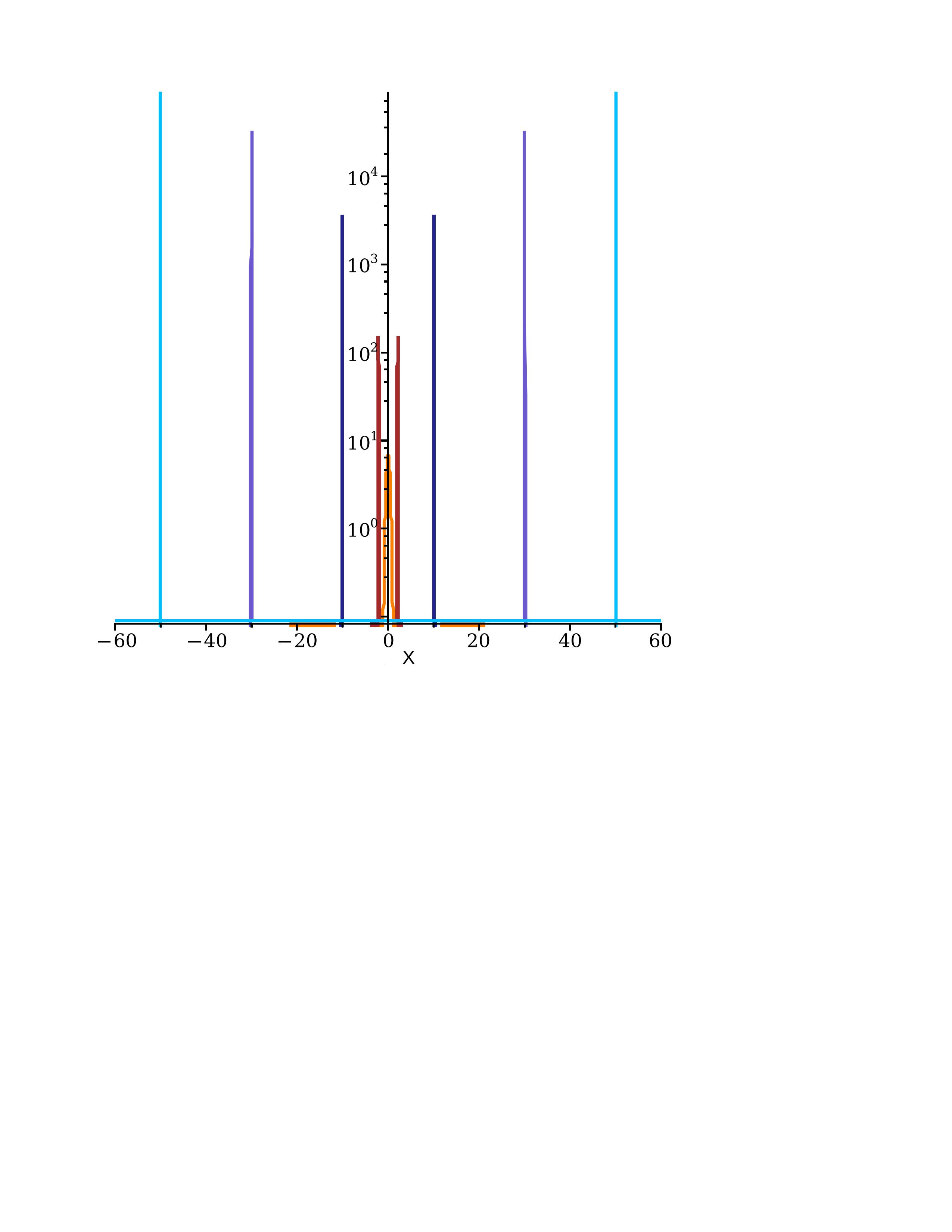}
\caption{Energy density $\ln\mathcal{E}$ for $b=0.7$ (Left) and $b=0.5$ (Right)}
\label{fig:ener.b=0.7.b=0.5}
\end{figure}

Plots of the conserved energy density on a logarithmic scale, $\ln\mathcal{E}$,
at times $t'=0,2,10,30,50$ are provided
in Figs.~\ref{fig:ener.b=0.998.b=0.95} and ~\ref{fig:ener.b=0.7.b=0.5}.
The energy density quickly concentrates at the position of the flanks. 

\section{Discussion and concluding remarks}\label{sec:remarks}

The long-time evolution of sphalerons has been studied
in a general quartic Klein-Gordon model with a false vacuum. 
Several new results are obtained. 

\begin{itemize}
\item
Numerical evolution of sphalerons under a growing perturbation is shown to yield
an accelerating kink-antikink pair 
whose height is equal to the value of the true vacuum 
and whose width increasingly expands. 
\item
An analytical approximation is derived by 
a nonlinear collective coordinate modulation of the sphaleron with three parametric functions. 
\item
The approximation for long times asymptotically
has the same features as the numerical evolution,
and indicates that the energy density concentrates at the flanks 
as they accelerate to approach the light-cone speed.
\item
At early times,
the approximation is close to the initial evolution of the sphaleron
where the perturbation is given by an initial kick. 
\end{itemize}

Mathematically, the approximate analytical solution
exhibits a gradient blow up for long times
and thus belongs to energy space but is not in $H^1$.

Previous studies of sphalerons in $\phi^4$ models with a false vacuum
have primarily focused on their existence and instability, 
as well as qualitatively identifying two different channels for evolution of the instability
\cite{Ave.Baz.Los.Men,Man-review,Nav-Obr.Que}.
There has been no analytical work done on the channel describing
formation of an expanding region of true vacuum and its detailed properties,
specifically the accelerating separation of the kink–antikink pair
and the concentration of energy at the flanks.
The present work fills this gap and provides 
a new collective coordinate approach to studying the dynamics. 

For future investigation,
it would be of particular interest to derive an analytical approximation
for the large-time behaviour of the sphaleron in the other a channel where the perturbation
leads to a collapse into a long-lived oscillon \cite{Man.Rom.2023}. 

The fact that a perturbed sphaleron solution can evolve into two different final states 
naturally leads to the question of what is the outcome of a collision between two sphalerons. 
In a subsequent paper,
the scattering of two sphalerons with a false vacuum will be studied.

\appendix

\section{Derivations}
\label{appendix:derivations}

\subsection{Effective action}\label{appendix:action}

To evaluate the nonlinear KG action principle \eqref{KG.action.potential}
for the collective coordinate profile \eqref{ansatz},
we first change the integration variable $x = \xi/B$ for convenience.
This yields 
\begin{equation}\label{S.moduli}
S[Z] = \int_{-\infty}^{\infty}\big( {-}\tfrac{1}{2} \dot Z\, g(Z)\, \dot Z^\t + V_\text{dyn}(Z)  \big)\,dt
\end{equation}
with $Z=(A(t),B(t),C(t))$ denoting the variables in the profile,
where 
\begin{equation}
g(Z) = \int_{-\infty}^{\infty} \partial_Z \phi \partial_Z \phi \,d\xi
\end{equation}
comprises the 3x3 metric, 
and
\begin{equation}
V_\text{dyn}(Z) = \int_{-\infty}^{\infty} \big( \tfrac{1}{2} (\partial_\xi \phi)^2 + V(\phi) \big)\,d\xi
\end{equation}
is the dynamical potential. 
Both the metric and the  potential are given by 
integrals involving combinations of powers of $\tanh(\xi \pm C)$ functions.
In particular, we have 
\begin{equation}
g_{AA} = \frac{1}{4B} J_1,
\
g_{BB} = \frac{A^2}{B^3} J_9,
\
g_{CC} = \frac{A^2}{4B} J_6,
\
g_{AB} = -\frac{A}{4B^2} J_7,
\
g_{AC} = \frac{A}{4B} J_3, 
\
g_{BC} = \frac{A^2}{B^2} J_8 , 
\end{equation}
and
\begin{equation}
V_\text{dyn.} = 
\frac{A^2}{2B} J_1 
- \frac{A^3}{2B}\coth(2 a) J_2
+\frac{A^4}{8B} J_4
+\frac{A^2 B}{8} J_5 , 
\end{equation}
where
\begin{subequations}
\begin{align}
J_1 & =   \int_{-\infty}^{\infty}\big(\tanh(\xi+C) -\tanh(\xi-C)\big)^2\,d\xi
= 8 C \coth(2 C) - 4 , 
\\
J_2 & = \int_{-\infty}^{\infty}\big(\tanh(\xi+C) -\tanh(\xi-C)\big)^3\,d\xi
=  8 C (3\coth(2 C)^2  -1) -12 \coth(2 C) , 
\\
J_3 & = \int_{-\infty}^{\infty}\big(\tanh(\xi+C) -\tanh(\xi-C)\big)\big(2- \tanh(\xi+C)^2 - \tanh(\xi-C)^2\big)\,d\xi
\nonumber\\& 
= 8 C(1 -\coth(2 C)^2) + 4 \coth(2 C) , 
\\
J_4 & = \int_{-\infty}^{\infty}\big(\tanh(\xi+C) -\tanh(\xi-C)\big)^4\,d\xi
\nonumber\\& 
= 16 C \coth(C) (5\coth(2 C)^2 -3)  -40 \coth(2 C)^2 +\tfrac{32}{3} , 
\\
J_5  & =  \int_{-\infty}^{\infty}\big(\tanh(\xi+C)^2 -\tanh(\xi-C)^2\big)^2\,d\xi
\nonumber\\& 
= 16 C \coth(2 C) (1 -\coth(2 C)^2)  + 8 \coth(2 C)^2  -\tfrac{16}{3}  , 
\\
J_6 & = \int_{-\infty}^{\infty} \big(2- \tanh(\xi+C)^2 - \tanh(\xi-C)^2\big)^2\,d\xi
\nonumber\\& 
= 16 C \coth(2 C) (\coth(2 C)^2 - 1) -8 \coth(2 C)^2 +\tfrac{32}{3} , 
\\
J_7 & = \int_{-\infty}^{\infty}\big(\tanh(\xi+C)-\tanh(\xi-C)\big)^2\big(\tanh(\xi+C) +\tanh(\xi-C)\big)\xi\,d\xi
\nonumber\\& 
= 4 C \coth(2 C) -2 , 
\\
J_8  & = \int_{-\infty}^{\infty}\big( \tanh(\xi+C)^2(\tanh(\xi +C)^2 -2) -\tanh(\xi-C)^2(\tanh(\xi -C)^2 -2)\big)^2\xi\,d\xi
\nonumber\\& 
= -\tfrac{8}{3} C , 
\\
J_9 & = \int_{-\infty}^{\infty}\big(\tanh(\xi+C)^2 -\tanh(\xi-C)^2\big)^2\xi^2\,d\xi
\nonumber\\&
= \tfrac{4}{3} (\pi^2  + 4 C^2)C \coth(2 C)( 1 -\coth(2 C)^2) 
+\tfrac{2}{9}(\pi^2 + 12 C^2) (3\coth(2 C)^2 -2)
-\tfrac{4}{3} . 
\end{align}
\end{subequations}
Thus, the effective action principle is explicitly given by 
\begin{equation}\label{S.A.B.C}
\begin{aligned}
S[A,B,C] = 
\int_{-\infty}^\infty \bigg( &
\Big(\frac{A^2}{2B} +\frac{A'^2}{8B}\Big) J_1
- \frac{A^3}{2B}\coth(2 a) J_2
+\frac{A^4}{8B} J_4
+\frac{A^2B}{8} J_5
-\frac{A A'B'}{4B^2} J_7
\\&\qquad
+ \frac{A A' C'}{4B} J_3
+\frac{A^2 C'^2}{8B} J_6
+\frac{A^2 B'C'}{B^2} J_8
+\frac{A^2 B'^2}{2B^3} J_9
\bigg)\,dt . 
\end{aligned}
\end{equation}

\subsection{Variational equations}\label{appendix:eom}

We first substitute the relation \eqref{C.BX.rel} into the action principle \eqref{S.A.B.C}
to get $\tilde S[A,B,X] = S[A,B,BX]$, 
and then we take the variation derivative of $\tilde S[A,B,X]$ 
with respect to the variables $A$, $B$, $X$.
This yields the variational equations
\begin{align}
& \begin{aligned}
\frac{\delta \tilde S}{\delta A} = &
\Big(
4 A B^2 X (X'^2 - 1)
+ 8 A B X^2 B' X'
+ \tfrac{16}{3}  A X^3 B'^2 + \tfrac{1}{3}\pi^2 A X B'^2 /B^2
+ 5 \alpha A^3 X
\Big)
\coth(2 B X)^3
\\&\quad
+ \Big(
2 A B (1 - X X'' - X'^2)  
- 4 B X A' X'
- \tfrac{3}{2} \alpha (a^2 + 2) A^2 X 
-2 A X (X B'' +2 B' X') 
\\&\quad
- 4 X^2 A' B'
- \tfrac{5}{2} \alpha A^3/B
- \tfrac{1}{6} \pi^2 A B'^2/B^3
\Big)
\coth(2 B X)^2
+ \Big(
4 A B^2 X (1 - X'^2)
\\&\quad
- 8 A B X^2 B' X' 
- 3 \alpha A^3 X
+\alpha a^2 A X
-\tfrac{16}{3} A X^3 B'^2
+ A X'' + 2 (X A')' 
\\&\quad
+ \tfrac{3}{4}  \alpha (a^2 + 2) A^2/B 
- \tfrac{1}{3} \pi^2 A X B'^2/ B^2
\Big)
\coth(2 B X)
+ A B \big( \tfrac{4}{3}(X'^2 -1) + 2 X X'' \big)
\\&\quad
+ 4 B X A' X'
+ 4 X^2 A' B'
+  \tfrac{1}{2} \alpha (a^2 + 2) A^2 X
+ 2 A X (X B'' + 2 B' X')
\\&\quad
+ (\tfrac{2}{3} \alpha A^3 - \tfrac{1}{2} \alpha a^2 A - A'')/B
+ (A' B' + \tfrac{1}{2} A B'')/B^2
+ \tfrac{1}{9}(\pi^2 - 6) A B'^2/B^3
=0 , 
\end{aligned}  
\\
& \begin{aligned}
\frac{\delta \tilde S}{\delta B} = &
\Big(
12 A^2 B^2 X^2 (1 -X'^2) 
- 16 A^2 B X^3 B' X' 
- 8 A^2 X^4 B'^2
- \tfrac{15}{2} \alpha A^4 X^2 
+ 2 \pi^2 A^2 X X'/B
\\&\quad
+ \pi^2 A^2 X^2 B'^2/B^2
\Big)
\coth(2 B X)^4
+ \Big(
4 A^2 B X \big( X X'' + 2 (X'^2 - 1) \big)
\\&\quad
+8 A^2 X^2 (B' X' + \tfrac{1}{3} X B'')
+ 8 A B X^2 A' X'
+ \tfrac{16}{3} A X^3 A' B' 
+ 2 \alpha (a^2 + 2) A^3 X^2 
\\&\quad
+ \tfrac{5}{2} \alpha A^4 X/B
- \pi^2 A^2 (\tfrac{1}{3} X B'' - B' X')
- \tfrac{2}{3} \pi^2 A X B' A'/B^2
\Big)
\coth(2 B X)^3
\\&\quad
+ \Big(
16 A^2 B^2 X^2 (X'^2 - 1) 
-2 A X (A'' X +2 A' X')
+\tfrac{64}{3} A^2 B X^3 B' X' 
\\&\quad
+ A^2 \big( \tfrac{32}{3} X^4 B'^2 - \alpha a^2 X^2 - 2 X'' X - X'^2 + 1 \big) 
+ 9 \alpha A^4 X^2
- \tfrac{1}{2} \alpha (a^2 + 2) A^3 X/B
\\&\quad
- \tfrac{8}{3} \pi^2 A^2 X B' X'/B
+ A^2\big( \tfrac{5}{8} \alpha A^2 - \tfrac{4}{3} \pi^2 X^2 B'^2 \big)/B^2
+ \tfrac{1}{6} \pi^2 A (2 A' B' + A B'')/B^3
\\&\quad
- \tfrac{1}{4} \pi^2 A^2 B'^2/B^4
\Big)
\coth(2 B X)^2
-\Big(
8 A^2 X^2 (B' X' + \tfrac{1}{3} X B'')
+ \tfrac{16}{3} A X^3 A' B'
\\&\quad
+ 4 B \big( A^2 (X^2 X'' + 2X (X'^2 -1)) +2 A X^2 A' X' \big)
+ 2 \alpha (a^2 + 2) A^3 X^2
+ \tfrac{5}{2} \alpha A^4 X /B
\\&\quad
+ \big( \tfrac{1}{4}\alpha (a^2 + 2)  A^3 - \tfrac{1}{3} \pi^2 A^2 (XB'' + 3 B' X' ) 
- \tfrac{2}{3} \pi^2 A X A' B' \big)/B^2
\Big)
\coth(2 B X)
\\&\quad
-A^2 B^2 X^2 (X'^2 -1)
- \tfrac{16}{3} A^2 B X^3 B' X'
+ 2A X (X A'' + 2 A' X') 
\\&\quad
+ A^2 \big({-}\tfrac{8}{3} X^4 B'^2 + \alpha ( a^2 - \tfrac{3}{2} A^2) X^2
+ 2 X'' X + \tfrac{2}{3}(X'^2 -1)  \big)
\\&\quad
+ \big( \tfrac{1}{2} \alpha (a^2 + 2) A^3 X 
+ \tfrac{2}{3} \pi^2 A^2 X B' X' \big)/B
+ \big( \tfrac{1}{3} \pi^2 A^2 X^2 B'^2 
+ \alpha (\tfrac{1}{4} a^2 - \tfrac{1}{6} A^2) A^2
\\&\quad
+ \tfrac{1}{2} A A'' \big)/B^2
- \tfrac{1}{9}(\pi^2 + 3) A \big( A B'' + 2 A' B' \big)/B^3
+ \tfrac{1}{6} (\pi^2 + 3) A^2 B'^2/B^4
=0 , 
\end{aligned}  
\\
& \begin{aligned}
\frac{\delta \tilde S}{\delta X} = &
\Big(
12 A^2 B^3 X (1 - X'^2) - 24 A^2 B^2 X^2 B' X' 
-A^2 B X \big( \tfrac{15}{2} \alpha A^2 + 16 X^2 B'^2 \big)
\\&\quad
- \pi^2 A^2 X B'^2/B
\Big)
\coth(2 B X)^4
+ \Big(
A^2  B^2\big( 4 X X'' + 6 (X'^2 - 1) \big)
+ 8  A  B^2 X A' X' 
\\&\quad
+ 2 \alpha (a^2 + 2) A^3 B X
+ 4 A^2 B X (X B'' + 4 B' X')
+ 8 A B X^2 A' B' 
+ 8 A^2 X^2 B'^2
\\&\quad
+ \tfrac{15}{4} \alpha A^4
+ \tfrac{1}{2}\pi^2 A^2 B'^2/B^2
\Big)
\coth(2 B X)^3
+ \Big(
16 A^2 B^3 X (X'^2 - 1)
+ 32 A^2 B^2 X^2 B' X' 
\\&\quad
+ 9 \alpha A^4 B X + A^2 B (\tfrac{64}{3} X^3 B'^2 - \alpha a^2 X - 2 X'') 
- 2 A B (X A'' +2 A' X')
- \alpha (a^2 + 2) A^3
\\&\quad
-2 A^2 (X B')' 
- 4 A X A' B' + \tfrac{4}{3} \pi^2 A^2X B'^2 /B
\Big)
\coth(2 B X)^2
-\Big(
8 A B^2 X A' X'
\\&\quad
+A^2 B^2 \big( 4 X'' X + 6(X'^2 -1) \big) 
+ B X \big( 2 \alpha (a^2 + 2) A^3 + 4 (X B'' +4 B' X') A^2 
\\&\quad
+ 8 A X A' B' \big)
+\tfrac{13}{4} \alpha A^4
- \tfrac{1}{2} \alpha a^2 A^2 - A A''
+ 8 A^2 X^2 B'^2 
+ \tfrac{1}{2} \pi^2 A^2 B'^2/B^2
\Big)
\coth(2 B X)
\\&\quad
+ 4 A^2 B^3 X (1 -X'^2)
+ 4 A X A' B' 
- 8 A^2 B^2 X^2 B' X' 
+ \big(
2 A X A'' + \tfrac{8}{3} A^2 X'' 
\\&\quad
+ \tfrac{16}{3} A A' X' 
-\tfrac{16}{3} A^2 X^3 B'^2 
+ ( \alpha a^2 -\tfrac{3}{2} \alpha A^2 )A^2 X
\big) B
+ A^2(\tfrac{8}{3} B' X' + 2 X B'') 
\\&\quad
+ \tfrac{2}{3} \alpha  (a^2 + 2) A^3
- \tfrac{1}{3} \pi^2 A^2 X B'^2/ B
=0 . 
\end{aligned}      
\end{align}

These equations are equivalent to the variational equations given by
the variational derivatives of $S[A,B,C]$ with respect to the variables $A$, $B$, $C$.

\section{Energy}
\label{appendix:ener.correction}

We evaluate the nonlinear KG energy \eqref{ener} for the initial data \eqref{perturb.lump.initialdata.kick}
used to perturb the sphaleron. 
The energy can be expressed as the sum $E=E_\lump + E_\text{perturb}$
where $E_\lump$ is the energy of the sphaleron
and $E_\text{perturb}$ is the contribution due to the initial kick \eqref{kick.initialdata}. 

From expression \eqref{lump.E} for the sphaleron energy,
combined with the change of parameterization \eqref{a.b.rel},
we obtain 
\begin{equation}
E_\lump =
\frac{(1-b^2)(2+b^2)(4-b^2)}{(3b)^3} \ln\Big( \frac{(1-b)(2-b)}{(1+b)(2+b)}\Big)
+ \frac{2((1-b^2)^2 + 3)}{(3b)^2}
\end{equation}
Next we evaluate
\begin{equation}\label{ener.pertrub}
E_\text{perturb} = \int_{-\infty}^{\infty} \tfrac{1}{2} \phi_t^2\, dx
= \dfrac{\epsilon^2}{2\tau^2} \tanh(a)^2 \int_{-\infty}^{\infty} \eta_{-1}(x)^2 \, dx
\end{equation}
where $\tau$ and $\eta_{-1}(x)$ are given respectively by expressions \eqref{timescale.approx} and \eqref{groundstate.approx}.

In the case $a\gtrsim 1.00$,
the integral of $\eta_{-1}(x)^2$ can be brought to rational form
by means of the substitution $y= \tanh(x/2)$: 
\begin{equation}
\int_{-\infty}^{\infty} \eta_{-1}(x)^2 \, dx
= -2\int_{-\infty}^{\infty} \frac{(y^2 - 1)^3 (y^4 + 6y^2 + 1)^8}{(y^2 + 1)^{16} (y^4 + 2(2e^{-2a}-1) y^2 + 1)^2}\,dx
\end{equation}
This is straightforward to calculate explicitly,
or it can be approximated by expanding in powers of $e^{-2a}$.
The leading term is found to be
\begin{equation}
\int_{-\infty}^{\infty} \eta_{-1}(x)^2 \, dx
  \simeq -\tfrac{8211328}{15015} + 256(a+ \ln(2)) + O(e^{-2a})
\end{equation}

In the case $a\lesssim 1.00$, 
the integral of $\eta_{-1}(x)^2$ can be split into three terms:
\begin{equation}
\begin{aligned}
\int_{-\infty}^{\infty} \eta_{-1}(x)^2 \, dx
& = \int_{-\infty}^{\infty} \sech(x)^6\big( 1 + 2 a^2 \tanh(x)^2 \big)^2\,dx
+ \tfrac{64}{49} a^4 \int_{-\infty}^{\infty} \sech(x)^6 \ln(\sech(x))^2\,dx
\\&\qquad\qquad
-\tfrac{8}{7} a^2 \int_{-\infty}^{\infty} \sech(x)^6\big( 1 + 2 a^2 \tanh(x)^2 \big) \ln(\sech(x)) \,dx
\end{aligned}
\end{equation}
The first and third integrals can each be explicitly evaluated by simple substitutions,
while the middle integral can be calculated numerically.
This yields
\begin{equation}
\int_{-\infty}^{\infty} \eta_{-1}(x)^2 \, dx
= \tfrac{8}{15}\big(1  +\tfrac{248}{105} a^2 \big)
- \tfrac{128}{105} \ln(2) a^2\big( 1 + \tfrac{2}{7} a^2\big)
+ 0.441 a^4
\end{equation}

\section*{Acknowledgements}

SCA thanks Willy Hereman and Alexandr Chernyavskiy for valuable discussions. 
DS is grateful to the Department of Mathematics \& Statistics at Brock
for support during the initial phase of this work
and thanks David Amundsen for productive discussions that provided valuable insights and ideas. 
Both authors thank Andrzej Wereszczy\'nski for very helpful physics comments
and are indebted to Thomas Wolf for useful remarks on numerical solutions. 

SCA is supported by an NSERC Discovery grant.

\end{document}